\documentclass[runningheads]{llncs}

\usepackage[boxruled,linesnumbered]{algorithm2e}

\newlength{\TMPLENGTH}
\def\hatell{\settowidth{\TMPLENGTH}{$\ell$}\hspace*{.5\TMPLENGTH}\mathclap{\ell}\kern1.4pt\hat{\vphantom{\ell}}\kern-1.4pt\hspace*{.5\TMPLENGTH}}

\def\squareforqedDEF{\hbox{$\blacklozenge$}}
\def\squareforqedEXP{\hbox{$\lozenge$}}
\def\qedDEF{\ifmmode\squareforqedDEF\else{\unskip\nobreak\hfil
\penalty50\hskip1em\null\nobreak\hfil\squareforqedDEF
\parfillskip=0pt\finalhyphendemerits=0\endgraf}\fi}
\def\qedEXP{\ifmmode\squareforqedEXP\else{\unskip\nobreak\hfil
\penalty50\hskip1em\null\nobreak\hfil\squareforqedEXP
\parfillskip=0pt\finalhyphendemerits=0\endgraf}\fi}

\newcommand{\Paragraph}[2]{\par\vspace*{1ex plus .2ex minus .2ex}\noindent\textsl{\textbf{Step #1:\,#2}}\leavevmode\\[1ex plus .2ex minus .2ex]}

\def\deadline{2021-03-05;2021-03-12}
\def\pagelimit{16p(inclusive)}
\def\TOC#1#2{#1}

\usepackage[final]{listings}

\usepackage[most]{tcolorbox}
\tcbuselibrary{listingsutf8}

\usepackage{ifdraft}

\newcommand{\OFFICIALNAME}[1]{\textsc{#1}\xspace}
\newcommand{\TOOL}[1]{\OFFICIALNAME{#1}}

\newcommand{\AUTOGRAPH}{\TOOL{AutoGraph}}
\newcommand{\Zthree}{\TOOL{Z3}}

\usepackage[binary-units=true]{siunitx}
\DeclareSIUnit[]{\RPM}{RPM}

\usepackage{pgf,tikz,pgfplots}
\pgfplotsset{compat=1.16}
\usetikzlibrary{trees}
\usetikzlibrary{decorations}
\usetikzlibrary{arrows}
\usetikzlibrary{automata}
\usetikzlibrary{shadows}
\usetikzlibrary{positioning}
\usetikzlibrary{plotmarks}
\usetikzlibrary{backgrounds}
\usetikzlibrary{shapes}
\usetikzlibrary{calc}
\usetikzlibrary{matrix}
\usetikzlibrary{fit}
\usetikzlibrary{petri}
\usetikzlibrary{patterns}
\usetikzlibrary{arrows.meta}
\usetikzlibrary{decorations.pathreplacing}
\usetikzlibrary{decorations.markings}
\usetikzlibrary{decorations.pathmorphing}
\usetikzlibrary{intersections}
\usetikzlibrary{shapes.multipart}

\tikzset{    
    double line with arrow/.style args={#1,#2}{decorate,decoration={markings,mark=at position 0 with {\coordinate (ta-base-1) at (0,1pt);
    \coordinate (ta-base-2) at (0,-1pt);},
    mark=at position 1 with {\draw[#1] (ta-base-1) -- (0,1pt);
    \draw[#2] (ta-base-2) -- (0,-1pt);
    }}}
}

\newlength\triplesep
\newlength\triplelinewidth
\tikzset{
    triple/.style args={#1}{
        line width=\triplelinewidth,
        black,
        preaction={
            preaction={
                draw,
                line width=2\triplesep+3\triplelinewidth,
                black
            }, 
            draw,
            line width=2\triplesep+\triplelinewidth,
            #1
        }
    }
}

\colorlet{mycustomBGcolor}{red!10!white}
\tikzset{
    fill background/.style={background rectangle/.style={fill=mycustomBGcolor}, show background rectangle},
    >=latex,
    inclusion/.style={right hook->},
    satisfaction/.style={double,-{|[width=3mm]},double equal sign distance},
    epimorphism/.style={->>},
    isomorphism/.style={right hook->>},
    isomorphismSWAP/.style={left hook->>},
    implies/.style={double,double equal sign distance,-implies},
    isom/.style={double,double equal sign distance},
    equalarrow/.style={-,double line with arrow={-,-}},
    steparrow/.style={double,double equal sign distance,-implies},
    inclusionSWAP/.style={left hook->},
    separationA/.style={inner sep=0pt, minimum size=0pt},
    vertex/.style={inner sep=1.5pt, minimum size=1ex,rectangle,draw},
    shorten >=0cm,
    shorten <=0cm,
    vertexR/.style={inner sep=2pt, minimum height=4ex,rectangle,draw},
    vertexRX/.style={inner sep=0pt,inner xsep=3pt, minimum height=0pt,rectangle},
    condition/.style={inner xsep=-2pt,inner ysep=0pt},
}

\pgfdeclarelayer{bg}
\pgfdeclarelayer{bg1}
\pgfdeclarelayer{bg2}
\pgfdeclarelayer{bg3}
\pgfdeclarelayer{bg4}
\pgfsetlayers{bg4,bg3,bg2,bg1,bg,background,main}
\usepackage{tkz-euclide} \usetikzlibrary{quotes}

\newcommand{\AppendixREF}[1]{\hyperref[#1]{App.~\ref*{#1}}}
\newcommand{\SQUEEZE}[2][1pt]{\kern#1{#2}\kern#1}
\newcommand{\marksubgoal}[1]{\text{\ding{\the\numexpr171+#1}}}

\newcommand{\OMIT}[1]{}

\newcommand{\mycentering}[1]{\mbox{}\hfill #1\hfill\mbox{}}
\newenvironment{nscenter}
 {\parskip=0pt\par\nopagebreak\centering}
 {\par\noindent\ignorespacesafterend}

\let\saveLongrightarrow\Longrightarrow

\makeatletter
\newlength{\LongrightarrowWidth}
\newlength{\LongrightarrowLineWidth}
\renewcommand*{\Longrightarrow}[1][white]{\settowidth{\LongrightarrowWidth}{${}{\saveLongrightarrow}{}$}\settowidth{\LongrightarrowLineWidth}{.}\ensuremath{\mathrel{\hspace*{-.5pt}\begin{tikzpicture}[baseline,inner sep=0pt,outer sep=0pt,line width=.2\LongrightarrowLineWidth]
    \node[anchor=base] (START) at (0,0) {\vphantom{a}};
    \node[anchor=base] (END) at (\LongrightarrowWidth-.5pt,0) {\vphantom{a}};
    \draw[-implies,double=#1,double equal sign distance] (START) -- (END);
    \end{tikzpicture}\hspace*{-.5pt}}}}
\makeatother

\makeatother

\def\hyph{-\penalty0\hskip0pt\relax}

\newcommand{\ARROW}[5][\saveLongrightarrow]{\settowidth{\LongrightarrowWidth}{\begin{tabular}{@{}l@{}}${}{#1}{}$\\\ensuremath{\scriptstyle#3}\end{tabular}}\settowidth{\LongrightarrowLineWidth}{.}\ensuremath{\begin{tikzpicture}[baseline,inner sep=0pt,outer sep=0pt,line width=.1\LongrightarrowLineWidth]
    \node[inner sep=0pt,outer sep=0pt,anchor=south] (START) at (0,0) {$#2$};
    \node[inner sep=0pt,outer sep=0pt,node distance=.8\LongrightarrowWidth,right=of START.south east,anchor=south west] (END) {$#4$};
    \gettikzxy{(START.east)}{\InnerNodeAX}{\InnerNodeAY}
    \gettikzxy{(END.west)}{\InnerNodeBX}{\InnerNodeBY}
    \draw[] (START) edge[#5,shorten >= .05\LongrightarrowWidth,shorten <= .05\LongrightarrowWidth] node[above,outer sep=1pt] {\ensuremath{\scriptstyle#3}} ($(\InnerNodeBX,\InnerNodeAY)$);
    \end{tikzpicture}}}

\newcommand{\IFaia}{, if an item applies.}

\newcommand{\tildephi}{\kern2pt\tilde{\kern-2pt\smash{\phi}}}
\newcommand{\tildegamma}{\kern2pt\tilde{\kern-2pt\smash{\gamma}}}

\tikzset{
  matrixstyle/.style={
    matrix of nodes,
    nodes in empty cells,
    column sep      = .2ex,
    row sep         = .2ex,
    nodes={inner sep=0pt,outer sep=0pt,
anchor=west,
}
}}

\usepackage{amssymb}

\newcommand{\GTRULEvar}[1][]{\ensuremath{\rho_{#1}}\xspace}

\newcommand{\PTGTRULEvar}[1][]{\ensuremath{\sigma_{#1}}\xspace}

\newcommand{\PTGTRULEGetrules}[1][\PTGTRULEvar]{\ensuremath{\mathsf{rules}(#1)}\xspace}

\newcommand{\PTGTStoPTSNAME}{\ensuremath{\mathsf{PTGTStoPTS}}\xspace}
\newcommand{\PTGTStoPTS}[1]{\ensuremath{\PTGTStoPTSNAME(#1)}\xspace}

\newcommand{\PTGTStoPTANAME}{\ensuremath{\mathsf{PTGTStoPTA}}\xspace}
\newcommand{\PTGTStoPTA}[1]{\ensuremath{\PTGTStoPTANAME(#1)}\xspace}

\newcommand{\PTGTSGetatomicPropositions}[1][\IPTGTSvar]{\ensuremath{\mathsf{aps}(#1)}\xspace}

\newcommand{\PTCTLsemantics}[2]{\ensuremath{\langle\kern-3pt\langle#1,#2\rangle\kern-3pt\rangle}\xspace}

\newcommand{\PTSvar}[1][]{\ensuremath{P_{#1}}\xspace}
\newcommand{\PTSGetstates}[1][\PTSvar]{\ensuremath{\mathsf{states}(#1)}}
\newcommand{\PTSGetinitialState}[1][\PTSvar]{\ensuremath{\mathsf{istate}(#1)}}
\newcommand{\PTSGetactions}[1][\PTSvar]{\ensuremath{\mathsf{acts}(#1)}}
\newcommand{\PTSGetsteps}[1][\PTSvar]{\ensuremath{\mathsf{steps}(#1)}}
\newcommand{\PTSGetatomicPropositions}[1][\PTSvar]{\ensuremath{\mathsf{aps}(#1)}}
\newcommand{\PTSGetlabelling}[1][\PTSvar]{\ensuremath{\mathsf{lab}(#1)}}

\newcommand{\PTAtoPTSNAME}{\ensuremath{\mathsf{PTAtoPTS}}\xspace}
\newcommand{\PTAtoPTS}[1]{\ensuremath{\PTAtoPTSNAME(#1)}\xspace}

\newcommand{\PTCTLmax}[1]{\ensuremath{\mathcal{P}_{\OPmaxNAME\var{=?}}(#1)}\xspace}
\newcommand{\PTCTLmin}[1]{\ensuremath{\mathcal{P}_{\OPminNAME\var{=?}}(#1)}\xspace}
\newcommand{\PTCTLprob}[2]{\ensuremath{\mathcal{P}_{#1}(#2)}\xspace}

\newcommand{\PTCTLeventuallyUnbounded}[1]{\ensuremath{\mathsf{F}\;#1}\xspace}

\newcommand{\IPTAvar}[1][]{\ensuremath{A_{#1}}\xspace}

\newcommand{\IPTAGetlocations}[1][\IPTAvar]{\ensuremath{\mathsf{locs}(#1)}}

\newcommand{\IPTAGetclocks}[1][\IPTAvar]{\ensuremath{\mathsf{clocks}(#1)}}
\newcommand{\IPTAGetinvariants}[1][\IPTAvar]{\ensuremath{\mathsf{invs}(#1)}}

\newcommand{\PTAvar}[1][]{\ensuremath{A_{#1}}\xspace}
\newcommand{\PTAGetlocations}[1][\PTAvar]{\ensuremath{\mathsf{locs}(#1)}}
\newcommand{\PTAGetinitialLocation}[1][\PTAvar]{\ensuremath{\mathsf{iloc}(#1)}}
\newcommand{\PTAGetactions}[1][\PTAvar]{\ensuremath{\mathsf{acts}(#1)}}
\newcommand{\PTAGetclocks}[1][\PTAvar]{\ensuremath{\mathsf{clocks}(#1)}}
\newcommand{\PTAGetinvariants}[1][\PTAvar]{\ensuremath{\mathsf{invs}(#1)}}
\newcommand{\PTAGetedges}[1][\PTAvar]{\ensuremath{\mathsf{edges}(#1)}}
\newcommand{\PTAGetlabelling}[1][\PTAvar]{\ensuremath{\mathsf{lab}(#1)}}
\newcommand{\PTAGetatomicPropositions}[1][\PTAvar]{\ensuremath{\mathsf{aps}(#1)}}

\newcommand{\ClockConstraints}[1]{\ensuremath{\mathsf{CC}(#1)}\xspace}
\newcommand{\CCvar}[1][]{\ensuremath{\psi_{#1}}\xspace}
\newcommand{\ClockValuations}[1]{\ensuremath{\mathsf{CV}(#1)}\xspace}
\newcommand{\InitialClockValuation}[1]{\ensuremath{\mathsf{ICV}(#1)}\xspace}
\newcommand{\ClockReset}[2]{\ensuremath{#1[#2:=0]}\xspace}
\newcommand{\ClockIncrement}[2]{\ensuremath{#1+#2}\xspace}
\newcommand{\ClockSatisfaction}[2]{\ensuremath{#1\models #2}\xspace}
\newcommand{\DPD}[1]{\mathsf{DPD}(#1)}

\newcommand{\PTGTSvar}[1][]{\ensuremath{S_{#1}}\xspace}

\def\MULTISETleft{\mbox{\ensuremath{\{{}\kern-.51em{}\mid}}}
\def\MULTISETright{\mbox{\ensuremath{\mid{}\kern-.51em{}\}}}}
\newcommand{\MultiSetSum}[2]{\sum\MULTISETleft#1\mid#2\MULTISETright}

\newcommand{\var}[1]{\ensuremath{\mathit{#1}}\xspace}

\newcommand{\FUN}{\ARROW{\vphantom{A}}{}{\vphantom{A}}{->}}

\newcommand{\OPmaxNAME}{\ensuremath{\mathsf{max}}\xspace}

\newcommand{\OPminNAME}{\ensuremath{\mathsf{min}}\xspace}

\newcommand{\NAT}{\ensuremath{\mathbf{N}}\xspace}
\newcommand{\REAL}{\ensuremath{\mathbf{R}}\xspace}

\newcommand{\POWERSET}[1]{\ensuremath{\mathrm{2}^{#1}}\xspace}
\newcommand{\NONNEGATIVEREALS}{\ensuremath{\REAL^{\text{\begin{tikzpicture}[scale=.07]\draw (0,1) -- (2,1);\draw (1,2) -- (1,0);\end{tikzpicture}}}_0}\xspace}
\newcommand{\NONNEGATIVEREALSWITHOUTZERO}{\ensuremath{\REAL^{\text{\begin{tikzpicture}[scale=.07]\draw (0,1) -- (2,1);\draw (1,2) -- (1,0);\end{tikzpicture}}}}\xspace}

\newcommand{\INTERVALcc}[2]{[#1,#2]}

\newcommand{\INTERVALco}[2]{[#1,#2)}

\NewDocumentCommand\V{oo}{\ensuremath{#1_{#2}}\xspace}

\newcommand{\OPsymbolicGraphNAME}{\ensuremath{\mathcal{S}^{\mathrm{graphs}}}\xspace}
\newcommand{\OPsymbolicGraph}[1]{\ensuremath{\OPsymbolicGraphNAME_{\ifthenelse{\equal{#1}{}}{}{#1}}}\xspace}

\newcommand{\OPsymbolicGraphFiniteNAME}{\ensuremath{\mathcal{S}^{\mathrm{graphs}}}\xspace}
\newcommand{\OPsymbolicGraphFinite}[1]{\ensuremath{\OPsymbolicGraphFiniteNAME_{\mathrm{fin}\ifthenelse{\equal{#1}{}}{}{,#1}}}\xspace}

\newcommand{\OPtypedGraphFiniteNAME}{\ensuremath{\mathcal{S}^{\mathrm{graphs}}}\xspace}
\newcommand{\OPtypedGraphFinite}[1]{\ensuremath{\OPtypedGraphFiniteNAME_{\mathrm{fin}\ifthenelse{\equal{#1}{}}{}{,#1}}}\xspace}

\newcommand{\TypeGraph}{\ensuremath{\mathit{TG}}\xspace}
\newcommand{\emptygraph}{\ensuremath{\emptyset}\xspace}

\newcommand{\OPGraphSelect}[2]{\ensuremath{{#1}{.}{\operatorname{#2}}}}
\newcommand{\GraphAC}[1][G]{\OPGraphSelect{#1}{ac}\xspace}

\newcommand{\MOR}{\ARROW{\vphantom{A}}{}{\vphantom{A}}{->}}

\newcommand{\MORmono}{\ARROW{\vphantom{A}}{}{\vphantom{A}}{inclusion}}

\newcommand{\MORinitialONAME}{\ensuremath{\operatorname{i}}\xspace}
\newcommand{\MORinitialO}[1]{\ensuremath{\MORinitialONAME(#1)}\xspace}

\newcommand{\RulesSetFinite}[1]{\ensuremath{\mathcal{S}_{\mathrm{fin}\ifthenelse{\equal{#1}{}}{}{,#1}}^{\mathrm{rules}}}\xspace}

\ExplSyntaxOn
\NewDocumentCommand\OPL{mm}{\tl_if_blank:nTF{#1#2}{\ensuremath{\mathsf{left}}}{\ensuremath{\mathsf{left}(#1,#2)}}\xspace}
\NewDocumentCommand\OPR{mm}{\tl_if_blank:nTF{#1#2}{\ensuremath{\mathsf{right}}}{\ensuremath{\mathsf{right}(#1,#2)}}\xspace}
\ExplSyntaxOff

\newcommand{\TimedGraphSequenceNAMEintext}{\ifmmode\mathrm{TGS}\else\textrm{TGS}\fi\xspace}
\newcommand{\TimedGraphSequencesNAMEintext}{\ifmmode\mathrm{TGSs}\else\textrm{TGSs}\fi\xspace}

\newcommand{\TimedGraphSequence}[2]{\ensuremath{\Pi^{\ifthenelse{\equal{#1}{}}{}{#1}}_{\ifthenelse{\equal{#2}{}}{}{#2}}}\xspace}
\newcommand{\TimedGraphSequenceFinite}[2]{\ensuremath{\Pi^{\mathrm{fin},\ifthenelse{\equal{#1}{}}{}{#1}}_{\ifthenelse{\equal{#2}{}}{}{#2}}}\xspace}
\newcommand{\TimedGraphSequenceInfinite}[2]{\ensuremath{\Pi^{\infty,\ifthenelse{\equal{#1}{}}{}{#1}}_{\ifthenelse{\equal{#2}{}}{}{#2}}}\xspace}

\newcommand{\GraphSequence}[1]{\ensuremath{\Pi_{\ifthenelse{\equal{#1}{}}{}{#1}}}\xspace}
\newcommand{\GraphSequenceFinite}[1]{\ensuremath{\Pi^{\mathrm{fin}}_{\ifthenelse{\equal{#1}{}}{}{#1}}}\xspace}
\newcommand{\GraphSequenceInfinite}[1]{\ensuremath{\Pi^{\infty}_{\ifthenelse{\equal{#1}{}}{}{#1}}}\xspace}

\newcommand{\PIpropagationSimpleNAME}{\ensuremath{\operatorname{PM}}\xspace}
\newcommand{\PIpropagationSimple}[4]{\ensuremath{\PIpropagationSimpleNAME(#1,#2,#3,#4)}\xspace}

\newcommand{\VarExtractTime}[1][]{\ensuremath{\mathit{time}\ifthenelse{\equal{#1}{}}{}{(#1)}}\xspace}
\newcommand{\ExtractTime}[1][]{\ensuremath{\operatorname{time}\ifthenelse{\equal{#1}{}}{}{(#1)}}\xspace}

\newcommand{\AL}{\ifmmode\mathrm{AL}\else\textrm{AL}\fi\xspace}
\newcommand{\AC}{\ifmmode\mathrm{AC}\else\textrm{AC}\fi\xspace}
\newcommand{\ACs}{\ifmmode\mathrm{ACs}\else\textrm{ACs}\fi\xspace}

\newcommand{\ACvar}{\gamma}

\newcommand{\SetTerms}[3]{\ensuremath{\mathcal{S}^{\mathrm{terms}}_{#1\ifthenelse{\isempty{#3}}{}{,#3}}\ifthenelse{\isempty{#2}}{}{(#2)}}\xspace}

\newcommand{\sort}[1]{\ensuremath{\mathsf{#1}}\xspace}

\newcommand{\TypeInteger}{\sort{int}}
\newcommand{\TypeReal}{\sort{real}}

\newcommand{\ACformulasNAME}{\ensuremath{\mathcal{S}^{\mathrm{AC}}}\xspace}
\newcommand{\ACformulas}[1]{\ensuremath{\ACformulasNAME_{#1}\xspace}}

\newcommand{\ACformulasDS}[1][]{\ACformulas{\mathrm{DS}\ifthenelse{\equal{#1}{}}{}{(#1)}}}

\newcommand{\ACtrueNAME}{\ensuremath{\top}\xspace}
\newcommand{\ACtrue}{\ensuremath{\ACtrueNAME}\xspace}

\newcommand{\ACfalseNAME}{\ensuremath{\bot}\xspace}
\newcommand{\ACfalse}{\ensuremath{\ACfalseNAME}\xspace}

\newcommand{\ACandNAME}{\ensuremath{\wedge}\xspace}

\newcommand{\GL}{\ifmmode\mathrm{GL}\else\textrm{GL}\fi\xspace}
\newcommand{\GC}{\ifmmode\mathrm{GC}\else\textrm{GC}\fi\xspace}
\newcommand{\GCs}{\ifmmode\mathrm{GCs}\else\textrm{GCs}\fi\xspace}

\newcommand{\GCvar}[1][]{\phi\ifthenelse{\equal{#1}{}}{}{_{\var{#1}}}}

\newcommand{\GCformulasNAME}{\ensuremath{\mathsf{GC}}\xspace}
\newcommand{\GCformulas}[2][]{\ensuremath{\GCformulasNAME(\ifthenelse{\equal{#1}{}}{}{#1,}#2)}\xspace}

\newcommand{\GCandNAME}{\ensuremath{\wedge}\xspace}

\newcommand{\GCandBinaryArgs}[2]{\ensuremath{#1\GCandNAME#2}\xspace}

\newcommand{\GCnegNAME}{\ensuremath{\neg}\xspace}
\newcommand{\GCneg}[1]{\ensuremath{\GCnegNAME#1}\xspace}

\newcommand{\GCexistsNAME}{\ensuremath{\exists}\xspace}
\newcommand{\GCexists}[2]{\ensuremath{\GCexistsNAME(#1,#2)}\xspace}
\newcommand{\GCexistsNAMEintext}{\emph{exists}\xspace}

\newcommand{\GCrestrictNAME}{\ensuremath{\nu}\xspace}
\newcommand{\GCrestrict}[2]{\ensuremath{\GCrestrictNAME(#1,#2)}\xspace}
\newcommand{\GCrestrictNAMEintext}{\emph{restrict}\xspace}

\newcommand{\GCtrueNAME}{\ensuremath{\top}\xspace}
\newcommand{\GCtrue}{\ensuremath{\GCtrueNAME}\xspace}

\newcommand{\GCfalseNAME}{\ensuremath{\bot}\xspace}
\newcommand{\GCfalse}{\ensuremath{\GCfalseNAME}\xspace}

\newcommand{\GCorNAME}{\ensuremath{\vee}\xspace}

\newcommand{\GCforallNAME}{\ensuremath{\forall}\xspace}

\newcommand{\GCsatNAME}{\ensuremath{\models}\xspace}
\newcommand{\GCsatINNER}[2]{\ensuremath{#1\GCsatNAME#2}\xspace}
\newcommand{\GCsatOUTER}[2]{\ensuremath{#1\GCsatNAME#2}\xspace}

\newcommand{\GCNOTsatNAME}{\ensuremath{\not\models}\xspace}
\newcommand{\GCNOTsatINNER}[2]{\ensuremath{#1\GCNOTsatNAME#2}\xspace}
\newcommand{\GCNOTsatOUTER}[2]{\ensuremath{#1\GCNOTsatNAME#2}\xspace}

\newcommand{\BGL}{\ifmmode\mathrm{BGL}\else\textrm{BGL}\fi\xspace}
\newcommand{\BGC}{\ifmmode\mathrm{BGC}\else\textrm{BGC}\fi\xspace}
\newcommand{\BGCs}{\ifmmode\mathrm{BGCs}\else\textrm{BGCs}\fi\xspace}

\newcommand{\BGCformulasNAME}{\ensuremath{\mathcal{S}^{\mathrm{BGC}}}\xspace}
\newcommand{\BGCformulas}[2][]{\ensuremath{\BGCformulasNAME_{\ifthenelse{\equal{#1}{}}{}{#1,}#2}}\xspace}

\newcommand{\MTGL}{\ifmmode\mathrm{MTGL}\else\textrm{MTGL}\fi\xspace}
\newcommand{\MTGC}{\ifmmode\mathrm{MTGC}\else\textrm{MTGC}\fi\xspace}
\newcommand{\MTGCs}{\ifmmode\mathrm{MTGCs}\else\textrm{MTGCs}\fi\xspace}

\newcommand{\MTGCvar}[1][]{\theta\ifthenelse{\equal{#1}{}}{}{_{\var{#1}}}}
\newcommand{\PMTGCvar}[1][]{\chi\ifthenelse{\equal{#1}{}}{}{_{\var{#1}}}}

\newcommand{\MTGCformulasNAME}{\ensuremath{\mathsf{MTGC}}\xspace}
\newcommand{\MTGCformulas}[2][]{\ensuremath{\MTGCformulasNAME(\ifthenelse{\equal{#1}{}}{}{#1,}#2)}\xspace}

\newcommand{\MTGCandNAME}{\ensuremath{\wedge}\xspace}

\newcommand{\MTGCandBinaryArgs}[2]{\ensuremath{#1\MTGCandNAME#2}\xspace}

\newcommand{\MTGCnegNAME}{\ensuremath{\neg}\xspace}
\newcommand{\MTGCneg}[1]{\ensuremath{\MTGCnegNAME#1}\xspace}

\makeatletter
\newcommand{\dotDelta}{{\vphantom{\Delta}\mathpalette\d@tD@lta\relax}}
\newcommand{\d@tD@lta}[2]{\ooalign{\hidewidth$\m@th#1\mkern-1mu\cdot$\hidewidth\cr$\m@th#1\Delta$\cr}}
\makeatother

\newcommand{\DeltaKindNew}{\ensuremath{\mathrm{N}}\xspace}

\newcommand{\MTGCtrueNAME}{\ensuremath{\top}\xspace}
\newcommand{\MTGCtrue}{\ensuremath{\MTGCtrueNAME}\xspace}

\newcommand{\MTGCexistsNAME}{\ensuremath{\exists}\xspace}

\newcommand{\MTGCexistsSimpleNAME}{\ensuremath{\exists}\xspace}
\newcommand{\MTGCexistsSimple}[2]{\ensuremath{\MTGCexistsSimpleNAME(#1,#2)}\xspace}
\newcommand{\MTGCexistsSimpleNAMEintext}{\emph{exists}\xspace}

\newcommand{\MTGCforallNewSimpleNAME}{\ensuremath{\forall^{\DeltaKindNew}}\xspace}
\newcommand{\MTGCforallSimple}[2]{\ensuremath{\MTGCforallNewSimpleNAME(#1,#2)}\xspace}
\newcommand{\MTGCforallNewSimpleNAMEintext}{\emph{forall-new}\xspace}

\newcommand{\MTGCexistsNewSimpleNAME}{\ensuremath{\exists^{\DeltaKindNew}}\xspace}
\newcommand{\MTGCexistsNewSimple}[2]{\ensuremath{\MTGCexistsNewSimpleNAME(#1,#2)}\xspace}
\newcommand{\MTGCexistsNewSimpleNAMEintext}{\emph{exists-new}\xspace}

\newcommand{\MTGCforallNAME}{\ensuremath{\forall}\xspace}

\newcommand{\MTGCrestrictNAME}{\ensuremath{\nu}\xspace}

\newcommand{\MTGCrestrictSimpleNAME}{\ensuremath{\nu}\xspace}
\newcommand{\MTGCrestrictSimple}[2]{\ensuremath{\MTGCrestrictSimpleNAME(#1,#2)}\xspace}
\newcommand{\MTGCrestrictSimpleNAMEintext}{\emph{restrict}\xspace}

\newcommand{\MTGCuntilNAME}{\ensuremath{\operatorname{U}}\xspace}

\newcommand{\MTGCuntilSimpleNAME}{\ensuremath{\operatorname{U}}\xspace}
\newcommand{\MTGCuntilSimple}[3]{\ensuremath{#2\mathrel{\MTGCuntilSimpleNAME_{#1}}#3}\xspace}
\newcommand{\MTGCuntilSimpleNAMEintext}{\emph{until}\xspace}

\newcommand{\MTGCsatTGSNAME}{\ensuremath{\models}\xspace}
\newcommand{\MTGCsatTGSINNER}[4]{\ensuremath{(#1,#2,#3)\MTGCsatTGSNAME#4}\xspace}
\newcommand{\MTGCsatTGSOUTER}[2]{\ensuremath{#1\MTGCsatTGSNAME#2}\xspace}

\newcommand{\MTGCNOTsatNAME}{\ensuremath{\not\models}\xspace}
\newcommand{\MTGCNOTsatINNER}[3]{\ensuremath{(#1,#2,#3)\MTGCNOTsatNAME}\xspace}

\newcommand{\GH}{\ifmmode\mathrm{GH}\else\textrm{GH}\fi\xspace}
\newcommand{\GHs}{\ifmmode\mathrm{GHs}\else\textrm{GHs}\fi\xspace}

\newcommand{\OPencodeNAME}{\ensuremath{\mathsf{encode}}\xspace}
\newcommand{\OPencode}[1]{\ensuremath{\OPencodeNAME(#1)}\xspace}
\newcommand{\OPfoldNAME}{\ensuremath{\mathsf{fold}}\xspace}
\newcommand{\OPfold}[1]{\ensuremath{\OPfoldNAME(#1)}\xspace}

\ExplSyntaxOn
\NewDocumentCommand\OPalive{mm}{\tl_if_blank:nTF{#1#2}{\ensuremath{\operatorname{alive}}}{\ensuremath{\operatorname{alive}(#1,#2)}}\xspace}
\NewDocumentCommand\OPearliest{mm}{\tl_if_blank:nTF{#1#2}{\ensuremath{\operatorname{earliest}}}{\ensuremath{\operatorname{earliest}(#1,#2)}}\xspace}
\ExplSyntaxOff

\ExplSyntaxOn
\NewDocumentCommand\OPgetPrefix{mm}{\tl_if_blank:nTF{#1#2}{\ensuremath{\operatorname{getPrefix}}}{\ensuremath{\operatorname{getPrefix}(#1,#2)}}\xspace}
\NewDocumentCommand\OPterminated{m}{\tl_if_blank:nTF{#1}{\ensuremath{\operatorname{terminated}}}{\ensuremath{\operatorname{terminated}(#1)}}\xspace}
\NewDocumentCommand\OPviolations{mmm}{\tl_if_blank:nTF{#1#2#3}{\ensuremath{\operatorname{true-violations}}}{\ensuremath{\operatorname{true-violations}(#1,#2,#3)}}\xspace}
\ExplSyntaxOff

\DeclareSIUnit\timeunit{timeunit}
\DeclareSIUnit\timeunits{timeunits}

\newcommand{\MFOTL}{\ifmmode\mathrm{MFOTL}\else\textrm{MFOTL}\fi\xspace}
\newcommand{\MFOTC}{\ifmmode\mathrm{MFOTC}\else\textrm{MFOTC}\fi\xspace}
\newcommand{\MFOTCs}{\ifmmode\mathrm{MFOTCs}\else\textrm{MFOTCs}\fi\xspace}

\newcommand{\MFOTCformulasNAME}{\ensuremath{\mathcal{S}^{\mathrm{MFOTC}}}\xspace}
\newcommand{\MFOTCformulas}[2][]{\ensuremath{\MFOTCformulasNAME_{#2\ifthenelse{\equal{#1}{}}{}{,#1}}}\xspace}

\newcommand{\GCST}{\ifmmode\mathrm{ST}\else\textrm{ST}\fi\xspace}
\newcommand{\GCSTs}{\ifmmode\mathrm{STs}\else\textrm{STs}\fi\xspace}
\newcommand{\GCSTvar}[1][]{\begingroup\color{blue}\gamma\endgroup\ifthenelse{\equal{#1}{}}{}{_{\var{#1}}}}

\newcommand{\GCSTformulasNAME}{\ensuremath{\mathcal{S}^{\mathrm{GCST}}}\xspace}
\newcommand{\GCSTformulas}[2][]{\ensuremath{\GCSTformulasNAME_{\ifthenelse{\equal{#1}{}}{}{#1,}#2}}\xspace}

\def\GCSTexistsIgnoreFalse{1}
\newcommand{\GCSTexistsNAME}{\ensuremath{\exists}\xspace}
\newcommand{\GCSTexists}[4]{\ensuremath{\GCSTexistsNAME(#1,#2,#3\if\GCSTexistsIgnoreFalse1\else,#4\fi)}\xspace}

\newcommand{\GCSTforallNAME}{\ensuremath{\forall}\xspace}
\newcommand{\GCSTforall}[4]{\ensuremath{\GCSTforallNAME(#1,#2,#3,\if\GCSTexistsIgnoreFalse1\else,#4\fi)}\xspace}

\ExplSyntaxOn
\NewDocumentCommand\OPsubconditions{m}{\ensuremath{\operatorname{\operatorname{sub}}\tl_if_blank:nTF{#1}{}{(#1)}}\xspace}
\NewDocumentCommand\AutoGraphAlgo{m}{\ensuremath{\operatorname{\mathcal{A}}\tl_if_blank:nTF{#1}{}{(#1)}}\xspace}
\NewDocumentCommand\AutoGraphModels{m}{\ensuremath{\operatorname{\mathcal{M}}\tl_if_blank:nTF{#1}{}{(#1)}}\xspace}
\NewDocumentCommand\AlgoStateBasedA{mm}{\ensuremath{\operatorname{\mathcal{R}epair\c_math_subscript_token{\mathrm{sb,1}}}\tl_if_blank:nTF{#1#2}{}{(#1,#2)}}\xspace}
\NewDocumentCommand\OPrestrictionTree{mm}{\ensuremath{\operatorname{RT}\tl_if_blank:nTF{#1#2}{}{(#1,#2)}}\xspace}
\NewDocumentCommand\AlgoStateBasedB{mm}{\ensuremath{\operatorname{\mathcal{R}epair\c_math_subscript_token{\mathrm{sb,2}}}\tl_if_blank:nTF{#1#2}{}{(#1,#2)}}\xspace}
\NewDocumentCommand\AlgoStateBasedREC{mmmm}{\ensuremath{\operatorname{\mathcal{R}epair\c_math_subscript_token{\mathrm{rec}}}\tl_if_blank:nTF{#1#2#3#4}{}{(#1,#2,#3,#4)}}\xspace}
\NewDocumentCommand\OPconstructGCST{mm}{\ensuremath{\operatorname{cst}\tl_if_blank:nTF{#1#2}{}{(#1,#2)}}\xspace}
\NewDocumentCommand\OPconstructGCSTForGCPAT{mm}{\ensuremath{\operatorname{cstPAT}\tl_if_blank:nTF{#1#2}{}{(#1,#2)}}\xspace}
\NewDocumentCommand\OPpropagateMatch{mm}{\ensuremath{\operatorname{ppgMatch}\tl_if_blank:nTF{#1#2}{}{(#1,#2)}}\xspace}
\NewDocumentCommand\OPcountMatches{m}{\ensuremath{\operatorname{cntMatches}\tl_if_blank:nTF{#1}{}{(#1)}}\xspace}
\NewDocumentCommand\OPpropagateBackwards{mm}{\ensuremath{\operatorname{ppgB}\tl_if_blank:nTF{#1#2}{}{(#1,#2)}}\xspace}
\NewDocumentCommand\OPpropagateForwards{mm}{\ensuremath{\operatorname{ppgF}\tl_if_blank:nTF{#1#2}{}{(#1,#2)}}\xspace}
\NewDocumentCommand\OPpropagateUpdate{mm}{\ensuremath{\operatorname{ppgU}\tl_if_blank:nTF{#1#2}{}{(#1,#2)}}\xspace}
\NewDocumentCommand\OPbinaryDiversityINNER{mmmmmmmmm}{\ensuremath{\operatorname{bd}\tl_if_blank:nTF{#1#2#3#4#5#6#7#8#9}{}{(#1,\allowbreak#2,\allowbreak#3,\allowbreak#4,\allowbreak#5,\allowbreak#6,\allowbreak#7,\allowbreak#8,\allowbreak#9)}}\xspace}
\NewDocumentCommand\OPbinaryDiversity{mmmm}{\ensuremath{\operatorname{binaryDiversity}\tl_if_blank:nTF{#1#2#3}{}{(#1,\allowbreak#2,\allowbreak#3,\allowbreak#4)}}\xspace}
\NewDocumentCommand\OPsetDiversity{mmmm}{\ensuremath{\operatorname{setDiversity}\tl_if_blank:nTF{#1#2#3#4}{}{(#1,\allowbreak#2,\allowbreak#3,\allowbreak#4)}}\xspace}
\NewDocumentCommand\OPSTviolations{mm}{\ensuremath{\operatorname{violations}\tl_if_blank:nTF{#1#2}{}{(#1,#2)}}\xspace}
\NewDocumentCommand\OPqualitativeCoverage{mm}{\ensuremath{\operatorname{qualCoverage}\tl_if_blank:nTF{#1#2}{}{(#1,#2)}}\xspace}
\NewDocumentCommand\GCPATcoverable{m}{\ensuremath{\operatorname{coverable}\tl_if_blank:nTF{#1}{}{(#1)}}\xspace}
\NewDocumentCommand\GCPATcovered{mm}{\ensuremath{\operatorname{covered}\tl_if_blank:nTF{#1#2}{}{(#1,#2)}}\xspace}

\NewDocumentCommand\AlgoDeltaBasedONERuleAdd{mmmm}{\ensuremath{\operatorname{\mathcal{R}epair\c_math_subscript_token{\mathrm{add}}}\tl_if_blank:nTF{#1#2#3#4}{}{(#1,#2,#3,#4)}}\xspace}
\NewDocumentCommand\AlgoDeltaBasedONERuleDel{mmm}{\ensuremath{\operatorname{\mathcal{R}epair\c_math_subscript_token{\mathrm{del}}}\tl_if_blank:nTF{#1#2#3}{}{(#1,#2,#3)}}\xspace}
\NewDocumentCommand\AlgoDeltaBasedONE{mm}{\ensuremath{\operatorname{\mathcal{R}epair\c_math_subscript_token{\mathrm{db1}}}\tl_if_blank:nTF{#1#2}{}{(#1,#2)}}\xspace}
\NewDocumentCommand\AlgoDeltaBased{mmm}{\ensuremath{\operatorname{\mathcal{R}epair\c_math_subscript_token{\mathrm{db}}}\tl_if_blank:nTF{#1#2#3}{}{(#1,#2,#3)}}\xspace}
\ExplSyntaxOff

\newcommand{\DRAWexample}[6][0]{
    \begin{tikzpicture}
    \def\xpos{0}
    \def\strut{\vphantom{$a_1b_1a_2b_2$}}
    \ifthenelse{\equal{#4}{1}}{\node (a1) at (\xpos,0) {\strut$a_1$};\edef\xpos{\xpos+.5}}{}    
    \ifthenelse{\equal{#6}{1}}{\edef\xpos{\xpos+.5}}{}
    \ifthenelse{\equal{#5}{1}}{\node (b1) at (\xpos,0) {\strut$b_1$};\edef\xpos{\xpos+.5}}{}
    \ifthenelse{\equal{#6}{1}}{
        \ifthenelse{\equal{#1}{1}}{
            \draw (a1) edge[->] node[above] {$e_1$} (b1);
        }{
            \draw (a1) edge[->] (b1);
        }
    }{}    
\ifthenelse{\equal{#2}{1}}{\node (a2) at (\xpos,0) {\strut$a_2$};}{}
\ifthenelse{\equal{#3}{1}}{
        \ifthenelse{\equal{#1}{1}}{
            \draw (a2) edge[loop right,->] node[right] {$e_3$} (a2);
        }{
            \draw (a2) edge[loop right,->] (a2);
        }
    }{}
    \end{tikzpicture}
}

\newcommand{\DRAWresolvedexample}[2][0]{
    \ifthenelse{\equal{#2}{1}}{
        \begin{tikzpicture}
        \def\strut{\vphantom{$a_1b_1a_2b_2$}}
        \node (a1) at (0,0) {\strut$a_1$};
        \node (b1) at (1,0) {\strut$b_1$};
        \ifthenelse{\equal{#1}{1}}{
            \draw (a1) edge[->] node[above] {$e_1$} (b1);
        }{
            \draw (a1) edge[->] (b1);
        }        
        \node (a2) at (2,0) {\strut$a_2$};
        \ifthenelse{\equal{#1}{1}}{
            \draw (a2) edge[->] node[above] {$e_2$} (b1);
        }{
            \draw (a2) edge[->] (b1);
        }
        \end{tikzpicture}
    }{}
    \ifthenelse{\equal{#2}{2}}{
        \begin{tikzpicture}
        \def\strut{\vphantom{$a_1b_1a_2b_2$}}
        \node (a1) at (0,0) {\strut$a_1$};
        \node (b1) at (1,0) {\strut$b_1$};
        \draw (a1) edge[->] (b1);
        \node (b2) at (1.5,0) {\strut$b_2$};
        \node (a2) at (2.5,0) {\strut$a_2$};        
        \draw (a2) edge[->] (b2);
        \end{tikzpicture}
    }{}
    \ifthenelse{\equal{#2}{3}}{
        \begin{tikzpicture}
        \def\strut{\vphantom{$a_1b_1a_2b_2$}}
        \node (a1) at (0,0) {\strut$a_1$};
        \node (b1) at (1,0) {\strut$b_1$};
        \draw (a1) edge[->] (b1);
        \end{tikzpicture}
    }{}
    \ifthenelse{\equal{#2}{4}}{
        \begin{tikzpicture}
        \def\strut{\vphantom{$a_1b_1a_2b_2$}}
        \node (a1) at (0,0) {\strut$a_2$};
        \node (b1) at (1,0) {\strut$b_1$};
        \draw (a1) edge[->] (b1);
        \end{tikzpicture}
    }{}
    \ifthenelse{\equal{#2}{5}}{
        \begin{tikzpicture}
        \def\strut{\vphantom{$a_1b_1a_2b_2$}}
        \node (a1) at (0,0) {\strut$a_1$};
        \node (b1) at (1,0) {\strut$b_1$};
        \ifthenelse{\equal{#1}{1}}{
            \draw (a1) edge[->] node[above] {$e_1$} (b1);
        }{
            \draw (a1) edge[->] (b1);
        }
        \node (b2) at (1.5,0) {\strut$b_2$};
        \node (a2) at (2.5,0) {\strut$a_2$};        
        \ifthenelse{\equal{#1}{1}}{
            \draw (a2) edge[->] node[above] {$e_2$} (b2);
        }{
            \draw (a2) edge[->] (b2);
        }
        \ifthenelse{\equal{#1}{1}}{
            \draw (a2) edge[->,loop right] node[above] {$e_3$} (a2);
        }{
            \draw (a2) edge[->,loop right] (a2);
        }
        \end{tikzpicture}
    }{}
    \ifthenelse{\equal{#2}{6}}{
        \begin{tikzpicture}
        \def\strut{\vphantom{$a_1b_1a_2b_2$}}
        \node (a1) at (0,0) {\strut$a_1$};
        \node (b1) at (1,0) {\strut$b_1$};
        \ifthenelse{\equal{#1}{1}}{
            \draw (a1) edge[->] node[above] {$e_1$} (b1);
        }{
            \draw (a1) edge[->] (b1);
        }
        \node (b2) at (1.5,0) {\strut$b_2$};
        \end{tikzpicture}
    }{}    
}

\newcommand{\DRAWpsi}[1]{
    \ifthenelse{\equal{#1}{1}}{
        \begin{tikzpicture}[baseline,anchor=base,inner sep=0pt]
        \begin{scope}[inner sep=1pt]
        \def\strut{\vphantom{$ab$}}
        \node (a1) at (0,0) {\strut$a$};
        \end{scope}
        \end{tikzpicture}
    }{}
    \ifthenelse{\equal{#1}{2}}{
        \begin{tikzpicture}[baseline,anchor=base,inner sep=0pt]
        \begin{scope}[inner sep=1pt]
        \def\strut{\vphantom{$a_1b_1a_2b_2$}}
        \node (a1) at (0,0) {\strut$a$};
        \node (b1) at (1,0) {\strut$b$};
        \draw (a1) edge[->] node[above,inner sep=1pt] {$e$} (b1);
        \end{scope}
        \end{tikzpicture}
    }{}    
    \ifthenelse{\equal{#1}{3}}{
        \begin{tikzpicture}[baseline,anchor=base,inner sep=0pt]
        \begin{scope}[inner sep=1pt]
        \def\strut{\vphantom{$a_1b_1a_2b_2$}}
        \node (a1) at (0,0) {\strut$a$};
        \draw (a1) edge[loop right,->] node[right,inner sep=1pt] {$e$} (a1);
        \end{scope}
        \end{tikzpicture}
    }{}    
    \ifthenelse{\equal{#1}{4}}{
        \begin{tikzpicture}[baseline,anchor=base,inner sep=0pt]
        \begin{scope}[inner sep=1pt]
        \def\strut{\vphantom{$a_1b_1a_2b_2$}}
        \node (a1) at (0,0) {\strut$a_1$};
        \node (b1) at (1,0) {\strut$a_2$};
        \node (c1) at (2,0) {\strut$a_3$};
        \draw (a1) edge[->] node[above,inner sep=1pt] {$e_1$} (b1);
        \draw (b1) edge[->] node[above,inner sep=1pt] {$e_2$} (c1);
        \end{scope}
        \end{tikzpicture}
    }{}
    \ifthenelse{\equal{#1}{5}}{
        \begin{tikzpicture}[baseline,anchor=base,inner sep=0pt]
        \begin{scope}[inner sep=1pt]
        \def\strut{\vphantom{$a_1b_1a_2b_2$}}
        \node (a1) at (0,0) {\strut$a_1$};
        \node (b1) at (1,0) {\strut$a_2$};
        \node (c1) at (2,0) {\strut$a_3$};
        \draw (a1) edge[->] node[above,inner sep=1pt] {$e_1$} (b1);
        \draw (b1) edge[->] node[above,inner sep=1pt] {$e_2$} (c1);
        \draw (c1.north) edge[<-,bend left=10,out=315,in=-90,looseness=.3] node[above,inner sep=1pt] {$e_3$} (a1.north);
        \end{scope}
        \end{tikzpicture}
    }{}
    \ifthenelse{\equal{#1}{6}}{
        \begin{tikzpicture}[baseline,anchor=base,inner sep=0pt]
        \begin{scope}[inner sep=1pt]
        \def\strut{\vphantom{$a_1b_1a_2b_2$}}
        \node (a1) at (0,0) {\strut$a$};
        \node (b1) at (.35,0) {\strut$b$};
        \end{scope}
        \end{tikzpicture}
    }{}
    \ifthenelse{\equal{#1}{7}}{
        \begin{tikzpicture}[baseline,anchor=base,inner sep=0pt]
        \begin{scope}[inner sep=1pt]
        \def\strut{\vphantom{$a_1b_1a_2b_2$}}
        \node (a1) at (0,0) {\strut$a$};
        \node (b1) at (-1,0) {\strut$b$};
        \draw (a1) edge[->] node[above,inner sep=1pt] {$e'$} (b1);
        \draw (a1) edge[loop right,->] node[right,inner sep=1pt] {$e$} (a1);
        \end{scope}
        \end{tikzpicture}
    }{}
    \ifthenelse{\equal{#1}{8}}{
        \begin{tikzpicture}[baseline,anchor=base,inner sep=0pt]
        \begin{scope}[inner sep=1pt]
        \def\strut{\vphantom{$a_1b_1a_2b_2$}}
        \node (a1) at (0,0) {\strut$a$};
        \node (b1) at (-.5,0) {\strut$b$};
\draw (a1) edge[loop right,->] node[right,inner sep=1pt] {$e$} (a1);
        \end{scope}
        \end{tikzpicture}
    }{}    
    \ifthenelse{\equal{#1}{9}}{
        \begin{tikzpicture}[baseline,anchor=base,inner sep=0pt]
        \begin{scope}[inner sep=1pt]
        \def\strut{\vphantom{$a_1b_1a_2b_2$}}
        \node (b1) at (-.5,0) {\strut$b$};
        \end{scope}
        \end{tikzpicture}
    }{}        
}

\newcommand{\SHOULDSAT}[1]{}

\newcommand{\DRAWpsiMatch}[4]{
    \ifthenelse{\equal{#1}{1}}{
        \begin{tikzpicture}[baseline,anchor=base,inner sep=0pt]
        \begin{scope}[inner sep=1pt]
        \def\strut{\vphantom{$ab$}}
        \node (a1) at (0,0) {\strut$#2$};
        \end{scope}
        \end{tikzpicture}
    }{}
    \ifthenelse{\equal{#1}{2}}{
        \begin{tikzpicture}[baseline,anchor=base,inner sep=0pt]
        \begin{scope}[inner sep=1pt]
        \def\strut{\vphantom{$a_1b_1a_2b_2$}}
        \node (a1) at (0,0) {\strut$#2$};
        \node (b1) at (1,0) {\strut$#3$};
        \draw (a1) edge[->] node[above,inner sep=1pt] {$#4$} (b1);
        \end{scope}
        \end{tikzpicture}
    }{}    
    \ifthenelse{\equal{#1}{3}}{
        \begin{tikzpicture}[baseline,anchor=base,inner sep=0pt]
        \begin{scope}[inner sep=1pt]
        \def\strut{\vphantom{$a_1b_1a_2b_2$}}
        \node (a1) at (0,0) {\strut$#2$};
        \draw (a1) edge[loop right,->] node[right,inner sep=1pt] {$#3$} (a1);
        \end{scope}
        \end{tikzpicture}
    }{}    
}

\newcommand{\EXAMPLEgammaG}[1][]{\ensuremath{\mathbf{\pmb\gamma_{u\ifthenelse{\equal{#1}{}}{}{,#1}}}}\xspace}
\newcommand{\EXAMPLEgammaD}[1][]{\ensuremath{\mathbf{\pmb\gamma^D_{u\ifthenelse{\equal{#1}{}}{}{,#1}}}}\xspace}
\newcommand{\EXAMPLEgammaGprime}[1][]{\ensuremath{\mathbf{\pmb\gamma'_{u\ifthenelse{\equal{#1}{}}{}{,#1}}}}\xspace}

\newcommand{\shownum}[1]{}

\newcommand{\GLPAT}{\ifmmode\mathrm{GLPAT}\else\textrm{GLPAT}\fi\xspace}
\newcommand{\GCPAT}{\ifmmode\mathrm{GCPAT}\else\textrm{GCPAT}\fi\xspace}
\newcommand{\GCPATs}{\ifmmode\mathrm{GCPATs}\else\textrm{GCPATs}\fi\xspace}

\newcommand{\GCPATvar}[1][]{\begingroup\color{blue}\xi\endgroup\ifthenelse{\equal{#1}{}}{}{_{\var{#1}}}}

\newcommand{\GCPATformulasNAME}{\ensuremath{\mathcal{S}^{\mathrm{GCPAT}}}\xspace}
\newcommand{\GCPATformulas}[2][]{\ensuremath{\GCPATformulasNAME_{\ifthenelse{\equal{#1}{}}{}{#1,}#2}}\xspace}

\newcommand{\GCPATexistsNONENAME}{\ensuremath{\exists_0}\xspace}

\newcommand{\GCPATexistsNONEomit}[2]{\ensuremath{\GCPATexistsNONENAME}\xspace}

\newcommand{\MYNODE}[4][]{\node[#1] (#2#3) at (#4) {$\vphantom{b_{1,2}}#2_{#3}$};}
\def\GRAPHFOUND{}
\newcommand{\GRAPH}[1]{\typeout{loading graph #1}\def\GRAPHFOUND{0}\ifthenelse{\equal{#1}{}}{\begin{tikzpicture}[minimum height=0cm,inner sep=1pt,outer sep=1pt,anchor=base,baseline]
        \node[](A1) at (0,0) {$\phantom{A}$};
        \end{tikzpicture}\def\GRAPHFOUND{1}}{}\ifthenelse{\equal{#1}{xt1}}{\begin{tikzpicture}[minimum height=0cm,inner sep=1pt,outer sep=1pt,anchor=base,baseline]
        \node[](A1) at (0,0) {$t_1$};
        \end{tikzpicture}\def\GRAPHFOUND{1}}{}\ifthenelse{\equal{#1}{t1}}{\begin{tikzpicture}[minimum height=0cm,inner sep=1pt,outer sep=1pt,anchor=base,baseline]
        \node[mybeameralertRED={3-}](A1) at (0,0) {$t_1$};
        \end{tikzpicture}\def\GRAPHFOUND{1}}{}\ifthenelse{\equal{#1}{t1t2}}{\begin{tikzpicture}[minimum height=0cm,inner sep=1pt,outer sep=1pt,anchor=base,baseline]
        \node[mybeameralertRED={4-}](A1) at (0,0) {$t_1$};
        \node[node distance=.5ex,below=of A1](A4) {$t_2$};
        \end{tikzpicture}\def\GRAPHFOUND{1}}{}\ifthenelse{\equal{#1}{xt1t2}}{\begin{tikzpicture}[minimum height=0cm,inner sep=1pt,outer sep=1pt,anchor=base,baseline]
        \node[](A1) at (0,0) {$t_1$};
        \node[node distance=.5ex,below=of A1](A4) {$t_2$};
        \end{tikzpicture}\def\GRAPHFOUND{1}}{}\ifthenelse{\equal{#1}{t1->c1t2}}{\begin{tikzpicture}[minimum height=0cm,inner sep=1pt,outer sep=1pt,anchor=base,baseline]
        \node[mybeameralertRED={5-}](A1) at (0,0) {$t_1$};
        \node[mybeameralertRED={7-},node distance=2ex,right=of A1](A3) {$c_1$};
        \node[node distance=.5ex,below=of A1](A4) {$t_2$};
        \draw(A1)edge[->,mybeameralertedgeX={7-}](A3);
        \end{tikzpicture}\def\GRAPHFOUND{1}}{}\ifthenelse{\equal{#1}{xt1->c1t2}}{\begin{tikzpicture}[minimum height=0cm,inner sep=1pt,outer sep=1pt,anchor=base,baseline]
        \node[](A1) at (0,0) {$t_1$};
        \node[node distance=2ex,right=of A1](A3) {$c_1$};
        \node[node distance=.5ex,below=of A1](A4) {$t_2$};
        \draw(A1)edge[->](A3);
        \end{tikzpicture}\def\GRAPHFOUND{1}}{}\ifthenelse{\equal{#1}{t1->c1t2->c2}}{\begin{tikzpicture}[minimum height=0cm,inner sep=1pt,outer sep=1pt,anchor=base,baseline]
        \node[](A1) at (0,0) {$t_1$};
        \node[node distance=2ex,right=of A1](A3) {$c_1$};
        \node[node distance=.5ex,below=of A1](A4) {$t_2$};
        \node[node distance=2ex,right=of A4](A5) {$c_2$};
        \draw(A1)edge[->](A3);
        \draw(A4)edge[->](A5);
        \end{tikzpicture}\def\GRAPHFOUND{1}}{}\ifthenelse{\equal{#1}{xt1->c1t2->c2}}{\begin{tikzpicture}[minimum height=0cm,inner sep=1pt,outer sep=1pt,anchor=base,baseline]
        \node[](A1) at (0,0) {$t_1$};
        \node[node distance=2ex,right=of A1](A3) {$c_1$};
        \node[node distance=.5ex,below=of A1](A4) {$t_2$};
        \node[node distance=2ex,right=of A4](A5) {$c_2$};
        \draw(A1)edge[->](A3);
        \draw(A4)edge[->](A5);
        \end{tikzpicture}\def\GRAPHFOUND{1}}{}\ifthenelse{\equal{#1}{foldedt1->c1t2->c2}}{\begin{tikzpicture}[minimum height=0cm,inner sep=1pt,outer sep=1pt,anchor=base,baseline]
		\node[box=colorTask] (A1) at (0,0) {\TaskNodeGH{t_1}{CD}{}{1}{-1}};
        \node[box=colorCompleted,node distance=3.4cm,right=of A1](A3) {\CompletedNodeGH{c_1}{CD}{6}{-1}};
        \node[box=colorTask,node distance=1ex,below=of A1](A4) {\TaskNodeGH{t_2}{CD}{}{3}{-1}};
        \gettikzxy{(A3.south)}{\NodeAX}{\NodeAY}
        \gettikzxy{(A4.east)}{\NodeBX}{\NodeBY}
        \node[box=colorCompleted,anchor=center](A5) at (\NodeAX,\NodeBY) {\CompletedNodeGH{c_2}{CD}{9}{-1}};
        \draw[->] ($(A1.north east)+(0ex,-1ex)$) to
				node[below,pos=.5,boxX=2,inner sep=0pt,outer sep=1ex] {\strut $e_1$\Type{is}\nodepart{second}\strut\AttributeValue{cts}{6}\\\strut\AttributeValue{dts}{-1}}
				($(A3.north west)+(0ex,-1ex)$);
        \draw[->,yshift=-1ex] ($(A4.north east)+(0ex,-1ex)$) to
				node[below,pos=.5,boxX=2,inner sep=0pt,outer sep=1ex] {\strut $e_2$\Type{is}\nodepart{second}\strut\AttributeValue{cts}{9}\\\strut\AttributeValue{dts}{-1}}
				($(A5.north west)+(0ex,-1ex)$);
        \end{tikzpicture}\def\GRAPHFOUND{1}}{}\ifthenelse{\equal{#1}{TGtaskcompleted}}{\begin{tikzpicture}[minimum height=0cm,inner sep=1pt,outer sep=1pt,anchor=base,baseline]
        \node[draw](A1) at (0,0) {${:}\mathrm{Task}$};
        \node[draw,node distance=2ex,right=of A1](A4) {${:}\mathrm{Completed}$};
        \draw(A1)edge[->](A4);
        \begin{pgfonlayer}{background}
        \node[fill=white,fit=(A1) (A4)] (FOO) {};
        \end{pgfonlayer}
        \end{tikzpicture}\def\GRAPHFOUND{1}}{}\ifthenelse{\equal{#1}{t}}{\begin{tikzpicture}[minimum height=0cm,inner sep=1pt,outer sep=1pt,anchor=base,baseline]
        \node[](A1) at (0,0) {$t$};
        \begin{pgfonlayer}{background}
        \node[fill=white,fit=(A1)] (FOO) {};
        \end{pgfonlayer}
        \end{tikzpicture}\def\GRAPHFOUND{1}}{}\ifthenelse{\equal{#1}{xt}}{\begin{tikzpicture}[minimum height=0cm,inner sep=1pt,outer sep=1pt,anchor=base,baseline]
        \node[](A1) at (0,0) {$t$};
        \end{tikzpicture}\def\GRAPHFOUND{1}}{}\ifthenelse{\equal{#1}{t->c}}{\begin{tikzpicture}[minimum height=0cm,inner sep=1pt,outer sep=1pt,anchor=base,baseline]
        \node[](A1) at (0,0) {$t$};
        \node[node distance=2ex,right=of A1](A4) {$c$};
        \draw(A1)edge[->](A4);
        \begin{pgfonlayer}{background}
        \node[fill=white,fit=(A1) (A4)] (FOO) {};
        \end{pgfonlayer}
        \end{tikzpicture}\def\GRAPHFOUND{1}}{}\ifthenelse{\equal{#1}{xt->c}}{\begin{tikzpicture}[minimum height=0cm,inner sep=1pt,outer sep=1pt,anchor=base,baseline]
        \node[](A1) at (0,0) {$t$};
        \node[node distance=2ex,right=of A1](A4) {$c$};
        \draw(A1)edge[->](A4);
        \end{tikzpicture}\def\GRAPHFOUND{1}}{}\ifthenelse{\equal{#1}{a2->bL}}{\begin{tikzpicture}[minimum height=0cm,inner sep=1pt,outer sep=1pt,anchor=base,baseline]
        \node[](A1) at (0,0) {$a_2$};
        \node[node distance=2ex,right=of A1](A4) {$b$};
        \draw(A1)edge[->](A4);
        \draw(A4)edge[->,loop right](A4);
        \end{tikzpicture}\def\GRAPHFOUND{1}}{}\ifthenelse{\equal{#1}{a1L}}{\begin{tikzpicture}[minimum height=0cm,inner sep=1pt,outer sep=1pt,anchor=base,baseline]
        \node[](A1) at (0,0) {$a_1$};
        \draw(A1)edge[->,loop right](A1);
        \end{tikzpicture}\def\GRAPHFOUND{1}}{}\ifthenelse{\equal{#1}{a1}}{\begin{tikzpicture}[minimum height=0cm,inner sep=1pt,outer sep=1pt,anchor=base,baseline]
        \node[](A1) at (0,0) {$a_1$};
        \end{tikzpicture}\def\GRAPHFOUND{1}}{}\ifthenelse{\equal{#1}{b1}}{\begin{tikzpicture}[minimum height=0cm,inner sep=1pt,outer sep=1pt,anchor=base,baseline]
        \node[](A1) at (0,0) {$b_1$};
        \end{tikzpicture}\def\GRAPHFOUND{1}}{}\ifthenelse{\equal{#1}{b2}}{\begin{tikzpicture}[minimum height=0cm,inner sep=1pt,outer sep=1pt,anchor=base,baseline]
        \node[](A1) at (0,0) {$b_2$};
        \end{tikzpicture}\def\GRAPHFOUND{1}}{}\ifthenelse{\equal{#1}{a1a2L}}{\begin{tikzpicture}[minimum height=0cm,inner sep=1pt,outer sep=1pt,anchor=base,baseline]
        \node[](A1) at (0,0) {$a_1$};
        \node[node distance=.5ex,below=of A1](A2) {$a_2$};
        \draw(A2)edge[->,loop right](A2);
        \end{tikzpicture}\def\GRAPHFOUND{1}}{}\ifthenelse{\equal{#1}{a1a2}}{\begin{tikzpicture}[minimum height=0cm,inner sep=1pt,outer sep=1pt,anchor=base,baseline]
        \node[](A1) at (0,0) {$a_1$};
        \node[node distance=.5ex,below=of A1](A2) {$a_2$};
        \end{tikzpicture}\def\GRAPHFOUND{1}}{}\ifthenelse{\equal{#1}{a2->b}}{\begin{tikzpicture}[minimum height=0cm,inner sep=1pt,outer sep=1pt,anchor=base,baseline]
        \node[](A1) at (0,0) {$a_2$};
        \node[node distance=2ex,right=of A1](A4) {$b$};
        \draw(A1)edge[->](A4);
        \end{tikzpicture}\def\GRAPHFOUND{1}}{}\ifthenelse{\equal{#1}{a2L}}{\begin{tikzpicture}[minimum height=0cm,inner sep=1pt,outer sep=1pt,anchor=base,baseline]
            \node[](A1) at (0,0) {$a_2$};
            \draw(A1)edge[->,loop right](A1);
            \end{tikzpicture}\def\GRAPHFOUND{1}}{}\ifthenelse{\equal{#1}{a0L}}{\begin{tikzpicture}[minimum height=0cm,inner sep=1pt,outer sep=1pt,anchor=base,baseline]
        \node[](A1) at (0,0) {$a_0$};
        \draw(A1)edge[->,loop right](A1);
        \end{tikzpicture}\def\GRAPHFOUND{1}}{}\ifthenelse{\equal{#1}{a0}}{\begin{tikzpicture}[minimum height=0cm,inner sep=1pt,outer sep=1pt,anchor=base,baseline]
        \node[](A1) at (0,0) {$a_0$};
        \end{tikzpicture}\def\GRAPHFOUND{1}}{}\ifthenelse{\equal{#1}{a1a3La2->b}}{\begin{tikzpicture}[minimum height=0cm,inner sep=1pt,outer sep=1pt,anchor=base,baseline]
        \node[](A1) at (0,0) {$a_1$};
        \node[node distance=.5ex,right=of A1](A4) {$a_3$};
        \draw(A4)edge[->,loop right](A4);
        \node[node distance=.5ex,below=of A1](A2) {$a_2$};
        \node[node distance=2ex,right=of A2](A3) {$b$};
        \draw(A2)edge[->](A3);
        \end{tikzpicture}\def\GRAPHFOUND{1}}{}\ifthenelse{\equal{#1}{aa}}{\begin{tikzpicture}[minimum height=0cm,inner sep=1pt,outer sep=1pt,anchor=base,baseline]
        \node[](A1) at (0,0) {$a$};
        \node[node distance=.5ex,right=of A1](A4) {$a$};
        \end{tikzpicture}\def\GRAPHFOUND{1}}{}\ifthenelse{\equal{#1}{bb}}{\begin{tikzpicture}[minimum height=0cm,inner sep=1pt,outer sep=1pt,anchor=base,baseline]
        \node[](A1) at (0,0) {$b$};
        \node[node distance=.5ex,right=of A1](A4) {$b$};
        \end{tikzpicture}\def\GRAPHFOUND{1}}{}\ifthenelse{\equal{#1}{aLa->b}}{\begin{tikzpicture}[minimum height=0cm,inner sep=1pt,outer sep=1pt,anchor=base,baseline]
        \node[](A1) at (0,0) {$\vphantom{aa'b}a'$};
        \draw(A1)edge[->,loop left](A1);
        \node[node distance=.5ex,right=of A1](A2) {$\vphantom{aa'b}a$};
        \node[node distance=2ex,right=of A2](A3) {$\vphantom{aa'b}b$};
        \draw(A2)edge[->](A3);
        \end{tikzpicture}\def\GRAPHFOUND{1}}{}\ifthenelse{\equal{#1}{aaL}}{\begin{tikzpicture}[minimum height=0cm,inner sep=1pt,outer sep=1pt,anchor=base,baseline]
        \node[](A1) at (0,0) {$a$};
        \node[node distance=.5ex,right=of A1](A2) {$a$};
        \draw(A2)edge[->,loop right](A2);
        \end{tikzpicture}\def\GRAPHFOUND{1}}{}\ifthenelse{\equal{#1}{a1a2LHOR}}{\begin{tikzpicture}[minimum height=0cm,inner sep=1pt,outer sep=1pt,anchor=base,baseline]
        \node[](A1) at (0,0) {$a_1$};
        \node[node distance=.5ex,right=of A1](A2) {$a_2$};
        \draw(A2)edge[->,loop right](A2);
    \end{tikzpicture}\def\GRAPHFOUND{1}}{}\ifthenelse{\equal{#1}{a1La2L}}{\begin{tikzpicture}[minimum height=0cm,inner sep=1pt,outer sep=1pt,anchor=base,baseline]
        \node[](A1) at (0,0) {$a_1$};
        \node[node distance=.5ex,right=of A1](A2) {$a_2$};
        \draw(A2)edge[->,loop right](A2);
        \draw(A1)edge[->,loop left](A1);
    \end{tikzpicture}\def\GRAPHFOUND{1}}{}\ifthenelse{\equal{#1}{a1La2La3}}{\begin{tikzpicture}[minimum height=0cm,inner sep=1pt,outer sep=1pt,anchor=base,baseline]
        \node[](A1) at (0,0) {$a_1$};
        \node[node distance=.5ex,right=of A1](A2) {$a_2$};
        \draw(A2)edge[->,loop right](A2);
        \draw(A1)edge[->,loop left](A1);
        \node[node distance=3.5ex,right=of A2](A3) {$a_3$};
    \end{tikzpicture}\def\GRAPHFOUND{1}}{}\ifthenelse{\equal{#1}{bL}}{\begin{tikzpicture}[minimum height=0cm,inner sep=1pt,outer sep=1pt,anchor=base,baseline]
        \node[](A1) at (0,0) {$b$};
        \draw(A1)edge[->,loop right](A1);
        \end{tikzpicture}\def\GRAPHFOUND{1}}{}\ifthenelse{\equal{#1}{aL->bL}}{\begin{tikzpicture}[minimum height=0cm,inner sep=1pt,outer sep=1pt,anchor=base,baseline]
        \node[](A1) at (0,0) {$a$};
        \node[node distance=2ex,right=of A1](A3) {$b$};
        \draw(A1)edge[->](A3);
        \draw(A3)edge[->,loop right](A3);
        \draw(A1)edge[->,loop left](A1);
        \end{tikzpicture}\def\GRAPHFOUND{1}}{}\ifthenelse{\equal{#1}{aLL->b}}{\begin{tikzpicture}[minimum height=0cm,inner sep=1pt,outer sep=1pt,anchor=base,baseline]
        \node[](A1) at (0,0) {$a$};
        \node[node distance=2ex,right=of A1](A3) {$b$};
        \draw(A1)edge[->](A3);
        \draw(A1)edge[->,loop left](A1);
        \draw(A1)edge[->,loop above](A1);
        \end{tikzpicture}\def\GRAPHFOUND{1}}{}\ifthenelse{\equal{#1}{aL->b}}{\begin{tikzpicture}[minimum height=0cm,inner sep=1pt,outer sep=1pt,anchor=base,baseline]
        \node[](A1) at (0,0) {$a$};
        \node[node distance=2ex,right=of A1](A3) {$b$};
        \draw(A1)edge[->](A3);
        \draw(A1)edge[->,loop left](A1);
        \end{tikzpicture}\def\GRAPHFOUND{1}}{}\ifthenelse{\equal{#1}{a1a2->bL}}{\begin{tikzpicture}[minimum height=0cm,inner sep=1pt,outer sep=1pt,anchor=base,baseline]
        \node[](A1) at (0,0) {$a_1$};
        \node[node distance=.5ex,below=of A1](A2) {$a_2$};
        \node[node distance=2ex,right=of A2](A3) {$b$};
        \draw(A2)edge[->](A3);
        \draw(A3)edge[->,loop right,line width=1.5pt,red](A3);
        \end{tikzpicture}\def\GRAPHFOUND{1}}{}\ifthenelse{\equal{#1}{a1a2->b}}{\begin{tikzpicture}[minimum height=0cm,inner sep=1pt,outer sep=1pt,anchor=base,baseline]
        \node[](A1) at (0,0) {$a_1$};
        \node[node distance=.5ex,below=of A1](A2) {$a_2$};
        \node[node distance=2ex,right=of A2](A3) {$b$};
        \draw(A2)edge[->](A3);
        \end{tikzpicture}\def\GRAPHFOUND{1}}{}\ifthenelse{\equal{#1}{a1a2L->b}}{\begin{tikzpicture}[minimum height=0cm,inner sep=1pt,outer sep=1pt,anchor=base,baseline]
        \node[](A1) at (0,0) {$a_1$};
        \node[node distance=.5ex,below=of A1](A2) {$a_2$};
        \node[node distance=2ex,right=of A2](A3) {$b$};
        \draw(A2)edge[->](A3);
        \draw(A2)edge[<-,loop left,line width=1.5pt,red](A2);
        \end{tikzpicture}\def\GRAPHFOUND{1}}{}\ifthenelse{\equal{#1}{a<-b->a}}{\begin{tikzpicture}[minimum height=0cm,inner sep=1pt,outer sep=1pt,anchor=base,baseline]
        \node[](A1) at (0,0) {$a$};
        \node[node distance=2ex,right=of A1](A2) {$a$};
        \node[node distance=2ex,right=of A2](A3) {$a$};
        \draw(A2)edge[->](A1);
        \draw(A2)edge[->](A3);
        \end{tikzpicture}\def\GRAPHFOUND{1}}{}\ifthenelse{\equal{#1}{a2LL}}{\begin{tikzpicture}[minimum height=0cm,inner sep=1pt,outer sep=1pt,anchor=base,baseline]
        \node[](A1) at (0,0) {$a_2$};
        \draw(A1)edge[<-,loop left](A1);
        \draw(A1)edge[->,loop right](A1);
        \end{tikzpicture}\def\GRAPHFOUND{1}}{}\ifthenelse{\equal{#1}{a2}}{\begin{tikzpicture}[minimum height=0cm,inner sep=1pt,outer sep=1pt,anchor=base,baseline]
        \node[](A1) at (0,0) {$a_2$};
        \end{tikzpicture}\def\GRAPHFOUND{1}}{}\ifthenelse{\equal{#1}{a1a2a3a4a5}}{\begin{tikzpicture}[minimum height=0cm,inner sep=1pt,outer sep=1pt,anchor=base,baseline]
        \node[](A1) at (0,0) {$a_1$};
        \node[node distance=.5ex,right=of A1](A2) {$a_2$};
        \node[node distance=.5ex,right=of A2](A3) {$a_3$};
        \node[node distance=.5ex,right=of A3](A4) {$a_4$};
        \node[node distance=.5ex,right=of A4](A5) {$a_5$};
        \end{tikzpicture}\def\GRAPHFOUND{1}}{}\ifthenelse{\equal{#1}{a1->a2}}{\begin{tikzpicture}[minimum height=0cm,inner sep=1pt,outer sep=1pt,anchor=base,baseline]
        \node[](A1) at (0,0) {$\vphantom{b}a_1$};
        \node[node distance=2ex,right=of A1](A2)  {$\vphantom{b}a_2$};
        \draw(A1)edge[->](A2);
        \end{tikzpicture}\def\GRAPHFOUND{1}}{}\ifthenelse{\equal{#1}{a1->b2}}{\begin{tikzpicture}[minimum height=0cm,inner sep=1pt,outer sep=1pt,anchor=base,baseline]
        \node[](A1) at (0,0) {$\vphantom{b}a_1$};
        \node[node distance=2ex,right=of A1](A2)  {$\vphantom{b}b_2$};
        \draw(A1)edge[->](A2);
        \end{tikzpicture}\def\GRAPHFOUND{1}}{}\ifthenelse{\equal{#1}{a0->b0}}{\begin{tikzpicture}[minimum height=0cm,inner sep=1pt,outer sep=1pt,anchor=base,baseline]
        \node[](A1) at (0,0) {$\vphantom{b}a_0$};
        \node[node distance=2ex,right=of A1](A2)  {$\vphantom{b}b_0$};
        \draw(A1)edge[->](A2);
        \end{tikzpicture}\def\GRAPHFOUND{1}}{}\ifthenelse{\equal{#1}{a1->b0}}{\begin{tikzpicture}[minimum height=0cm,inner sep=1pt,outer sep=1pt,anchor=base,baseline]
        \node[](A1) at (0,0) {$\vphantom{b}a_1$};
        \node[node distance=2ex,right=of A1](A2)  {$\vphantom{b}b_0$};
        \draw(A1)edge[->](A2);
        \end{tikzpicture}\def\GRAPHFOUND{1}}{}\ifthenelse{\equal{#1}{a1->c0}}{\begin{tikzpicture}[minimum height=0cm,inner sep=1pt,outer sep=1pt,anchor=base,baseline]
        \node[](A1) at (0,0) {$\vphantom{b}a_1$};
        \node[node distance=2ex,right=of A1](A2)  {$\vphantom{b}c_0$};
        \draw(A1)edge[->](A2);
        \end{tikzpicture}\def\GRAPHFOUND{1}}{}\ifthenelse{\equal{#1}{a0->c0}}{\begin{tikzpicture}[minimum height=0cm,inner sep=1pt,outer sep=1pt,anchor=base,baseline]
        \node[](A1) at (0,0) {$\vphantom{b}a_0$};
        \node[node distance=2ex,right=of A1](A2)  {$\vphantom{b}c_0$};
        \draw(A1)edge[->](A2);
        \end{tikzpicture}\def\GRAPHFOUND{1}}{}\ifthenelse{\equal{#1}{a1->c3}}{\begin{tikzpicture}[minimum height=0cm,inner sep=1pt,outer sep=1pt,anchor=base,baseline]
        \node[](A1) at (0,0) {$\vphantom{b}a_1$};
        \node[node distance=2ex,right=of A1](A2)  {$\vphantom{b}c_3$};
        \draw(A1)edge[->](A2);
        \end{tikzpicture}\def\GRAPHFOUND{1}}{}\ifthenelse{\equal{#1}{a}}{\begin{tikzpicture}[minimum height=0cm,inner sep=1pt,outer sep=1pt,anchor=base,baseline]
        \node(A1) at (0,0) {$a$};
        \end{tikzpicture}\def\GRAPHFOUND{1}}{}\ifthenelse{\equal{#1}{b}}{\begin{tikzpicture}[minimum height=0cm,inner sep=1pt,outer sep=1pt,anchor=base,baseline]
        \node(A1) at (0,0) {$b$};
        \end{tikzpicture}\def\GRAPHFOUND{1}}{}\ifthenelse{\equal{#1}{c}}{\begin{tikzpicture}[minimum height=0cm,inner sep=1pt,outer sep=1pt,anchor=base,baseline]
        \node(A1) at (0,0) {$c$};
        \end{tikzpicture}\def\GRAPHFOUND{1}}{}\ifthenelse{\equal{#1}{aL}}{\begin{tikzpicture}[minimum height=0cm,inner sep=1pt,outer sep=1pt,anchor=base,baseline]
        \node(A1) at (0,0) {$a$};
        \draw(A1)edge[loop right](A1);
        \end{tikzpicture}\def\GRAPHFOUND{1}}{}\ifthenelse{\equal{#1}{aLleft}}{\begin{tikzpicture}[minimum height=0cm,inner sep=1pt,outer sep=1pt,anchor=base,baseline]
        \node(A1) at (0,0) {$a$};
        \draw(A1)edge[loop left](A1);
        \end{tikzpicture}\def\GRAPHFOUND{1}}{}\ifthenelse{\equal{#1}{a->b}}{\begin{tikzpicture}[minimum height=0cm,inner sep=1pt,outer sep=1pt,anchor=base,baseline]
        \node[](A1) at (0,0) {$\strut a$};
        \node[node distance=2ex,right=of A1] (A2) {$\strut b$};
        \draw(A1)edge[->](A2);
        \end{tikzpicture}\def\GRAPHFOUND{1}}{}\ifthenelse{\equal{#1}{a->bL->c}}{\begin{tikzpicture}[minimum height=0cm,inner sep=1pt,outer sep=1pt,anchor=base,baseline]
        \node[](A1) at (0,0) {$\strut a$};
        \node[node distance=2ex,right=of A1] (A2) {$\strut b$};
        \node[node distance=2ex,right=of A2] (A3) {$\strut c$};
        \draw(A1)edge[->](A2);
        \draw(A2)edge[loop right,->](A2);
        \draw(A2)edge[->](A3);
        \end{tikzpicture}\def\GRAPHFOUND{1}}{}\ifthenelse{\equal{#1}{a->bL}}{\begin{tikzpicture}[minimum height=0cm,inner sep=1pt,outer sep=1pt,anchor=base,baseline]
        \node[](A1) at (0,0) {$\strut a$};
        \node[node distance=2ex,right=of A1] (A2) {$\strut b$};
        \draw(A1)edge[->](A2);
        \draw(A2)edge[loop right,->](A2);
        \end{tikzpicture}\def\GRAPHFOUND{1}}{}\ifthenelse{\equal{#1}{b->c}}{\begin{tikzpicture}[minimum height=0cm,inner sep=1pt,outer sep=1pt,anchor=base,baseline]
        \node[](A1) at (0,0) {$\strut b$};
        \node[node distance=2ex,right=of A1] (A2) {$\strut c$};
        \draw(A1)edge[->](A2);
        \end{tikzpicture}\def\GRAPHFOUND{1}}{}\ifthenelse{\equal{#1}{a->b->c}}{\begin{tikzpicture}[minimum height=0cm,inner sep=1pt,outer sep=1pt,anchor=base,baseline]
        \node[](A1) at (0,0) {$\strut a$};
        \node[node distance=2ex,right=of A1] (A2) {$\strut b$};
        \node[node distance=2ex,right=of A2] (A3) {$\strut c$};
        \draw(A1)edge[->](A2);
        \draw(A2)edge[->](A3);
        \end{tikzpicture}\def\GRAPHFOUND{1}}{}\ifthenelse{\equal{#1}{b<-a->c}}{\begin{tikzpicture}[minimum height=0cm,inner sep=1pt,outer sep=1pt,anchor=base,baseline]
        \node[](A1) at (0,0) {$\strut a$};
        \node[node distance=2ex,left=of A1] (A2) {$\strut b$};
        \draw(A1)edge[->](A2);
        \node[node distance=2ex,right=of A1] (A3) {$\strut c$};
        \draw(A1)edge[->](A3);        
        \end{tikzpicture}\def\GRAPHFOUND{1}}{}\ifthenelse{\equal{#1}{b2<-a1->c3}}{\begin{tikzpicture}[minimum height=0cm,inner sep=1pt,outer sep=1pt,anchor=base,baseline]
        \node[](A1) at (0,0) {$\strut a_1$};
        \node[node distance=2ex,left=of A1] (A2) {$\strut b_2$};
        \draw(A1)edge[->](A2);
        \node[node distance=2ex,right=of A1] (A3) {$\strut c_3$};
        \draw(A1)edge[->](A3);
        \end{tikzpicture}\def\GRAPHFOUND{1}}{}\ifthenelse{\equal{#1}{a->ba->c}}{\begin{tikzpicture}[minimum height=0cm,inner sep=1pt,outer sep=1pt,anchor=base,baseline]
        \node[](A1) at (0,0) {$\strut a$};
        \node[node distance=2ex,right=of A1] (A2) {$\strut b$};
        \node[node distance=1ex,right=of A2] (A3) {$\strut a$};
        \node[node distance=2ex,right=of A3] (A4) {$\strut c$};        
        \draw(A1)edge[->](A2);
        \draw(A3)edge[->](A4);
        \end{tikzpicture}\def\GRAPHFOUND{1}}{}\ifthenelse{\equal{#1}{a->bLb->c}}{\begin{tikzpicture}[minimum height=0cm,inner sep=1pt,outer sep=1pt,anchor=base,baseline]
        \node[](A1) at (0,0) {$\strut a$};
        \node[node distance=2ex,right=of A1] (A2) {$\strut b$};
        \node[node distance=2.5ex,right=of A2] (A3) {$\strut b$};
        \node[node distance=2ex,right=of A3] (A4) {$\strut c$};        
        \draw(A1)edge[->](A2);
        \draw(A3)edge[->](A4);
        \draw(A2)edge[loop right,->](A2);
        \end{tikzpicture}\def\GRAPHFOUND{1}}{}\ifthenelse{\equal{#1}{ab}}{\begin{tikzpicture}[minimum height=0cm,inner sep=1pt,outer sep=1pt,anchor=base,baseline]
        \node[](A1) at (0,0) {$\strut a$};
        \node[node distance=.15ex,right=of A1] (A2) {$\strut b$};
        \end{tikzpicture}\def\GRAPHFOUND{1}}{}\ifthenelse{\equal{#1}{a->bb->c}}{\begin{tikzpicture}[minimum height=0cm,inner sep=1pt,outer sep=1pt,anchor=base,baseline]
        \node[](A1) at (0,0) {$\strut a$};
        \node[node distance=2ex,right=of A1] (A2) {$\strut b$};
        \node[node distance=1ex,right=of A2] (A3) {$\strut b$};
        \node[node distance=2ex,right=of A3] (A4) {$\strut c$};        
        \draw(A1)edge[->](A2);
        \draw(A3)edge[->](A4);
        \end{tikzpicture}\def\GRAPHFOUND{1}}{}\ifthenelse{\equal{#1}{a->c}}{\begin{tikzpicture}[minimum height=0cm,inner sep=1pt,outer sep=1pt,anchor=base,baseline]
        \node[](A1) at (0,0) {$\strut a$};
        \node[node distance=2ex,right=of A1] (A2) {$\strut c$};
        \draw(A1)edge[->](A2);
        \end{tikzpicture}\def\GRAPHFOUND{1}}{}\ifthenelse{\equal{#1}{a'->a->b}}{\begin{tikzpicture}[minimum height=0cm,inner sep=1pt,outer sep=1pt,anchor=base,baseline]
        \node[](A1) at (0,0) {$\vphantom{a'b}a'$};
        \node[node distance=2ex,right=of A1] (A2) {$\vphantom{a'b}a$};
        \node[node distance=2ex,right=of A2] (A3) {$\vphantom{a'b}b$};
        \draw(A1)edge[->](A2);
        \draw(A2)edge[->](A3);
        \end{tikzpicture}\def\GRAPHFOUND{1}}{}\ifthenelse{\equal{#1}{a'->a}}{\begin{tikzpicture}[minimum height=0cm,inner sep=1pt,outer sep=1pt,anchor=base,baseline]
        \node[](A1) at (0,0) {$\vphantom{a'b}a'$};
        \node[node distance=2ex,right=of A1] (A2) {$\vphantom{a'b}a$};
        \draw(A1)edge[->](A2);
        \end{tikzpicture}\def\GRAPHFOUND{1}}{}\ifthenelse{\equal{#1}{MIN01}}{\begin{tikzpicture}[minimum height=0cm,scale=1,inner sep=1pt,outer sep=1pt,anchor=base,baseline,minimum width=0pt,minimum height=0pt]
        \MYNODE{a}{1}{0,0}
        \MYNODE{a}{2}{1,0}
        \MYNODE{a}{3}{2,0}
        \MYNODE{a}{4}{2.5,1}
        \MYNODE{b}{1}{1,-1}
        \MYNODE{b}{2}{3,0}
        \draw(a1)edge[->](b1);
        \draw(a2)edge[->](b1);
        \draw(a3)edge[->](b1);
        \draw(a4)edge[->](a3);
        \draw(a4)edge[->](b2);
        \end{tikzpicture}\def\GRAPHFOUND{1}}{}\ifthenelse{\equal{#1}{MIN02}}{\begin{tikzpicture}[minimum height=0cm,scale=1,inner sep=1pt,outer sep=1pt,anchor=base,baseline,minimum width=0pt,minimum height=0pt]
        \MYNODE{b}{1}{0,0}
        \MYNODE{a}{1}{1,0}
        \MYNODE{b}{2}{2,0}
        \MYNODE{a}{2}{.5,1}
        \MYNODE{a}{3}{1.5,1}
        \MYNODE{a}{4}{2.5,1}
        \MYNODE{b}{3}{1,-1}
        \draw(a2)edge[->](b1);
        \draw(a2)edge[->](a1);
        \draw(a1)edge[->](b3);
        \draw(a3)edge[->](b2);
        \draw(a4)edge[->](b2);
        \end{tikzpicture}\def\GRAPHFOUND{1}}{}\ifthenelse{\equal{#1}{MIN03}}{\begin{tikzpicture}[minimum height=0cm,scale=1,inner sep=1pt,outer sep=1pt,anchor=base,baseline,minimum width=0pt,minimum height=0pt]
        \MYNODE{b}{1}{0,0}
        \MYNODE{a}{2}{1,0}
        \MYNODE{a}{1}{.5,1}
        \MYNODE{b}{2}{1,-1}
        \draw(a1)edge[->](b1);
        \draw(a1)edge[->](a2);
        \draw(a2)edge[->](b2);
        \end{tikzpicture}\def\GRAPHFOUND{1}}{}\ifthenelse{\equal{#1}{MIN03inter}}{\begin{tikzpicture}[minimum height=0cm,scale=1,inner sep=1pt,outer sep=1pt,anchor=base,baseline,minimum width=0pt,minimum height=0pt]
        \MYNODE{a}{2}{1,0}
        \MYNODE{a}{1}{.5,1}
        \MYNODE{b}{2}{1,-1}
        \draw(a1)edge[->](a2);
        \draw(a2)edge[->](b2);
        \end{tikzpicture}\def\GRAPHFOUND{1}}{}\ifthenelse{\equal{#1}{MIN03right}}{\begin{tikzpicture}[minimum height=0cm,scale=1,inner sep=1pt,outer sep=1pt,anchor=base,baseline,minimum width=0pt,minimum height=0pt]
        \MYNODE{a}{2}{1,0}
        \MYNODE{a}{1}{.5,1}
        \MYNODE{b}{2}{1,-1}
        \draw(a1)edge[->](a2);
        \draw(a1)edge[loop right,->](a1);
        \draw(a2)edge[->](b2);
        \end{tikzpicture}\def\GRAPHFOUND{1}}{}\ifthenelse{\equal{#1}{MIN04}}{\begin{tikzpicture}[minimum height=0cm,scale=1,inner sep=1pt,outer sep=1pt,anchor=base,baseline,minimum width=0pt,minimum height=0pt]
        \MYNODE{a}{1}{0,0}
        \MYNODE{a}{2}{1,0}
        \MYNODE{b}{1}{.5,-1}
        \draw(a1)edge[->](a2);
        \draw(a1)edge[->](b1);
        \draw(a2)edge[->](b1);
        \end{tikzpicture}\def\GRAPHFOUND{1}}{}\ifthenelse{\equal{#1}{NoAttributeExample01}}{\begin{tikzpicture}[minimum height=0cm,scale=1,inner sep=1pt,outer sep=1pt,anchor=base,baseline,minimum width=0pt,minimum height=0pt]
        \MYNODE[mybeameralert3={2-6}]{a}{1}{0,0}
        \MYNODE[mybeameralert3={2-3}]{b}{1}{-.5,-1}
        \MYNODE[mybeameralert3={4-6}]{b}{2}{.5,-1}
        \MYNODE[mybeameralert3={5}]{c}{1}{.5,-2}
        \draw(a1)edge[->,mybeameralertedge={2-3}](b1);
        \draw(a1)edge[->,mybeameralertedge={4-6}](b2);
        \draw(b2)edge[->,mybeameralertedge={5}](c1);
        \draw(b2)edge[->,mybeameralertedge={6},loop right](b2);
        \end{tikzpicture}\def\GRAPHFOUND{1}}{}\ifthenelse{\equal{#1}{NoAttributeExample01Min1}}{\begin{tikzpicture}[minimum height=0cm,scale=1,inner sep=1pt,outer sep=1pt,anchor=base,baseline,minimum width=0pt,minimum height=0pt]
        \MYNODE[]{a}{1}{0,0}
        \MYNODE[]{b}{1}{1,0}
        \MYNODE[]{c}{1}{2,0}
        \draw(a1)edge[->](b1);
        \draw(b1)edge[->](c1);
        \end{tikzpicture}\def\GRAPHFOUND{1}}{}\ifthenelse{\equal{#1}{NoAttributeExample01Min2}}{\begin{tikzpicture}[minimum height=0cm,scale=1,inner sep=1pt,outer sep=1pt,anchor=base,baseline,minimum width=0pt,minimum height=0pt]
        \MYNODE[]{a}{1}{0,0}
        \MYNODE[]{b}{1}{1,0}
        \draw(a1)edge[->,](b1);
        \draw(b1)edge[->,loop right](b1);
        \end{tikzpicture}\def\GRAPHFOUND{1}}{}\ifthenelse{\equal{#1}{NoAttributeExample01Min3}}{\begin{tikzpicture}[minimum height=0cm,scale=1,inner sep=1pt,outer sep=1pt,anchor=base,baseline,minimum width=0pt,minimum height=0pt]
        \MYNODE[]{a}{1}{0,0}
        \MYNODE[]{b}{1}{1,0}
        \draw(a1)edge[->,bend left=10](b1);
        \draw(a1)edge[->,bend left=-10](b1);
        \draw(b1)edge[->,loop right](b1);
        \end{tikzpicture}\def\GRAPHFOUND{1}}{}\ifthenelse{\equal{#1}{NoAttributeExample01MinNeg1}}{\begin{tikzpicture}[minimum height=0cm,scale=1,inner sep=1pt,outer sep=1pt,anchor=base,baseline,minimum width=0pt,minimum height=0pt]
        \MYNODE[]{a}{1}{1,0}
        \MYNODE[]{b}{1}{0,0}
        \draw(b1)edge[->](a1);
        \end{tikzpicture}\def\GRAPHFOUND{1}}{}\ifthenelse{\equal{#1}{NoAttributeExample01MinNeg2}}{\begin{tikzpicture}[minimum height=0cm,scale=1,inner sep=1pt,outer sep=1pt,anchor=base,baseline,minimum width=0pt,minimum height=0pt]
        \MYNODE[]{a}{1}{1,0}
        \MYNODE[]{b}{2}{0,0}
        \draw(b1)edge[->](a1);
        \end{tikzpicture}\def\GRAPHFOUND{1}}{}\ifthenelse{\equal{#1}{NoAttributeExample01Min4}}{\begin{tikzpicture}[minimum height=0cm,scale=1,inner sep=1pt,outer sep=1pt,anchor=base,baseline,minimum width=0pt,minimum height=0pt]
        \MYNODE[]{a}{1}{0,0}
        \MYNODE[]{b}{1}{1,0}
        \MYNODE[]{b}{2}{-1,0}
        \MYNODE[]{c}{1}{2,0}
        \draw(a1)edge[->](b1);
        \draw(b1)edge[->](c1);
        \draw(a1)edge[->](b2);
        \end{tikzpicture}\def\GRAPHFOUND{1}}{}\ifthenelse{\equal{#1}{NoAttributeExample01Min5}}{\begin{tikzpicture}[minimum height=0cm,scale=1,inner sep=1pt,outer sep=1pt,anchor=base,baseline,minimum width=0pt,minimum height=0pt]
        \MYNODE[]{a}{1}{0,0}
        \MYNODE[]{b}{1}{1,0}
        \MYNODE[]{b}{2}{-1,0}
        \draw(a1)edge[->,](b1);
        \draw(a1)edge[->,](b2);
        \draw(b1)edge[->,loop right](b1);
        \end{tikzpicture}\def\GRAPHFOUND{1}}{}\ifthenelse{\equal{#1}{NoAttributeExample01Min6}}{\begin{tikzpicture}[minimum height=0cm,scale=1,inner sep=1pt,outer sep=1pt,anchor=base,baseline,minimum width=0pt,minimum height=0pt]
        \MYNODE[]{a}{1}{0,0}
        \MYNODE[]{b}{1}{1,0}
        \MYNODE[]{c}{1}{2,0}
        \draw(a1)edge[->,bend left=10](b1);
        \draw(a1)edge[->,bend left=-10](b1);
        \draw(b1)edge[->](c1);
        \end{tikzpicture}\def\GRAPHFOUND{1}}{}\ifthenelse{\equal{#1}{NoAttributeExample01Cond01}}{\begin{tikzpicture}[minimum height=0cm,scale=1,inner sep=1pt,outer sep=1pt,anchor=base,baseline,minimum width=0pt,minimum height=0pt]
        \MYNODE[mybeameralert3={2-3}]{a}{0}{0,0}
        \MYNODE[mybeameralert3={2-3}]{b}{0}{1,0}
        \draw(a0)edge[->,mybeameralertedge={2-3}](b0);
        \end{tikzpicture}\def\GRAPHFOUND{1}}{}\ifthenelse{\equal{#1}{NoAttributeExample01Cond05}}{\begin{tikzpicture}[minimum height=0cm,scale=1,inner sep=1pt,outer sep=1pt,anchor=base,baseline,minimum width=0pt,minimum height=0pt]
        \MYNODE[mybeameralert3={4-6}]{a}{0}{0,0}
        \MYNODE[mybeameralert3={4-6}]{b}{0}{1,0}
        \draw(a0)edge[->,mybeameralertedge={4-6}](b0);
        \end{tikzpicture}\def\GRAPHFOUND{1}}{}\ifthenelse{\equal{#1}{NoAttributeExample01Cond02}}{\begin{tikzpicture}[minimum height=0cm,scale=1,inner sep=1pt,outer sep=1pt,anchor=base,baseline,minimum width=0pt,minimum height=0pt]
        \MYNODE[mybeameralert3={3}]{b}{0}{0,0}
        \MYNODE[mybeameralert3={3}]{a}{0}{1,0}
        \draw(b0)edge[->,mybeameralertedge={3}](a0);
        \end{tikzpicture}\def\GRAPHFOUND{1}}{}\ifthenelse{\equal{#1}{NoAttributeExample01Cond03}}{\begin{tikzpicture}[minimum height=0cm,scale=1,inner sep=1pt,outer sep=1pt,anchor=base,baseline,minimum width=0pt,minimum height=0pt]
        \MYNODE[mybeameralert3={5}]{b}{0}{0,0}
        \MYNODE[mybeameralert3={5}]{c}{0}{1,0}
        \draw(b0)edge[->,mybeameralertedge={5}](c0);
        \end{tikzpicture}\def\GRAPHFOUND{1}}{}\ifthenelse{\equal{#1}{NoAttributeExample01Cond04}}{\begin{tikzpicture}[minimum height=0cm,scale=1,inner sep=1pt,outer sep=1pt,anchor=base,baseline,minimum width=0pt,minimum height=0pt]
        \MYNODE[mybeameralert3={6}]{b}{0}{0,0}
        \draw(b0)edge[->,mybeameralertedge={6},loop right](b0);
        \end{tikzpicture}\def\GRAPHFOUND{1}}{}\ifthenelse{\equal{#1}{BasicAttributeExample01}}{\begin{tikzpicture}[mtgc_inner]
        \SetNodeWidthName{BasicAttributeExample01}
        \node[box=colorUser] (Graph1N0) at (0,0) {\UserNode{u_1}{}{Julia}{30}};
        \node[box=colorUser,node distance=\eqboxheight{emptyedgelabel}+1.5ex,below=of Graph1N0.south west,anchor=north west] (Graph1N1) {\UserNode{u_2}{}{Bob}{15}};
        \node[box=colorUser,node distance=1ex,below=of Graph1N1.south west,anchor=north west] (Graph1N2) {\UserNode{u_3}{}{Adam}{35}};
        \draw ($(Graph1N0.south west)+(1ex,0ex)$) edge[->,line width=1pt]
            node[boxX=2,right] (emptyLabelE1)
                {\MyEdge{e_1}{knows}{}}
            ($(Graph1N1.north west)+(1ex,0ex)$);
        \ENLARGE{emptyLabelE1}{emptyedgelabel}
        \end{tikzpicture}\def\GRAPHFOUND{1}}{}\ifthenelse{\equal{#1}{BasicAttributeExample01Min1}}{\begin{tikzpicture}[mtgc_inner]
        \SetNodeWidthName{BasicAttributeExample01}
        \node[box=colorUser] (Graph1N0) at (0,0) {\UserNode{u_1}{}{Julia}{}};
        \node[box=colorUser,node distance=\eqboxheight{emptyedgelabel}+1.5ex,below=of Graph1N0.south west,anchor=north west] (Graph1N1) {\UserNode{u_2}{}{}{}};
        \draw ($(Graph1N0.south west)+(1ex,0ex)$) edge[->,line width=1pt]
            node[boxX=2,right] (emptyLabelE1)
                {\MyEdge{e_1}{knows}{}}
            ($(Graph1N1.north west)+(1ex,0ex)$);
        \ENLARGE{emptyLabelE1}{emptyedgelabel}
        \end{tikzpicture}\def\GRAPHFOUND{1}}{}\ifthenelse{\equal{#1}{BasicAttributeExample01Min2}}{\begin{tikzpicture}[mtgc_inner]
        \SetNodeWidthName{BasicAttributeExample01}
        \node[box=colorUser] (Graph1N0) at (0,0) {\UserNode{u_1}{}{Julia}{x\geq 20}};
        \node[box=colorUser,node distance=\eqboxheight{emptyedgelabel}+1.5ex,below=of Graph1N0.south west,anchor=north west] (Graph1N1) {\UserNode{u_2}{}{Bob}{y<x}};
        \draw ($(Graph1N0.south west)+(1ex,0ex)$) edge[->,line width=1pt]
            node[boxX=2,right] (emptyLabelE1)
                {\MyEdge{e_1}{knows}{}}
            ($(Graph1N1.north west)+(1ex,0ex)$);
        \ENLARGE{emptyLabelE1}{emptyedgelabel}
        \end{tikzpicture}\def\GRAPHFOUND{1}}{}\ifthenelse{\equal{#1}{BasicAttributeExample01Cond01}}{\begin{tikzpicture}[line width=1pt]
        \SetNodeWidthName{BasicAttributeExample01}

        \node[mtgc_outer] (node0) at (0,0) {\begin{tikzpicture}
            \node[conditionGraphNode] (Graph1) at (0,0) {\VerticalExtend{BasicAttributeExample01Cond01Line1height}{\begin{tikzpicture}[mtgc_inner]
                \node[box=colorUser] (Graph1N0) at (0,0) {\UserNode{u_4}{}{Julia}{}};
                \end{tikzpicture}}};
            \OPENINGBRACKET{Graph1}
            \node[outer xsep=0ex,inner xsep=0pt,node distance=.8ex,left=of Graph1,anchor=east] (LEFT)
                {\large$\GCexistsNAME$};
            \node[outer xsep=0ex,inner xsep=0pt,node distance=0ex,right=of Graph1,anchor=west] (RIGHT)
                {\strut$\COMMA$};
            \end{tikzpicture}};

        \node[mtgc_outer,node distance=0ex,right=of node0,anchor=west] (node1) {\begin{tikzpicture}
            \node[conditionGraphNode] (Graph1) at (0,0) {\eqboxwh{BasicAttributeExample01Cond01Line1height}{\begin{tikzpicture}[mtgc_inner]
                \node[box=colorUser] (Graph1N0) at (0,0) {\UserNode{u_4}{}{}{}};
                \node[box=colorUser,node distance=\eqboxheight{emptyedgelabel}+1.5ex,below=of Graph1N0.south west,anchor=north west] (Graph1N1) {\UserNode{u_5}{}{Adam}{}};
                \draw ($(Graph1N0.south west)+(1ex,0ex)$) edge[->,line width=1pt]
                    node[boxX=2,right] (emptyLabelE1)
                        {\MyEdge{e_2}{knows}{}}
                    ($(Graph1N1.north west)+(1ex,0ex)$);
                \ENLARGE{emptyLabelE1}{emptyedgelabel}
                \end{tikzpicture}}
            };
            \OPENINGBRACKET{Graph1}
            \node[outer xsep=0ex,inner xsep=0pt,node distance=.8ex,left=of Graph1,anchor=east] (LEFT)
                {\large\strut$\GCnegNAME\GCexistsNAME$};
            \node[outer xsep=0ex,inner xsep=0pt,node distance=0ex,right=of Graph1,anchor=west] (RIGHT)
                {\strut$\COMMA\GCtrue$};
            \CLOSINGBRACKET{Graph1}{RIGHT}{}
            \CLOSINGBRACKET[1ex]{Graph1}{RIGHT}{}
            \end{tikzpicture}};

        \uncover<2->{
        \node[xshift=-1.8ex,mtgc_outer,node distance=0ex,below right=of node0.south west,anchor=north west] (node2) {\begin{tikzpicture}
            \node[conditionGraphNode] (Graph1) at (0,0) {\VerticalExtend{BasicAttributeExample01Cond01Line2height}{\begin{tikzpicture}[mtgc_inner]
                \node[box=colorUser] (Graph1N0) at (0,0) {\UserNode{u_6}{}{}{}};
                \end{tikzpicture}}
            };
            \OPENINGBRACKET{Graph1}
            \node[outer xsep=0ex,inner xsep=0pt,node distance=.8ex,left=of Graph1,anchor=east] (LEFT)
                {\large\strut$\GCandNAME\GCexistsNAME$};
            \node[outer xsep=0ex,inner xsep=0pt,node distance=0ex,right=of Graph1,anchor=west] (RIGHT)
                {\strut$\COMMA$};
            \end{tikzpicture}};

        \node[mtgc_outer,node distance=0ex,right=of node2,anchor=west] (node3) {\begin{tikzpicture}
            \node[conditionGraphNode] (Graph1) at (0,0) {\eqboxwh{BasicAttributeExample01Cond01Line2height}{\begin{tikzpicture}[mtgc_inner]
                \node[box=colorUser] (Graph1N0) at (0,0) {\UserNode{u_6}{}{}{}};
                \node[box=colorUser,node distance=\eqboxheight{emptyedgelabel}+1.5ex,below=of Graph1N0.south west,anchor=north west] (Graph1N1) {\UserNode{u_7}{}{}{}};
                \draw ($(Graph1N0.south west)+(1ex,0ex)$) edge[->,line width=1pt]
                    node[boxX=2,right] (emptyLabelE1)
                        {\MyEdge{e_3}{knows}{}}
                    ($(Graph1N1.north west)+(1ex,0ex)$);
                \ENLARGE{emptyLabelE1}{emptyedgelabel}
                \end{tikzpicture}}
            };
            \OPENINGBRACKET{Graph1}
            \node[outer xsep=0ex,inner xsep=0pt,node distance=.8ex,left=of Graph1,anchor=east] (LEFT)
                {\large\strut$\GCexistsNAME$};
            \node[outer xsep=0ex,inner xsep=0pt,node distance=0ex,right=of Graph1,anchor=west] (RIGHT)
                {\strut$\COMMA\GCtrue$};
            \CLOSINGBRACKET{Graph1}{RIGHT}{}
            \CLOSINGBRACKET[1ex]{Graph1}{RIGHT}{}
            \end{tikzpicture}};
        }
        \end{tikzpicture}\def\GRAPHFOUND{1}}{}\ifthenelse{\equal{#1}{BasicAttributeExample01Cond02}}{\begin{tikzpicture}[line width=1pt]
        \SetNodeWidthName{BasicAttributeExample01}

        \node[mtgc_outer] (node0) at (0,0) {\begin{tikzpicture}
            \node[conditionGraphNode] (Graph1) at (0,0) {\VerticalExtend{BasicAttributeExample01Cond02Line1height}{\begin{tikzpicture}[mtgc_inner]
                \node[box=colorUser] (Graph1N0) at (0,0) {\UserNode{u_4}{}{Julia}{x\geq 20}};
                \end{tikzpicture}}};
            \OPENINGBRACKET{Graph1}
            \node[outer xsep=0ex,inner xsep=0pt,node distance=.8ex,left=of Graph1,anchor=east] (LEFT)
                {\large$\GCexistsNAME$};
            \node[outer xsep=0ex,inner xsep=0pt,node distance=0ex,right=of Graph1,anchor=west] (RIGHT)
                {\strut$\COMMA$};
            \end{tikzpicture}};

        \node[mtgc_outer,node distance=0ex,right=of node0,anchor=west] (node1) {\begin{tikzpicture}
            \node[conditionGraphNode] (Graph1) at (0,0) {\eqboxwh{BasicAttributeExample01Cond02Line1height}{\begin{tikzpicture}[mtgc_inner]
                \node[box=colorUser] (Graph1N0) at (0,0) {\UserNode{u_4}{}{}{}};
                \node[box=colorUser,node distance=\eqboxheight{emptyedgelabel}+1.5ex,below=of Graph1N0.south west,anchor=north west] (Graph1N1) {\UserNode{u_5}{}{Bob}{y<x}};
                \draw ($(Graph1N0.south west)+(1ex,0ex)$) edge[->,line width=1pt]
                    node[boxX=2,right] (emptyLabelE1)
                        {\MyEdge{e_2}{knows}{}}
                    ($(Graph1N1.north west)+(1ex,0ex)$);
                \ENLARGE{emptyLabelE1}{emptyedgelabel}
                \end{tikzpicture}}
            };
            \OPENINGBRACKET{Graph1}
            \node[outer xsep=0ex,inner xsep=0pt,node distance=.8ex,left=of Graph1,anchor=east] (LEFT)
                {\large\strut$\GCexistsNAME$};
            \node[outer xsep=0ex,inner xsep=0pt,node distance=0ex,right=of Graph1,anchor=west] (RIGHT)
                {\strut$\COMMA\GCtrue$};
            \CLOSINGBRACKET{Graph1}{RIGHT}{}
            \CLOSINGBRACKET[1ex]{Graph1}{RIGHT}{}
            \end{tikzpicture}};

        \uncover<2->{
        \node[xshift=-1.8ex,mtgc_outer,node distance=0ex,below right=of node0.south west,anchor=north west] (node2) {\begin{tikzpicture}
            \node[conditionGraphNode] (Graph1) at (0,0) {\VerticalExtend{BasicAttributeExample01Cond02Line2height}{\begin{tikzpicture}[mtgc_inner]
                \node[box=colorUser] (Graph1N0) at (0,0) {\UserNode{u_6}{}{}{x}};
                \end{tikzpicture}}
            };
            \OPENINGBRACKET{Graph1}
            \node[outer xsep=0ex,inner xsep=0pt,node distance=.8ex,left=of Graph1,anchor=east] (LEFT)
                {\large\strut$\GCandNAME\GCexistsNAME$};
            \node[outer xsep=0ex,inner xsep=0pt,node distance=0ex,right=of Graph1,anchor=west] (RIGHT)
                {\strut$\COMMA$};
            \end{tikzpicture}};

        \node[mtgc_outer,node distance=0ex,right=of node2,anchor=west] (node3) {\begin{tikzpicture}
            \node[conditionGraphNode] (Graph1) at (0,0) {\eqboxwh{BasicAttributeExample01Cond02Line2height}{\begin{tikzpicture}[mtgc_inner]
                \node[box=colorUser] (Graph1N0) at (0,0) {\UserNode{u_7}{}{}{z>x}};
                \end{tikzpicture}}
            };
            \OPENINGBRACKET{Graph1}
            \node[outer xsep=0ex,inner xsep=0pt,node distance=.8ex,left=of Graph1,anchor=east] (LEFT)
                {\large\strut$\GCexistsNAME$};
            \node[outer xsep=0ex,inner xsep=0pt,node distance=0ex,right=of Graph1,anchor=west] (RIGHT)
                {\strut$\COMMA\GCtrue$};
            \CLOSINGBRACKET{Graph1}{RIGHT}{}
            \CLOSINGBRACKET[1ex]{Graph1}{RIGHT}{}
            \end{tikzpicture}};
        }
        \end{tikzpicture}\def\GRAPHFOUND{1}}{}\ifthenelse{\equal{\GRAPHFOUND}{0}}{\stop\typeout{Graph #1 not found.}}{}
}
 
\makeatletter
\newlength{\maxbox@tempdima}
\newlength{\maxbox@tempdimb}
\newlength{\maxbox@tempdimc}
\def\maxbox@taglist{}
\newif\ifmaxbox@must@rerun
\RequirePackage{array}
\newsavebox{\maxbox@tabular@box}
\newsavebox{\maxbox@list@box}
\newlength{\maxbox@list@indent}
\RequirePackage{environ}
\newcommand*{\maxbox@storefont}{\xdef\maxbox@restorefont{\noexpand\usefont{\f@encoding}{\f@family}{\f@series}{\f@shape}\noexpand\fontsize{\f@size}{\f@baselineskip}\noexpand\selectfont
  }}
\newcommand{\maxbox@settowidth}[2]{\begingroup
    \global\setbox\maxbox@tabular@box=\hbox{\def\maxbox@endings{}\ifx\enditemize\endlist
        \g@addto@macro\maxbox@endings{\let\enditemize=\endlist}\fi
      \ifx\endenumerate\endlist
        \g@addto@macro\maxbox@endings{\let\endenumerate=\endlist}\fi
      \ifx\enddescription\endlist
        \g@addto@macro\maxbox@endings{\let\enddescription=\endlist}\fi
      \renewenvironment{list}[2]{\ifnum \@listdepth >5\relax
          \@toodeep
        \else
          \global\advance\@listdepth\@ne
        \fi
        \rightmargin\z@
        \listparindent\z@
        \itemindent\z@
        \csname @list\romannumeral\the\@listdepth\endcsname
        ##2\relax
        \renewcommand*{\item}[1][]{\mbox{}\\
          \box\maxbox@list@box\mbox{} \\
          \sbox\@tempboxa{\makelabel{####1}}\ifdim\wd\@tempboxa>\labelwidth
            \advance\maxbox@list@indent by -\labelwidth
            \advance\maxbox@list@indent by \wd\@tempboxa
          \fi
          \hspace*{\maxbox@list@indent}}\hspace*{-\maxbox@list@indent}\advance\maxbox@list@indent by \leftmargin
        \advance\maxbox@list@indent by \rightmargin
        \advance\maxbox@list@indent by \itemindent
        \global\setbox\maxbox@list@box=\hbox\bgroup
        \begin{tabular}{@{}l@{}}}{\item[]\end{tabular}\egroup
        \global\advance\@listdepth\m@ne
      }\maxbox@endings
      \global\let\maxbox@par=\par
      \maxbox@storefont
      \begin{tabular}{@{}>{\maxbox@restorefont}l<{\maxbox@storefont}@{}}\global\@setpar{\\}#2\\ \box\maxbox@list@box
      \end{tabular}\global\@restorepar
    }\endgroup
  \settowidth{#1}{\box\maxbox@tabular@box}}
\long\def\maxbox@compute@width#1#2{\maxbox@settowidth{\maxbox@tempdimb}{#2}\settoheight{\maxbox@tempdimc}{\begin{minipage}[b]{\maxbox@tempdimb}#2\end{minipage}}\@ifundefined{maxbox@minwd@#1}{}{\ifdim\maxbox@tempdimb<\csname maxbox@minwd@#1\endcsname
      \maxbox@tempdimb=\csname maxbox@minwd@#1\endcsname
    \fi
  }\@ifundefined{maxbox@maxwd@#1}{}{\ifdim\maxbox@tempdimb>\csname maxbox@maxwd@#1\endcsname
      \maxbox@tempdimb=\csname maxbox@maxwd@#1\endcsname
    \fi
  }\expandafter
  \ifx\csname maxbox@thiswd@#1\endcsname\relax
    \global\maxbox@must@reruntrue
    \expandafter\xdef\csname maxbox@thiswd@#1\endcsname{\the\maxbox@tempdimb}\expandafter\xdef\csname maxbox@nextwd@#1\endcsname{\the\maxbox@tempdimb}\expandafter\xdef\csname maxbox@thisht@#1\endcsname{\the\maxbox@tempdimc}\expandafter\xdef\csname maxbox@nextht@#1\endcsname{\the\maxbox@tempdimc}\else
    \maxbox@tempdima=\csname maxbox@thiswd@#1\endcsname\relax
    \ifdim\maxbox@tempdima<\maxbox@tempdimb
      \expandafter\xdef\csname maxbox@thiswd@#1\endcsname{\the\maxbox@tempdimb}\global\maxbox@must@reruntrue
    \fi
    \maxbox@tempdima=\csname maxbox@thisht@#1\endcsname\relax
    \ifdim\maxbox@tempdima<\maxbox@tempdimc
      \expandafter\xdef\csname maxbox@thisht@#1\endcsname{\the\maxbox@tempdimc}\global\maxbox@must@reruntrue
    \fi
    \maxbox@tempdima=\csname maxbox@nextwd@#1\endcsname\relax
    \ifdim\maxbox@tempdima<\maxbox@tempdimb
      \expandafter\xdef\csname maxbox@nextwd@#1\endcsname{\the\maxbox@tempdimb}\fi
    \maxbox@tempdima=\csname maxbox@nextht@#1\endcsname\relax
    \ifdim\maxbox@tempdima<\maxbox@tempdimc
      \expandafter\xdef\csname maxbox@nextht@#1\endcsname{\the\maxbox@tempdimc}\fi
  \fi
  \@ifundefined{maxbox@seen@#1}{\expandafter\gdef\csname maxbox@seen@#1\endcsname{}\@cons\maxbox@taglist{{#1}}}{}\maxbox@tempdima=\csname maxbox@thiswd@#1\endcsname\relax
  \maxbox@produce@box{\maxbox@tempdima}{#2}}
\def\maxbox@set@min@width#1#2{\expandafter\ifx\csname maxbox@thiswd@#1\endcsname\relax
    \global\maxbox@must@reruntrue
    \expandafter\xdef\csname maxbox@thiswd@#1\endcsname{#2}\expandafter\xdef\csname maxbox@nextwd@#1\endcsname{#2}\else
    \maxbox@tempdima=\csname maxbox@thiswd@#1\endcsname\relax
    \maxbox@tempdimb=#2\relax
    \ifdim\maxbox@tempdima<\maxbox@tempdimb
      \expandafter\xdef\csname maxbox@thiswd@#1\endcsname{\the\maxbox@tempdimb}\fi
    \maxbox@tempdima=\csname maxbox@nextwd@#1\endcsname\relax
    \ifdim\maxbox@tempdima<\maxbox@tempdimb
      \expandafter\xdef\csname maxbox@nextwd@#1\endcsname{\the\maxbox@tempdimb}\fi
  \fi
  \@ifundefined{maxbox@seen@#1}{\expandafter\gdef\csname maxbox@seen@#1\endcsname{}\@cons\maxbox@taglist{{#1}}}{}}
\DeclareRobustCommand{\maxbox}{\@ifnextchar[{\maxbox@i}{\maxbox@iii[c][\relax][s]}}
\def\maxbox@i[#1]{\@ifnextchar[{\maxbox@ii[#1]}{\maxbox@iii[#1][\relax][s]}}
\def\maxbox@ii[#1][#2]{\@ifnextchar[{\maxbox@iii[#1][#2]}{\maxbox@iii[#1][#2][#1]}}
\def\maxbox@iii[#1][#2][#3]{\long\gdef\maxbox@produce@box##1##2{\parbox[#1][#2][#3]{##1}{##2}}\maxbox@compute@width
}
\DeclareRobustCommand{\eqminipage}{\@ifnextchar[{\eqminipage@i}{\eqminipage@iii[c][\relax][s]}}

\long\def\eqminipage@i[#1]{\@ifnextchar[{\eqminipage@ii[#1]}{\eqminipage@iii[#1][\relax][s]}}
\def\eqminipage@ii[#1][#2]{\@ifnextchar[{\eqminipage@iii[#1][#2]}{\eqminipage@iii[#1][#2][#1]}}
\def\eqminipage@iii[#1][#2][#3]#4{\long\def\eqminipage@iv##1{\long\gdef\maxbox@produce@box####1####2{\begin{minipage}[#1][#2][#3]{####1}####2\end{minipage}}\maxbox@compute@width{#4}{##1}}\Collect@Body\eqminipage@iv
}
\DeclareRobustCommand{\eqmakebox}{\@ifnextchar[{\eqlrbox@i\makebox}{\makebox}}
\DeclareRobustCommand{\eqframebox}{\@ifnextchar[{\eqlrbox@i\framebox}{\framebox}}
\DeclareRobustCommand{\eqsavebox}[1]{\@ifnextchar[{\eqlrbox@i{\savebox{#1}}}{\savebox{#1}}}
\def\eqlrbox@i#1[#2]{\@ifnextchar[{\eqlrbox@ii{#1}[#2]}{\eqlrbox@ii{#1}[#2][c]}}
\def\eqlrbox@ii#1[#2][#3]{\long\gdef\maxbox@produce@box##1##2{#1[##1][#3]{##2}}\maxbox@compute@width{#2}}
\def\maxbox@set@min@height#1#2{\expandafter\ifx\csname maxbox@thisht@#1\endcsname\relax
    \global\maxbox@must@reruntrue
    \expandafter\xdef\csname maxbox@thisht@#1\endcsname{#2}\expandafter\xdef\csname maxbox@nextht@#1\endcsname{#2}\else
    \maxbox@tempdima=\csname maxbox@thisht@#1\endcsname\relax
    \maxbox@tempdimb=#2\relax
    \ifdim\maxbox@tempdima<\maxbox@tempdimb
      \expandafter\xdef\csname maxbox@thisht@#1\endcsname{\the\maxbox@tempdimb}\fi
    \maxbox@tempdima=\csname maxbox@nextht@#1\endcsname\relax
    \ifdim\maxbox@tempdima<\maxbox@tempdimb
      \expandafter\xdef\csname maxbox@nextht@#1\endcsname{\the\maxbox@tempdimb}\fi
  \fi
  \@ifundefined{maxbox@seen@#1}{\expandafter\gdef\csname maxbox@seen@#1\endcsname{}\@cons\maxbox@taglist{{#1}}}{}}
\newcommand{\maxboxMinHeight}[2]{\maxbox@set@min@height{#1}{#2}}
\newcommand{\maxboxMinWidth}[2]{\maxbox@set@min@width{#1}{#2}}
\newcommand{\eqboxwh}[2]{\maxbox[b][\csname maxbox@thisht@#1\endcsname]{#1}{#2}}
\newcommand{\eqparbox}[3][b]{\maxbox[#1]{#2}{#3}}
\newcommand*{\eqboxwidth}[1]{\@ifundefined{maxbox@thiswd@#1}{0pt}{\csname maxbox@thiswd@#1\endcsname}}
\newcommand*{\eqboxheight}[1]{\@ifundefined{maxbox@thisht@#1}{0pt}{\csname maxbox@thisht@#1\endcsname}}
\newcommand{\eqsetminwidth}[2]{\@tempdima=#2\relax
  \expandafter\xdef\csname maxbox@minwd@#1\endcsname{\the\@tempdima}\maxbox@set@min@width{#1}{\csname maxbox@minwd@#1\endcsname}}
\newcommand{\eqsetmaxwidth}[2]{\@tempdima=#2\relax
  \expandafter\xdef\csname maxbox@maxwd@#1\endcsname{\the\@tempdima}}
\newcommand{\eqsetminwidthto}[2]{\maxbox@settowidth{\@tempdima}{#2}\expandafter\xdef\csname maxbox@minwd@#1\endcsname{\the\@tempdima}\maxbox@set@min@width{#1}{\csname maxbox@minwd@#1\endcsname}}
\newcommand{\eqsetmaxwidthto}[2]{\maxbox@settowidth{\@tempdima}{#2}\expandafter\xdef\csname maxbox@maxwd@#1\endcsname{\the\@tempdima}}
\AtEndDocument{\begingroup
    \def\@elt#1{\@ifundefined{maxbox@minwd@#1}{}{\@ifundefined{maxbox@maxwd@#1}{}{\ifdim\csname maxbox@minwd@#1\endcsname>\csname maxbox@maxwd@#1\endcsname
            \PackageWarning{maxbox}{For tag `#1',
              minimum width (\csname maxbox@minwd@#1\endcsname) >
              maximum width (\csname maxbox@maxwd@#1\endcsname)}\fi
        }}\maxbox@tempdima\csname maxbox@thiswd@#1\endcsname\relax
      \maxbox@tempdimb\csname maxbox@nextwd@#1\endcsname\relax
      \ifdim\maxbox@tempdima=\maxbox@tempdimb
      \else
        \@latex@warning@no@line{Rerun to correct the width of maxbox `#1'}\fi
      \immediate\write\@auxout{\string\expandafter\string\gdef\string\csname\space
        maxbox@thiswd@#1\string\endcsname{\csname maxbox@nextwd@#1\endcsname
        }^^J\string\expandafter\string\gdef\string\csname\space
         maxbox@nextwd@#1\string\endcsname{0pt}}\maxbox@tempdima\csname maxbox@thisht@#1\endcsname\relax
      \maxbox@tempdimb\csname maxbox@nextht@#1\endcsname\relax
      \ifdim\maxbox@tempdima=\maxbox@tempdimb
      \else
        \@latex@warning@no@line{Rerun to correct the height of maxbox `#1'}\fi
      \immediate\write\@auxout{\string\expandafter\string\gdef\string\csname\space
        maxbox@thisht@#1\string\endcsname{\csname maxbox@nextht@#1\endcsname
        }^^J\string\expandafter\string\gdef\string\csname\space
         maxbox@nextht@#1\string\endcsname{0pt}}\@ifundefined{maxbox@minwd@#1}{}{\immediate\write\@auxout{\string\expandafter\string\gdef\string\csname\space
          maxbox@minwd@#1\string\endcsname{\csname maxbox@minwd@#1\endcsname
          }}}\@ifundefined{maxbox@maxwd@#1}{}{\immediate\write\@auxout{\string\expandafter\string\gdef\string\csname\space
          maxbox@maxwd@#1\string\endcsname{\csname maxbox@maxwd@#1\endcsname
          }}}}\maxbox@taglist
  \endgroup
  \ifmaxbox@must@rerun
    \@latex@warning@no@line{Rerun to correct maxbox widths}
  \fi
}
 \spnewtheorem{notation}{Notation}{\bf}{\it}
\spnewtheorem{fact}{Fact}{\bf}{\it}

\newenvironment{myexample}[2]{\example[#1]\unskip\label{example:#2}}{\endexample}
\newenvironment{mydefinition}[2]{\typeout{mydefinition: #1: #2}\definition[#1]\unskip\label{definition:#2}}{\enddefinition}

\NewEnviron{mytheorem}[2]{\begin{theorem}[#1]\label{theorem:#2}\BODY \endgraf\noindent\textsl{See page \pageref{proof:theorem:#2} for the proof of this theorem.}
    \end{theorem}}

\NewEnviron{mytheoremNOPROOF}[2]{\begin{theorem}[#1]\label{theorem:#2}\BODY \end{theorem}}

\NewEnviron{mycorollary}[2]{\begin{corollary}[#1]\label{corollary:#2}\BODY \endgraf\noindent\textsl{See page \pageref{proof:corollary:#2} for the proof of this corollary.}
    \end{corollary}}

\NewEnviron{mycorollaryNOPROOF}[2]{\begin{corollary}[#1]\label{corollary:#2}\BODY \end{corollary}}

\NewEnviron{mylemma}[2]{\begin{lemma}[#1]\label{lemma:#2}\BODY \endgraf\noindent\textsl{See page \pageref{proof:lemma:#2} for the proof of this lemma.}
    \end{lemma}}\NewEnviron{mylemmaNOPROOF}[2]{\begin{lemma}[#1]\label{lemma:#2}\BODY \end{lemma}}\NewEnviron{myfactNOPROOF}[2]{\begin{fact}[#1]\label{fact:#2}\BODY \end{fact}}

\NewEnviron{myfigureWITHBOX}[2]{\begin{figure}[t]\fbox{\begin{minipage}{\linewidth-2\fboxsep-2\fboxrule}\centering \BODY \caption{#1}\label{figure:#2}\end{minipage}}\end{figure}}

\NewEnviron{myfigureWITHBOXHere}[2]{\begin{figure}[ht]\fbox{\begin{minipage}{\linewidth-2\fboxsep-2\fboxrule}\centering \BODY \caption{#1}\label{figure:#2}\end{minipage}}\end{figure}}

\NewEnviron{myfigureTwoWITHBOX}[2]{\begin{figure*}\fbox{\begin{minipage}{\linewidth-2\fboxsep-2\fboxrule}\BODY \caption{#1}\label{figure:#2}\end{minipage}}\end{figure*}}

\NewEnviron{mytableWITHBOX}[2]{\begin{table}[t]\fbox{\begin{minipage}{\linewidth-2\fboxsep-2\fboxrule}\vspace*{-\baselineskip}\caption{#1}\label{table:#2}\BODY \end{minipage}}\end{table}}

\NewEnviron{mytableTwoWITHBOX}[2]{\begin{table*}\fbox{\begin{minipage}{\linewidth-2\fboxsep-2\fboxrule}\BODY \caption{#1}\label{table:#2}\end{minipage}}\end{table*}}

\newenvironment{myproof}[2]{\proof[\myref{#2}, p. \pageref{#2}: #1]\label{proof:#2}}{\endproof}

\newcommand{\mysubfigure}[4][\linewidth]{\begin{subfigure}[t]{#1}\fbox{\begin{minipage}{\linewidth-2\fboxsep-2\fboxrule}\begin{nscenter}#4\end{nscenter}\end{minipage}}\caption{\label{figure:#3}#2}\end{subfigure}}%
\tikzset{
    invisible/.style={opacity=0,text opacity=0},
    visible on/.style={alt=#1{}{invisible}},
    alt/.code args={<#1>#2#3}{\alt<#1>{\pgfkeysalso{#2}}{\pgfkeysalso{#3}}
    },
    mybeameralertRED/.style={alt={<#1>{fill=white,text=crimson!80!white}{}}},
    mybeameralert/.style={alt={<#1>{mymainthemecolor!80!white,fill=mymainthemecolor!80!white,text=black}{}}},
    mybeameralert2/.style={alt={<#1>{crimson!80!white,fill=crimson!80!white,text=black}{}}},
    mybeameralert3/.style={alt={<#1>{text=crimson}{}}},
    mybeameruncover/.style={alt={<#1>{void}{}}},
    mybeameralertedge/.style={alt={<#1>{crimson}{}}},
    mybeameralertedgeX/.style={alt={<#1>{crimson,line width=1pt}{}}},
beameralert/.style={alt={<#1>{fill=red!30,rounded corners}{}},anchor=base},
    BeamerAlert/.style={alt={#1{fill=red!30,rounded corners}{}},anchor=base},
}

\tcbset{
    myboxalert/.style={only=<#1>{colframe=myalertcolframe,colback=myalertcolback,colbacktitle=myalertcolbacktitle}},
}

\newlength{\halfphantomLength}

\newlength{\nphantomLENGTH}

\newrobustcmd{\black}[1]{\begingroup\color{black}#1\endgroup\xspace}
\newrobustcmd{\blue}[1]{\begingroup\color{blue}#1\endgroup\xspace}
\newrobustcmd{\orange}[1]{\begingroup\color{red!50!yellow}#1\endgroup\xspace}
\newrobustcmd{\darkorange}[1]{\begingroup\color{red!50!yellow!70!black}#1\endgroup\xspace}
\newrobustcmd{\red}[1]{\begingroup\color{red}#1\endgroup\xspace}
\newrobustcmd{\green}[1]{\begingroup\color{green}#1\endgroup\xspace}
\newrobustcmd{\gray}[1]{\begingroup\color{gray}#1\endgroup\xspace}

\tikzset{
    backgroundShape/.style = {
        line width=1pt,black,fill=black,opacity=.2
    }
}

\makeatletter
\tikzset{circle split part fill/.style  args={#1,#2}{alias=tmp@name,
  postaction={insert path={
     \pgfextra{\pgfpointdiff{\pgfpointanchor{\pgf@node@name}{center}}{\pgfpointanchor{\pgf@node@name}{east}}\pgfmathsetmacro\insiderad{\pgf@x}
      \fill[#1] (\pgf@node@name.base) ([xshift=-\pgflinewidth]\pgf@node@name.east) arc
                          (0:180:\insiderad-\pgflinewidth)--cycle;
      \fill[#2] (\pgf@node@name.base) ([xshift=\pgflinewidth]\pgf@node@name.west)  arc
                           (180:360:\insiderad-\pgflinewidth)--cycle;
         }}}}}
 \makeatother

\tikzset{stretch split horizontal/.style={align=center,text width=(\pgfkeysvalueof{/pgf/minimum width}-
  (\pgfkeysvalueof{/pgf/rectangle split parts}-1)*\pgflinewidth)
  /(\pgfkeysvalueof{/pgf/rectangle split parts})-2*(\pgfkeysvalueof{/pgf/inner xsep})}}

\newcommand{\ENLARGE}[2]{
\gettikzxy{(#1)}{\NodePosX}{\NodePosY}
\gettikzxy{(#1.south west)}{\NodePosSWX}{\NodePosSWY}
\gettikzxy{(#1.north east)}{\NodePosNEX}{\NodePosNEY}
\node[opacity=.5](SOME)at(\NodePosX,\NodePosY){\phantom{\eqparbox[b]{#2}{\rule{\NodePosNEX-\NodePosSWX-2ex}{\NodePosNEY-\NodePosSWY-2ex}}}};
}

\usepackage{scalerel}
\usepackage{tikz}
\usetikzlibrary{svg.path}
\definecolor{orcidlogocol}{HTML}{A6CE39}
\tikzset{
orcidlogo/.pic={
    \fill[orcidlogocol] svg{M256,128c0,70.7-57.3,128-128,128C57.3,256,0,198.7,0,128C0,57.3,57.3,0,128,0C198.7,0,256,57.3,256,128z};
    \fill[white] svg{M86.3,186.2H70.9V79.1h15.4v48.4V186.2z}
                svg{M108.9,79.1h41.6c39.6,0,57,28.3,57,53.6c0,27.5-21.5,53.6-56.8,53.6h-41.8V79.1z M124.3,172.4h24.5c34.9,0,42.9-26.5,42.9-39.7c0-21.5-13.7-39.7-43.7-39.7h-23.7V172.4z}
                svg{M88.7,56.8c0,5.5-4.5,10.1-10.1,10.1c-5.6,0-10.1-4.6-10.1-10.1c0-5.6,4.5-10.1,10.1-10.1C84.2,46.7,88.7,51.3,88.7,56.8z};
}
}

\newcommand{\myorcidIDInner}[1]{\,\href{https://orcid.org/#1}{\mbox{\scalerel*{
\begin{tikzpicture}[yscale=-1,transform shape]
\pic{orcidlogo};
\end{tikzpicture}
}{|}}}}

\makeatletter
\def\@citecolor{blue}\def\@urlcolor{blue}\def\@linkcolor{blue}
\def\orcidID#1{\smash{\href{http://orcid.org/#1}{\protect\raisebox{-1.25pt}{\protect\includegraphics[draft=false]{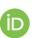}}}}}
\makeatother

\usepackage[T1]{fontenc}
\usepackage[utf8]{inputenc}
\usepackage[english]{babel}
\usepackage[babel]{csquotes}
\usepackage{xspace}

\usepackage{etex}
\usepackage{etoolbox}
\usepackage{morewrites}
\usepackage{amsfonts}
\usepackage{amsmath}
\usepackage{longtable}
\usepackage{booktabs}

\usepackage{centernot}
\usepackage{lastpage}
\usepackage{xifthen}
\usepackage[luatex]{pdflscape}
\usepackage{comment}

\usepackage[inline]{enumitem}
\newlist{compactitem}{itemize}{4} \setlist[compactitem]{nosep, leftmargin=*}
\setlist[compactitem,1]{label=\textbullet}
\setlist[compactitem,2]{label=$\circ$}
\setlist[compactitem,3]{label=$\triangleright$}
\setlist[compactitem,4]{label=$\triangleright\triangleright$}

\newlist{compactenum}{enumerate}{4} \setlist[compactenum]{label=(\arabic*),ref=(\arabic*),nosep, leftmargin=*}

\usepackage{graphicx}
\usepackage{afterpage}
\usepackage{adjustbox}
\usepackage{booktabs}

\usepackage{subcaption}
\usepackage{pifont} \usepackage[title]{appendix}
\usepackage{xifthen}
\usepackage{mathtools} \usepackage{xspace}
\usepackage{listings}

\usepackage[osf,sc]{mathpazo}
\usepackage{pifont}
\usepackage{textcomp}

\usepackage{varwidth}

\usepackage{mathpartir}

\usepackage{multirow}

\usepackage{mparhack}

\usepackage{hyphenat}
\def\hyph{-\penalty0\hskip0pt\relax}

\usepackage[xspace]{ellipsis}
\usepackage{ragged2e}

\usepackage{tocbasic}
\setuptoc{toc}{leveldown}
\makeatletter
\BeforeTOCHead{\cleardoublepage
    \edef\@tempa{\noexpand\pdfbookmark[0]{\list@fname}{\@currext}}\@tempa
}
\makeatother

\usepackage{colortbl}

\newcommand{\myrefDEF}[1]{Def.~\hyperref[definition:#1]{\ref*{definition:#1}}{$\vert$}\hyperref[definition:#1]{\pageref*{definition:#1}}}
\newcommand{\myref}[1]{\hyperref[#1]{\autoref*{#1}\unskip{$\vert$\allowbreak p.}\unskip\pageref*{#1}}}
\newcommand{\myrefBlock}[1]{\hyperref[#1]{block \ifcsname item:#1\endcsname\csname item:#1\endcsname\else ??\fi\unskip{$\vert$\allowbreak p.}\unskip\pageref*{#1}}}
\newcommand{\myrefEQ}[1]{\hyperref[#1]{(Eq.\,\ref*{#1})\unskip{$\vert$\allowbreak p.}\unskip\pageref*{#1}}}
\newcommand{\myrefEQnopage}[1]{\hyperref[#1]{(Eq.\,\ref*{#1})}}
\newcommand{\myrefPLAIN}[1]{\hyperref[#1]{\ref*{#1}\unskip{$\vert$p.}\unskip\pageref*{#1}}}

\ifdraft{
    \usepackage[
bookmarksnumbered=true,
        breaklinks=true,
        pdfsubject={none},
        unicode=true,
        pdfview=FitH,
        bookmarks,unicode,colorlinks=true,]{hyperref}
}{
}

\ifdraft{
    
    \newcommand{\POINT}[2][N.N.]{\marginpar{\scriptsize\blue{POINT[#1]:}#2}}
    \newcommand{\TODO}[2][N.N.]{\begin{tcolorbox}[breakable]\textbf{TODO[#1]:}#2\end{tcolorbox}}
    \newcommand{\DISCUSS}[2][N.N.]{\begin{tcolorbox}[breakable]\textbf{DISCUSS[#1]:}#2\end{tcolorbox}}
    \newcommand{\DONE}[2][N.N.]{\begin{tcolorbox}[breakable]\textbf{DONE[#1]:}#2\end{tcolorbox}}
    \newcommand{\REMARK}[2][N.N.]{\begin{tcolorbox}[breakable]\textbf{REMARK[#1]:}#2\end{tcolorbox}}
    \newcommand{\RESPONSIBLE}[2][N.N.]{\begin{tcolorbox}[breakable]\textbf{RESPONSIBLE[#1]:}#2\end{tcolorbox}}
    \newcommand{\DETAIL}[2][N.N.]{\begin{tcolorbox}[breakable]\textbf{DETAIL[#1]:} (Text dient nur der Erläuterung und fällt später weg!) #2\end{tcolorbox}}
    \newcommand{\REWRITE}[2][N.N.]{\begin{tcolorbox}[breakable]\textbf{REWRITE[#1]:}#2\end{tcolorbox}}
    \newcommand{\ORG}[1]{#1}
    \newcommand{\DFG}[1]{\begin{tcolorbox}\textbf{DFG:}#1\end{tcolorbox}}
    \newcommand{\LONG}[1]{#1}
    \newcommand{\ASSURANCE}[1]{#1}
}{
    
    \newcommand{\POINT}[2][N.N.]{}
    \newcommand{\TODO}[2][N.N.]{}
    \newcommand{\DISCUSS}[2][N.N.]{}
    \newcommand{\DONE}[2][N.N.]{}
    \newcommand{\REMARK}[2][N.N.]{}
    \newcommand{\RESPONSIBLE}[2][N.N.]{}
    \newcommand{\DETAIL}[2][N.N]{}
    \newcommand{\REWRITE}[2][N.N]{}
    \newcommand{\ORG}[1]{}
    \newcommand{\DFG}[1]{}
    \newcommand{\LONG}[1]{}
    \newcommand{\ASSURANCE}[1]{}
}

\ifdraft{\setlength{\overfullrule}{5pt}
 \usepackage{everypage}
 \usepackage[]{datetime2}
\AddEverypageHook{\tikz[overlay, remember picture]\node at ($(current page.north)+(0,-.9)$){\color{red}page=\thepage, compiled=\DTMnow; deadline=\ifdefined\deadline\deadline\fi; pagelimit=\ifdefined\pagelimit\pagelimit\fi};}\usepackage[pass,showframe]{geometry}}{}

\IfFileExists{chapter01.id}{\includecomment{BLOCKchapterONE}}{\excludecomment{BLOCKchapterONE}}
\IfFileExists{chapter02.id}{\includecomment{BLOCKchapterTWO}}{\excludecomment{BLOCKchapterTWO}}
\IfFileExists{chapter03.id}{\includecomment{BLOCKchapterTHREE}}{\excludecomment{BLOCKchapterTHREE}}
\IfFileExists{chapter04.id}{\includecomment{BLOCKchapterFOUR}}{\excludecomment{BLOCKchapterFOUR}}
\IfFileExists{chapter05.id}{\includecomment{BLOCKchapterFIVE}}{\excludecomment{BLOCKchapterFIVE}}
\IfFileExists{chapter06.id}{\includecomment{BLOCKchapterSIX}}{\excludecomment{BLOCKchapterSIX}}
\IfFileExists{chapter07.id}{\includecomment{BLOCKchapterSEVEN}}{\excludecomment{BLOCKchapterSEVEN}}
\IfFileExists{chapter08.id}{\includecomment{BLOCKchapterEIGHT}}{\excludecomment{BLOCKchapterEIGHT}}
\IfFileExists{chapter09.id}{\includecomment{BLOCKchapterNINE}}{\excludecomment{BLOCKchapterNINE}}
\IfFileExists{chapter10.id}{\includecomment{BLOCKchapterTEN}}{\excludecomment{BLOCKchapterTEN}}
\IfFileExists{chapter11.id}{\includecomment{BLOCKchapterELEVEN}}{\excludecomment{BLOCKchapterELEVEN}}
\IfFileExists{appendixA.id}{\includecomment{BLOCKappendixA}}{\excludecomment{BLOCKappendixA}}
\IfFileExists{appendixB.id}{\includecomment{BLOCKappendixB}}{\excludecomment{BLOCKappendixB}}
\IfFileExists{appendixC.id}{\includecomment{BLOCKappendixC}}{\excludecomment{BLOCKappendixC}}
\IfFileExists{appendixD.id}{\includecomment{BLOCKappendixD}}{\excludecomment{BLOCKappendixD}}
\IfFileExists{bibliography.id}{\includecomment{BLOCKbibliography}}{\excludecomment{BLOCKbibliography}}

\IfFileExists{auto.id}{}{\includecomment{BLOCKchapterONE}}
\IfFileExists{auto.id}{}{\includecomment{BLOCKchapterTWO}}
\IfFileExists{auto.id}{}{\includecomment{BLOCKchapterTHREE}}
\IfFileExists{auto.id}{}{\includecomment{BLOCKchapterFOUR}}
\IfFileExists{auto.id}{}{\includecomment{BLOCKchapterFIVE}}
\IfFileExists{auto.id}{}{\includecomment{BLOCKchapterSIX}}
\IfFileExists{auto.id}{}{\includecomment{BLOCKchapterSEVEN}}
\IfFileExists{auto.id}{}{\includecomment{BLOCKchapterEIGHT}}
\IfFileExists{auto.id}{}{\includecomment{BLOCKchapterNINE}}
\IfFileExists{auto.id}{}{\includecomment{BLOCKchapterTEN}}
\IfFileExists{auto.id}{}{\includecomment{BLOCKchapterELEVEN}}
\IfFileExists{auto.id}{}{\includecomment{BLOCKappendixA}}
\IfFileExists{auto.id}{}{\includecomment{BLOCKappendixB}}
\IfFileExists{auto.id}{}{\includecomment{BLOCKappendixC}}
\IfFileExists{auto.id}{}{\includecomment{BLOCKappendixD}}
\IfFileExists{auto.id}{}{\includecomment{BLOCKbibliography}}

\pgfkeys{
    /mychapter/.is family, /mychapter,
    title/.store in = \mychapterTitle,
    unbrokenTitle/.store in = \mychapterUnbrokenTitle,
    runningheaderTitle/.store in = \myshortchapterTitle,
    localTocTitle/.store in = \mylocalTocchapterTitle,
    tocTitle/.store in = \mytocchapterTitle,
    pdftocTitle/.store in = \mychapterpdftocTitle,
    label/.store in = \mychapterLabel,
}
\pgfkeys{
    /mysection/.is family, /mysection,
    title/.store in = \mysectionTitle,
    unbrokenTitle/.store in = \mysectionUnbrokenTitle,
    runningheaderTitle/.store in = \myshortsectionTitle,
    localTocTitle/.store in = \mylocalTocsectionTitle,
    tocTitle/.store in = \mytocsectionTitle,
    pdftocTitle/.store in = \mysectionpdftocTitle,
    label/.store in = \mysectionLabel,
}
\pgfkeys{
    /mysectionTODO/.is family, /mysectionTODO,
    title/.store in = \mysectionTODOTitle,
    unbrokenTitle/.store in = \mysectionTODOUnbrokenTitle,
    runningheaderTitle/.store in = \myshortsectionTODOTitle,
    localTocTitle/.store in = \mylocalTocsectionTODOTitle,
    tocTitle/.store in = \mytocsectionTODOTitle,
    pdftocTitle/.store in = \mysectionTODOpdftocTitle,
    label/.store in = \mysectionTODOLabel,
}
\pgfkeys{
    /mysubsection/.is family, /mysubsection,
    title/.store in = \mysubsectionTitle,
    unbrokenTitle/.store in = \mysubsectionUnbrokenTitle,
    runningheaderTitle/.store in = \myshortsubsectionTitle,
    localTocTitle/.store in = \mylocalTocsubsectionTitle,
    tocTitle/.store in = \mytocsubsectionTitle,
    pdftocTitle/.store in = \mysubsectionpdftocTitle,
    label/.store in = \mysubsectionLabel,
}
\pgfkeys{
    /mysubsubsection/.is family, /mysubsubsection,
    title/.store in = \mysubsubsectionTitle,
    unbrokenTitle/.store in = \mysubsubsectionUnbrokenTitle,
    runningheaderTitle/.store in = \myshortsubsubsectionTitle,
    localTocTitle/.store in = \mylocalTocsubsubsectionTitle,
    tocTitle/.store in = \mytocsubsubsectionTitle,
    pdftocTitle/.store in = \mysubsubsectionpdftocTitle,
    label/.store in = \mysubsubsectionLabel,
}

\makeatletter
\newcommand{\csxdefaux}[2]{\protected@write\@mainaux{}{\string\expandafter\string\gdef \string\csname\string\detokenize{#1}\string\endcsname{#2}}}\newcommand{\mychapter}[1]{\pgfkeys{/mychapter, #1}\ifdefined\mytocchapterTitle\else\def\mytocchapterTitle{\mychapterTitle}\fi \ifdefined\myshortchapterTitle\else\def\myshortchapterTitle{\mychapterTitle}\fi \ifdefined\mychapterpdftocTitle\else\def\mychapterpdftocTitle{\mychapterTitle}\fi \ifdefined\mylocalTocchapterTitle\else\def\mylocalTocchapterTitle{\mychapterTitle}\fi \ifdefined\mychapterUnbrokenTitle\else\def\mychapterUnbrokenTitle{\mychapterTitle}\fi \chapter[\protect\TOC{\mytocchapterTitle}{\texorpdfstring{\myshortchapterTitle}{\mychapterpdftocTitle}}]{\mychapterTitle}\label{chapter:\mychapterLabel}\expandafter\csxdefaux\expandafter{label2title:chapter:\mychapterLabel}{\mylocalTocchapterTitle}\expandafter\csxdefaux\expandafter{label2unbrokentitle:chapter:\mychapterLabel}{\mychapterUnbrokenTitle}\let\mychapterpdftocTitle\undefined \let\myshortchapterTitle\undefined \let\mytocchapterTitle\undefined \let\mylocalTocchapterTitle\undefined \let\mychapterUnbrokenTitle\undefined }\newcommand{\mysection}[1]{\pgfkeys{/mysection, #1}\ifdefined\mytocsectionTitle\else\def\mytocsectionTitle{\mysectionTitle}\fi \ifdefined\myshortsectionTitle\else\def\myshortsectionTitle{\mysectionTitle}\fi \ifdefined\mysectionpdftocTitle\else\def\mysectionpdftocTitle{\mysectionTitle}\fi \ifdefined\mysectionUnbrokenTitle\else\def\mysectionUnbrokenTitle{\mysectionTitle}\fi \ifdefined\mylocalTocsectionTitle\else\def\mylocalTocsectionTitle{\mysectionUnbrokenTitle}\fi \section[\protect\TOC{\mytocsectionTitle}{\texorpdfstring{\myshortsectionTitle}{\mysectionpdftocTitle}}]{\mysectionTitle}\label{section:\mysectionLabel}\expandafter\csxdefaux\expandafter{label2title:section:\mysectionLabel}{\mylocalTocsectionTitle}\expandafter\csxdefaux\expandafter{label2unbrokentitle:section:\mysectionLabel}{\mysectionUnbrokenTitle}\let\mysectionpdftocTitle\undefined \let\myshortsectionTitle\undefined \let\mytocsectionTitle\undefined \let\mylocalTocsectionTitle\undefined \let\mysectionUnbrokenTitle\undefined }\newcommand{\mysectionTODO}[1]{\pgfkeys{/mysectionTODO, #1}\ifdefined\mytocsectionTODOTitle\else\def\mytocsectionTODOTitle{\mysectionTODOTitle}\fi \ifdefined\myshortsectionTODOTitle\else\def\myshortsectionTODOTitle{\mysectionTODOTitle}\fi \ifdefined\mysectionTODOpdftocTitle\else\def\mysectionTODOpdftocTitle{\mysectionTODOTitle}\fi \ifdefined\mysectionTODOUnbrokenTitle\else\def\mysectionTODOUnbrokenTitle{\mysectionTODOTitle}\fi \ifdefined\mylocalTocsectionTODOTitle\else\def\mylocalTocsectionTODOTitle{\mysectionTODOUnbrokenTitle}\fi \TODO{\mysectionTODOTitle}\expandafter\csxdefaux\expandafter{label2title:sectionTODO:\mysectionTODOLabel}{\mylocalTocsectionTODOTitle}\expandafter\csxdefaux\expandafter{label2unbrokentitle:sectionTODO:\mysectionTODOLabel}{\mysectionTODOUnbrokenTitle}\let\mysectionTODOpdftocTitle\undefined \let\myshortsectionTODOTitle\undefined \let\mytocsectionTODOTitle\undefined \let\mylocalTocsectionTODOTitle\undefined \let\mysectionTODOUnbrokenTitle\undefined }\newcommand{\mysubsection}[1]{\pgfkeys{/mysubsection, #1}\ifdefined\mytocsubsectionTitle\else\def\mytocsubsectionTitle{\mysubsectionTitle}\fi \ifdefined\myshortsubsectionTitle\else\def\myshortsubsectionTitle{\mysubsectionTitle}\fi \ifdefined\mysubsectionpdftocTitle\else\def\mysubsectionpdftocTitle{\mysubsectionTitle}\fi \ifdefined\mysubsectionUnbrokenTitle\else\def\mysubsectionUnbrokenTitle{\mysubsectionTitle}\fi \ifdefined\mylocalTocsubsectionTitle\else\def\mylocalTocsubsectionTitle{\mysubsectionUnbrokenTitle}\fi \subsection[\protect\TOC{\mytocsubsectionTitle}{\texorpdfstring{\myshortsubsectionTitle}{\mysubsectionpdftocTitle}}]{\mysubsectionTitle}\label{subsection:\mysubsectionLabel}\expandafter\csxdefaux\expandafter{label2title:subsection:\mysubsectionLabel}{\mylocalTocsubsectionTitle}\expandafter\csxdefaux\expandafter{label2unbrokentitle:subsection:\mysubsectionLabel}{\mysubsectionUnbrokenTitle}\let\mysubsectionpdftocTitle\undefined \let\myshortsubsectionTitle\undefined \let\mytocsubsectionTitle\undefined \let\mylocalTocsubsectionTitle\undefined \let\mysubsectionUnbrokenTitle\undefined \let\mysubsectionUnbrokenTitle\undefined }\newcommand{\mysubsubsection}[1]{\pgfkeys{/mysubsubsection, #1}\ifdefined\mytocsubsubsectionTitle\else\def\mytocsubsubsectionTitle{\mysubsubsectionTitle}\fi \ifdefined\myshortsubsubsectionTitle\else\def\myshortsubsubsectionTitle{\mysubsubsectionTitle}\fi \ifdefined\mysubsubsectionpdftocTitle\else\def\mysubsubsectionpdftocTitle{\mysubsubsectionTitle}\fi \ifdefined\mysubsubsectionUnbrokenTitle\else\def\mysubsubsectionUnbrokenTitle{\mysubsubsectionTitle}\fi \ifdefined\mylocalTocsubsubsectionTitle\else\def\mylocalTocsubsubsectionTitle{\mysubsubsectionUnbrokenTitle}\fi \subsubsection[\protect\TOC{\mytocsubsubsectionTitle}{\texorpdfstring{\myshortsubsubsectionTitle
    }{\mysubsubsectionpdftocTitle}}]{\mysubsubsectionTitle}\label{subsubsection:\mysubsubsectionLabel}\expandafter\csxdefaux\expandafter{label2title:subsubsection:\mylocalTocsubsubsectionTitle}{\mysubsubsectionTitle}\expandafter\csxdefaux\expandafter{label2unbrokentitle:subsubsection:\mysectionLabel}{\mysubsubsectionUnbrokenTitle}\let\mysubsubsectionpdftocTitle\undefined \let\myshortsubsubsectionTitle\undefined \let\mytocsubsubsectionTitle\undefined \let\mylocalTocsubsubsectionTitle\undefined \let\mysubsubsectionUnbrokenTitle\undefined }

    [2020/05/28 v1.1 Create tikz-pictures based on typed attributed graphs]

\usepackage{tikz}
\usetikzlibrary{calc}
\usetikzlibrary{shapes.multipart}
\usetikzlibrary{positioning}

\usetikzlibrary{trees}
\usetikzlibrary{decorations}
\usetikzlibrary{arrows}
\usetikzlibrary{automata}
\usetikzlibrary{shadows}
\usetikzlibrary{positioning}
\usetikzlibrary{plotmarks}
\usetikzlibrary{backgrounds}
\usetikzlibrary{shapes}
\usetikzlibrary{calc}
\usetikzlibrary{matrix}
\usetikzlibrary{fit}
\usetikzlibrary{petri}
\usetikzlibrary{patterns}
\usetikzlibrary{arrows.meta}
\usetikzlibrary{decorations.pathreplacing}
\usetikzlibrary{decorations.markings}
\usetikzlibrary{decorations.pathmorphing}
\usetikzlibrary{intersections}
\usetikzlibrary{shapes.multipart}

\usepackage{xspace}
\usepackage{xifthen}
\usepackage{adjustbox}

\makeatletter
\newcommand{\gettikzxy}[3]{\tikz@scan@one@point\pgfutil@firstofone#1\relax
  \edef#2{\the\pgf@x}\edef#3{\the\pgf@y}}
\makeatother

\newlength{\TIKZGRAPHSedgelabelwidth}
\newlength{\TIKZGRAPHSedgelabelheight}

\newcommand{\NodeWidthName}{foo}

\newcommand{\SetNodeWidthName}[1]{\renewcommand{\NodeWidthName}{#1}\eqparbox{#1edgesWD}{}\eqparbox{#1edgesHT}{}\setlength{\TIKZGRAPHSedgelabelwidth}{\eqboxwidth{#1edgesWD}}\setlength{\TIKZGRAPHSedgelabelheight}{\eqboxheight{#1edgesHT}}}

\newcommand{\myTikzGraphsCentering}[1]{#1}

\newcommand{\NODEtype}[1]{\ensuremath{\mathit{{:}#1}}}
\newcommand{\NODEnameWithType}[2]{\ensuremath{\mathit{#1{:}#2}}}
\newcommand{\NODEcontentWithAttributes}[2]{\eqparbox[b]{\NodeWidthName nodes}{\myTikzGraphsCentering{#1}}\nodepart{second}\eqparbox[b]{\NodeWidthName nodes}{\myTikzGraphsCentering{#2}}}

\newcommand{\EDGEtype}[1]{\ensuremath{\mathit{{:}#1}}}
\newcommand{\EDGEnameWithType}[2]{\ensuremath{\mathit{#1{:}#2}}}
\newcommand{\EDGEcontentWithAttributes}[2]{\eqparbox[b]{\NodeWidthName edges}{\myTikzGraphsCentering{#1}}\nodepart{second}\eqparbox[b]{\NodeWidthName edges}{\myTikzGraphsCentering{#2}}}

\newcommand{\ATTRIBUTEnameWithSort}[2]{\ensuremath{\mathit{#1{:}#2}}}
\newcommand{\ATTRIBUTEnameWithValue}[2]{\ensuremath{\mathit{#1}{=}#2}}
\newcommand{\MULT}[2]{\ifthenelse{\equal{#1}{0} \and \equal{#2}{*}}{}{\;[#1..#2]}}

\tikzset{
    styleEdgePre/.style={
        line width=1pt,
    },
    styleEdge/.style={
        styleEdgePre,
        ->,
        >=latex
    },
    implies/.style={double,double equal sign distance,-implies},
}

\newcommand{\ConsiderEdgeLabel}[1]{
    \gettikzxy{(#1.north east)}{\NODEAX}{\NODEAY}
    \gettikzxy{(#1.south west)}{\NODEBX}{\NODEBY}
    \makeatletter
    \maxboxMinHeight{\NodeWidthName edgesHT}{\the\dimexpr\NODEAY-\NODEBY}
    \maxboxMinWidth{\NodeWidthName edgesWD}{\the\dimexpr\NODEAX-\NODEBX}
    \makeatother
}

\def\GraphEdgeDirectYSHIFT{0ex}
\def\GraphEdgeDirectXSHIFT{0ex}
\newcommand{\GraphEdgeDirect}[7]{
    \ifthenelse{\equal{#1}{S}}{\def\GEDsource{north }\def\GEDtarget{south }\def\GEDy{0ex}}{}
    \ifthenelse{\equal{#1}{N}}{\def\GEDsource{south }\def\GEDtarget{north }\def\GEDy{0ex}}{}
    \ifthenelse{\equal{#2}{L}}{\edef\GEDsource{\GEDsource west}\edef\GEDtarget{\GEDtarget west}\def\GEDx{1ex}\def\GEDlab{right}}{}
    \ifthenelse{\equal{#2}{R}}{\edef\GEDsource{\GEDsource east}\edef\GEDtarget{\GEDtarget east}\def\GEDx{-1ex}\def\GEDlab{left}}{}
\ifthenelse{\equal{#1}{W}}{\def\GEDsource{ west}\def\GEDtarget{ east}\def\GEDx{0ex}}{}
    \ifthenelse{\equal{#1}{E}}{\def\GEDsource{ east}\def\GEDtarget{ west}\def\GEDx{0ex}}{}
    \ifthenelse{\equal{#2}{B}}{\edef\GEDsource{north\GEDsource}\edef\GEDtarget{north\GEDtarget}\def\GEDy{-1ex}\def\GEDlab{below}}{}
    \ifthenelse{\equal{#2}{A}}{\edef\GEDsource{south\GEDsource}\edef\GEDtarget{south\GEDtarget}\def\GEDy{1ex}\def\GEDlab{above}}{}
\draw ($(#3.\GEDsource)+(\GEDx,\GEDy)+(\GraphEdgeDirectXSHIFT,\GraphEdgeDirectYSHIFT)$)
            edge[styleEdge]
            node[\GEDlab,#4,outer sep=1ex] (#5) {#6\unskip}
            ($(#7.\GEDtarget)+(\GEDx,\GEDy)+(\GraphEdgeDirectXSHIFT,\GraphEdgeDirectYSHIFT)$);
    \ConsiderEdgeLabel{#5}
\def\GraphEdgeDirectYSHIFT{0ex}
    \def\GraphEdgeDirectXSHIFT{0ex}
}

\newcommand{\GraphEdgeBendOnce}[7]{
    \ifthenelse{\equal{#1}{S}}{\def\GEDsource{north }\def\GEDtarget{south }\def\GEDy{0ex}}{}
    \ifthenelse{\equal{#1}{N}}{\def\GEDsource{south }\def\GEDtarget{north }\def\GEDy{0ex}}{}
    \ifthenelse{\equal{#2}{L}}{\edef\GEDsource{\GEDsource west}\edef\GEDtarget{\GEDtarget west}\def\GEDx{1ex}\def\GEDlab{right}}{}
    \ifthenelse{\equal{#2}{R}}{\edef\GEDsource{\GEDsource east}\edef\GEDtarget{\GEDtarget east}\def\GEDx{-1ex}\def\GEDlab{left}}{}
\ifthenelse{\equal{#1}{W}}{\def\GEDsource{ west}\def\GEDtarget{ east}\def\GEDx{0ex}}{}
    \ifthenelse{\equal{#1}{E}}{\def\GEDsource{ east}\def\GEDtarget{ west}\def\GEDx{0ex}}{}
    \ifthenelse{\equal{#2}{B}}{\edef\GEDsource{north\GEDsource}\edef\GEDtarget{north\GEDtarget}\def\GEDy{-1ex}\def\GEDlab{below}}{}
    \ifthenelse{\equal{#2}{A}}{\edef\GEDsource{south\GEDsource}\edef\GEDtarget{south\GEDtarget}\def\GEDy{1ex}\def\GEDlab{above}}{}
\draw[styleEdge] ($(#3.\GEDsource)+(\GEDx,\GEDy)+(\GraphEdgeDirectXSHIFT,\GraphEdgeDirectYSHIFT)$)
            -- ($0.5*(#3.\GEDsource)+0.5*(\GEDx,\GEDy)+0.5*(#7.\GEDtarget)+0.5*(\GEDx,\GEDy)$)
            node[\GEDlab,#4,outer sep=1ex] (#5) {#6\unskip}
            ($(#7.\GEDtarget)+(\GEDx,\GEDy)+(\GraphEdgeDirectXSHIFT,\GraphEdgeDirectYSHIFT)$);
    \ConsiderEdgeLabel{#5}
    \def\GraphEdgeDirectYSHIFT{0ex}
    \def\GraphEdgeDirectXSHIFT{0ex}
}

\newcommand{\GraphEdgeDirectStraight}[7]{
    \ifthenelse{\equal{#1}{S}}{\def\GEDsource{north }\def\GEDtarget{south }\def\GEDy{0ex}}{}
    \ifthenelse{\equal{#1}{N}}{\def\GEDsource{south }\def\GEDtarget{north }\def\GEDy{0ex}}{}
    \ifthenelse{\equal{#2}{L}}{\edef\GEDsource{\GEDsource west}\edef\GEDtarget{\GEDtarget west}\def\GEDx{1ex}\def\GEDlab{right}}{}
    \ifthenelse{\equal{#2}{R}}{\edef\GEDsource{\GEDsource east}\edef\GEDtarget{\GEDtarget east}\def\GEDx{-1ex}\def\GEDlab{left}}{}
\ifthenelse{\equal{#1}{W}}{\def\GEDsource{ west}\def\GEDtarget{ east}\def\GEDx{0ex}}{}
    \ifthenelse{\equal{#1}{E}}{\def\GEDsource{ east}\def\GEDtarget{ west}\def\GEDx{0ex}}{}
    \ifthenelse{\equal{#2}{B}}{\edef\GEDsource{north\GEDsource}\edef\GEDtarget{north\GEDtarget}\def\GEDy{-1ex}\def\GEDlab{below}}{}
    \ifthenelse{\equal{#2}{A}}{\edef\GEDsource{south\GEDsource}\edef\GEDtarget{south\GEDtarget}\def\GEDy{1ex}\def\GEDlab{above}}{}
\gettikzxy{(#3.\GEDsource)}{\NODEAX}{\NODEAY}
    \gettikzxy{(#7.\GEDtarget)}{\NODEBX}{\NODEBY}
\ifthenelse{\equal{#1}{S}}{\def\GEDtargetX{\NODEAX}\def\GEDtargetY{\NODEBY}}{}
    \ifthenelse{\equal{#1}{N}}{\def\GEDtargetX{\NODEAX}\def\GEDtargetY{\NODEBY}}{}
    \ifthenelse{\equal{#1}{E}}{\def\GEDtargetX{\NODEBX}\def\GEDtargetY{\NODEAY}}{}
    \ifthenelse{\equal{#1}{W}}{\def\GEDtargetX{\NODEBX}\def\GEDtargetY{\NODEAY}}{}
\draw ($(#3.\GEDsource)+(\GEDx,\GEDy)+(\GraphEdgeDirectXSHIFT,\GraphEdgeDirectYSHIFT)$)
            edge[styleEdge]
            node[\GEDlab,#4,outer sep=1ex] (#5) {#6\unskip}
            ($(\GEDtargetX,\GEDtargetY)+(\GEDx,\GEDy)+(\GraphEdgeDirectXSHIFT,\GraphEdgeDirectYSHIFT)$);
    \ConsiderEdgeLabel{#5}
    \def\GraphEdgeDirectYSHIFT{0ex}
    \def\GraphEdgeDirectXSHIFT{0ex}
}

\newcommand{\GraphEdgeLoopDist}{2ex}
\newcommand{\GraphEdgeLoopSep}{1ex}
\newcommand{\GraphEdgeLoop}[5]{
    \ifthenelse{\equal{#1}{S}}{\def\GEDsource{south west}\def\GEDtarget{south east}\def\GEDya{0ex}\def\GEDyb{0ex}\def\GEDxa{\GraphEdgeLoopSep}\def\GEDxb{-\GraphEdgeLoopSep}
        \def\GEDyaA{-\GraphEdgeLoopDist}\def\GEDybA{-\GraphEdgeLoopDist}\def\GEDxaA{0ex}\def\GEDxbA{0ex}\def\GEDlab{below}}{}
    \ifthenelse{\equal{#1}{N}}{\def\GEDsource{north west}\def\GEDtarget{north east}\def\GEDya{0ex}\def\GEDyb{0ex}\def\GEDxa{\GraphEdgeLoopSep}\def\GEDxb{-\GraphEdgeLoopSep}
        \def\GEDyaA{\GraphEdgeLoopDist}\def\GEDybA{\GraphEdgeLoopDist}\def\GEDxaA{0ex}\def\GEDxbA{0ex}\def\GEDlab{above}}{}
    \ifthenelse{\equal{#1}{E}}{\def\GEDsource{south east}\def\GEDtarget{north east}\def\GEDya{\GraphEdgeLoopSep}\def\GEDyb{-\GraphEdgeLoopSep}\def\GEDxa{0ex}\def\GEDxb{0ex}
        \def\GEDyaA{0ex}\def\GEDybA{0ex}\def\GEDxaA{\GraphEdgeLoopDist}\def\GEDxbA{\GraphEdgeLoopDist}\def\GEDlab{right}}{}
    \ifthenelse{\equal{#1}{W}}{\def\GEDsource{north west}\def\GEDtarget{south west}\def\GEDya{-\GraphEdgeLoopSep}\def\GEDyb{\GraphEdgeLoopSep}\def\GEDxa{0ex}\def\GEDxb{0ex}
        \def\GEDyaA{0ex}\def\GEDybA{0ex}\def\GEDxaA{-\GraphEdgeLoopDist}\def\GEDxbA{-\GraphEdgeLoopDist}\def\GEDlab{left}}{}
\draw[styleEdge]
            ($(#2.\GEDsource)+(\GEDxa,\GEDya)$)
            -- ($(#2.\GEDsource)+(\GEDxa,\GEDya)+(\GEDxaA,\GEDyaA)$)
            --
                node[\GEDlab,#3,outer sep=2pt] (#4) {#5\unskip}
                ($(#2.\GEDtarget)+(\GEDxb,\GEDyb)+(\GEDxbA,\GEDybA)$)
            -- ($(#2.\GEDtarget)+(\GEDxb,\GEDyb)$);
    \ConsiderEdgeLabel{#4}
}

\colorlet{colorConstraint}{red!20!white}
\colorlet{colorValuation}{blue!20!white}
\colorlet{colorFreeVariables}{blue!30!white}
\colorlet{colorGlobalVariables}{blue!30!white}

\tikzset{
	mtgc_outer/.style={
        outer ysep=0pt,inner ysep=1pt,outer xsep=1pt,inner xsep=0pt,minimum width=0pt,line width=1pt
	},
    conditionGraphNode/.style={
        outer xsep=.2ex,line width=1pt
    },
	mtgc_inner/.style={
        outer ysep=1pt,inner ysep=1pt,outer xsep=1pt,inner xsep=2pt,line width=1pt
	},
	globalVariablesNode/.style={
        fill=colorGlobalVariables,inner sep=2pt,outer xsep=1pt,line width=1pt
	},
	constraintNode/.style={
        fill=colorConstraint,inner sep=2pt,outer xsep=1pt,line width=1pt
	},
	valuationNode/.style={
        fill=colorValuation,inner sep=2pt,outer xsep=1pt,line width=1pt
	},
	freeVariablesNode/.style={
        fill=colorFreeVariables,inner sep=2pt,outer xsep=1pt,line width=1pt
	},	
}

\newcommand{\VerticalExtend}[3][t]{\adjustbox{valign=#1}{#3}\adjustbox{valign=#1}{\vphantom{\rule{10pt}{\eqboxheight{#2}}
        }}}

\tikzset{
    >=latex,
    inclusion/.style={right hook->},
    epimorphism/.style={->>},
    isomorphism/.style={right hook->>},
    isomorphismSWAP/.style={left hook->>},
    implies/.style={double,double equal sign distance,-implies},
    isom/.style={double,double equal sign distance},
    equalarrow/.style={-,double line with arrow={-,-}},
    steparrow/.style={double,double equal sign distance,-implies},
    inclusionSWAP/.style={left hook->},
    separationA/.style={inner sep=0pt, minimum size=0pt},
    vertex/.style={inner sep=1.5pt, minimum size=1ex,rectangle,draw},
    shorten >=0cm,
    shorten <=0cm,
}

\colorlet{graphbackground}{white}

\newcommand{\CLOSINGBRACKET}[4][0cm]{
\gettikzxy{(#2.north east)}{\NodeAX}{\NodeAY}
\gettikzxy{(#2.south east)}{\NodeBX}{\NodeBY}
\gettikzxy{(#3.east)}{\NodeCX}{\NodeCY}
\draw[-]
       ($(\NodeCX,\NodeAY)+(#1,0cm)$)
    to[bend left=30] ($(\NodeCX,\NodeAY)+(#1,0)+(.5ex,-.5ex)$) node[minimum size=0pt,inner sep=0pt,outer sep=0pt] (#4) {}
    to[] ($(\NodeCX,\NodeBY)+(#1,0)+(.5ex,.5ex)$)
    to[bend left=30] ($(\NodeCX,\NodeBY)+(#1,0)$);
}
\newcommand{\OPENINGBRACKET}[2][0ex]{
\gettikzxy{(#2.north west)}{\NodeAX}{\NodeAY}
\gettikzxy{(#2.south west)}{\NodeBX}{\NodeBY}
\gettikzxy{(#2.west)}{\NodeCX}{\NodeCY}
\draw[-]
       ($(\NodeCX+#1,\NodeAY)$)
    to[bend right=30] ($(\NodeCX,\NodeAY)+(-.5ex+#1,-.5ex)$)
    to[] ($(\NodeCX,\NodeBY)+(-.5ex+#1,.5ex)$)
    to[bend right=30] ($(\NodeCX+#1,\NodeBY)$);
}

\newcommand{\COMMA}{\ensuremath{\mathbf{,}}\xspace}
 \tikzset{
	styleNODERouter/.style={
		line width=1pt,
		minimum width=0cm,
		outer ysep=1pt,
		inner ysep=1pt,
		inner xsep=1pt,
		outer xsep=1pt,
		draw,
		shape=rectangle split,
		rectangle split parts=2,
		align=left,
		rectangle split part fill={blue!30,white},
		rectangle split every empty part={},
		rectangle split empty part height=0pt,
	}
}

\newcommand{\NODETGRouterShort}{\NODEcontentWithAttributes{\strut\NODEtype{Router\MULT{0}{*}}}{}}

\newcommand{\NODERouter}[5]{\NODEcontentWithAttributes{\strut\NODEnameWithType{#1}{Router}}{\def\NewLineNeeded{0}\ifthenelse{\equal{#2}{}}{}{\ifthenelse{\equal{\NewLineNeeded}{1}}{\\}{}\def\NewLineNeeded{1}\strut\ATTRIBUTEnameWithValue{cts}{#2}}\ifthenelse{\equal{#3}{}}{}{\ifthenelse{\equal{\NewLineNeeded}{1}}{\\}{}\def\NewLineNeeded{1}\strut\ATTRIBUTEnameWithValue{dts}{#3}}\ifthenelse{\equal{#4}{}}{}{\ifthenelse{\equal{\NewLineNeeded}{1}}{\\}{}\def\NewLineNeeded{1}\strut\ATTRIBUTEnameWithValue{cidx}{#4}}\ifthenelse{\equal{#5}{}}{}{\ifthenelse{\equal{\NewLineNeeded}{1}}{\\}{}\def\NewLineNeeded{1}\strut\ATTRIBUTEnameWithValue{didx}{#5}}}}

\pgfkeys{
	/NODERouter/.cd,
	name/.code = \xdef\myNODERouterNAME{\unexpanded{#1}},
	appendTitle/.code = \xdef\myNODERouterSFSexpand{\unexpanded{#1}},
	cts/.code = \xdef\myNODERouterATTcts{\unexpanded{#1}},
	dts/.code = \xdef\myNODERouterATTdts{\unexpanded{#1}},
	cidx/.code = \xdef\myNODERouterATTcidx{\unexpanded{#1}},
	didx/.code = \xdef\myNODERouterATTdidx{\unexpanded{#1}},
}
\newcommand*\myNODERouterNAME{}\newcommand*\myNODERouterSFSexpand{}\newcommand*\myNODERouterATTcts{}\newcommand*\myNODERouterATTdts{}\newcommand*\myNODERouterATTcidx{}\newcommand*\myNODERouterATTdidx{}

\newcommand{\NODERouterX}[1]{\gdef\myNODERouterNAME{}\gdef\myNODERouterSFSexpand{}\gdef\myNODERouterATTcts{}\gdef\myNODERouterATTdts{}\gdef\myNODERouterATTcidx{}\gdef\myNODERouterATTdidx{}\pgfqkeys{/NODERouter}{#1}\NODEcontentWithAttributes{\strut\NODEnameWithType{\myNODERouterNAME}{Router}\myNODERouterSFSexpand}{\def\NewLineNeeded{0}\ifthenelse{\equal{\myNODERouterATTcts}{}}{}{\ifthenelse{\equal{\NewLineNeeded}{1}}{\\}{}\def\NewLineNeeded{1}\strut\ATTRIBUTEnameWithValue{cts}{\myNODERouterATTcts}}\ifthenelse{\equal{\myNODERouterATTdts}{}}{}{\ifthenelse{\equal{\NewLineNeeded}{1}}{\\}{}\def\NewLineNeeded{1}\strut\ATTRIBUTEnameWithValue{dts}{\myNODERouterATTdts}}\ifthenelse{\equal{\myNODERouterATTcidx}{}}{}{\ifthenelse{\equal{\NewLineNeeded}{1}}{\\}{}\def\NewLineNeeded{1}\strut\ATTRIBUTEnameWithValue{cidx}{\myNODERouterATTcidx}}\ifthenelse{\equal{\myNODERouterATTdidx}{}}{}{\ifthenelse{\equal{\NewLineNeeded}{1}}{\\}{}\def\NewLineNeeded{1}\strut\ATTRIBUTEnameWithValue{didx}{\myNODERouterATTdidx}}}}

\tikzset{
	styleNODEMessage/.style={
		line width=1pt,
		minimum width=0cm,
		outer ysep=1pt,
		inner ysep=1pt,
		inner xsep=1pt,
		outer xsep=1pt,
		draw,
		shape=rectangle split,
		rectangle split parts=2,
		align=left,
		rectangle split part fill={red!20!yellow,white},
		rectangle split every empty part={},
		rectangle split empty part height=0pt,
	}
}

\newcommand{\NODETGMessageShort}{\NODEcontentWithAttributes{\strut\NODEtype{Message\MULT{0}{*}}}{\strut\ATTRIBUTEnameWithSort{clock}{\TypeReal}\\\strut\ATTRIBUTEnameWithSort{id}{\TypeInteger}}}

\newcommand{\NODEMessage}[7]{\NODEcontentWithAttributes{\strut\NODEnameWithType{#1}{Message}}{\def\NewLineNeeded{0}\ifthenelse{\equal{#2}{}}{}{\ifthenelse{\equal{\NewLineNeeded}{1}}{\\}{}\def\NewLineNeeded{1}\strut\ATTRIBUTEnameWithValue{clock}{#2}}\ifthenelse{\equal{#3}{}}{}{\ifthenelse{\equal{\NewLineNeeded}{1}}{\\}{}\def\NewLineNeeded{1}\strut\ATTRIBUTEnameWithValue{id}{#3}}\ifthenelse{\equal{#4}{}}{}{\ifthenelse{\equal{\NewLineNeeded}{1}}{\\}{}\def\NewLineNeeded{1}\strut\ATTRIBUTEnameWithValue{cts}{#4}}\ifthenelse{\equal{#5}{}}{}{\ifthenelse{\equal{\NewLineNeeded}{1}}{\\}{}\def\NewLineNeeded{1}\strut\ATTRIBUTEnameWithValue{dts}{#5}}\ifthenelse{\equal{#6}{}}{}{\ifthenelse{\equal{\NewLineNeeded}{1}}{\\}{}\def\NewLineNeeded{1}\strut\ATTRIBUTEnameWithValue{cidx}{#6}}\ifthenelse{\equal{#7}{}}{}{\ifthenelse{\equal{\NewLineNeeded}{1}}{\\}{}\def\NewLineNeeded{1}\strut\ATTRIBUTEnameWithValue{didx}{#7}}}}

\pgfkeys{
	/NODEMessage/.cd,
	name/.code = \xdef\myNODEMessageNAME{\unexpanded{#1}},
	appendTitle/.code = \xdef\myNODEMessageSFSexpand{\unexpanded{#1}},
	clock/.code = \xdef\myNODEMessageATTclock{\unexpanded{#1}},
	id/.code = \xdef\myNODEMessageATTid{\unexpanded{#1}},
	cts/.code = \xdef\myNODEMessageATTcts{\unexpanded{#1}},
	dts/.code = \xdef\myNODEMessageATTdts{\unexpanded{#1}},
	cidx/.code = \xdef\myNODEMessageATTcidx{\unexpanded{#1}},
	didx/.code = \xdef\myNODEMessageATTdidx{\unexpanded{#1}},
}
\newcommand*\myNODEMessageNAME{}\newcommand*\myNODEMessageSFSexpand{}\newcommand*\myNODEMessageATTclock{}\newcommand*\myNODEMessageATTid{}\newcommand*\myNODEMessageATTcts{}\newcommand*\myNODEMessageATTdts{}\newcommand*\myNODEMessageATTcidx{}\newcommand*\myNODEMessageATTdidx{}

\newcommand{\NODEMessageX}[1]{\gdef\myNODEMessageNAME{}\gdef\myNODEMessageSFSexpand{}\gdef\myNODEMessageATTclock{}\gdef\myNODEMessageATTid{}\gdef\myNODEMessageATTcts{}\gdef\myNODEMessageATTdts{}\gdef\myNODEMessageATTcidx{}\gdef\myNODEMessageATTdidx{}\pgfqkeys{/NODEMessage}{#1}\NODEcontentWithAttributes{\strut\NODEnameWithType{\myNODEMessageNAME}{Message}\myNODEMessageSFSexpand}{\def\NewLineNeeded{0}\ifthenelse{\equal{\myNODEMessageATTclock}{}}{}{\ifthenelse{\equal{\NewLineNeeded}{1}}{\\}{}\def\NewLineNeeded{1}\strut\ATTRIBUTEnameWithValue{clock}{\myNODEMessageATTclock}}\ifthenelse{\equal{\myNODEMessageATTid}{}}{}{\ifthenelse{\equal{\NewLineNeeded}{1}}{\\}{}\def\NewLineNeeded{1}\strut\ATTRIBUTEnameWithValue{id}{\myNODEMessageATTid}}\ifthenelse{\equal{\myNODEMessageATTcts}{}}{}{\ifthenelse{\equal{\NewLineNeeded}{1}}{\\}{}\def\NewLineNeeded{1}\strut\ATTRIBUTEnameWithValue{cts}{\myNODEMessageATTcts}}\ifthenelse{\equal{\myNODEMessageATTdts}{}}{}{\ifthenelse{\equal{\NewLineNeeded}{1}}{\\}{}\def\NewLineNeeded{1}\strut\ATTRIBUTEnameWithValue{dts}{\myNODEMessageATTdts}}\ifthenelse{\equal{\myNODEMessageATTcidx}{}}{}{\ifthenelse{\equal{\NewLineNeeded}{1}}{\\}{}\def\NewLineNeeded{1}\strut\ATTRIBUTEnameWithValue{cidx}{\myNODEMessageATTcidx}}\ifthenelse{\equal{\myNODEMessageATTdidx}{}}{}{\ifthenelse{\equal{\NewLineNeeded}{1}}{\\}{}\def\NewLineNeeded{1}\strut\ATTRIBUTEnameWithValue{didx}{\myNODEMessageATTdidx}}}}

\tikzset{
	styleNODESender/.style={
		line width=1pt,
		minimum width=0cm,
		outer ysep=1pt,
		inner ysep=1pt,
		inner xsep=1pt,
		outer xsep=1pt,
		draw,
		shape=rectangle split,
		rectangle split parts=2,
		align=left,
		rectangle split part fill={green!70!black!60!white,white},
		rectangle split every empty part={},
		rectangle split empty part height=0pt,
	}
}

\newcommand{\NODETGSenderShort}{\NODEcontentWithAttributes{\strut\NODEtype{Sender\MULT{0}{*}}}{\strut\ATTRIBUTEnameWithSort{num}{\TypeInteger}}}

\newcommand{\NODESender}[6]{\NODEcontentWithAttributes{\strut\NODEnameWithType{#1}{Sender}}{\def\NewLineNeeded{0}\ifthenelse{\equal{#2}{}}{}{\ifthenelse{\equal{\NewLineNeeded}{1}}{\\}{}\def\NewLineNeeded{1}\strut\ATTRIBUTEnameWithValue{num}{#2}}\ifthenelse{\equal{#3}{}}{}{\ifthenelse{\equal{\NewLineNeeded}{1}}{\\}{}\def\NewLineNeeded{1}\strut\ATTRIBUTEnameWithValue{cts}{#3}}\ifthenelse{\equal{#4}{}}{}{\ifthenelse{\equal{\NewLineNeeded}{1}}{\\}{}\def\NewLineNeeded{1}\strut\ATTRIBUTEnameWithValue{dts}{#4}}\ifthenelse{\equal{#5}{}}{}{\ifthenelse{\equal{\NewLineNeeded}{1}}{\\}{}\def\NewLineNeeded{1}\strut\ATTRIBUTEnameWithValue{cidx}{#5}}\ifthenelse{\equal{#6}{}}{}{\ifthenelse{\equal{\NewLineNeeded}{1}}{\\}{}\def\NewLineNeeded{1}\strut\ATTRIBUTEnameWithValue{didx}{#6}}}}

\pgfkeys{
	/NODESender/.cd,
	name/.code = \xdef\myNODESenderNAME{\unexpanded{#1}},
	appendTitle/.code = \xdef\myNODESenderSFSexpand{\unexpanded{#1}},
	num/.code = \xdef\myNODESenderATTnum{\unexpanded{#1}},
	cts/.code = \xdef\myNODESenderATTcts{\unexpanded{#1}},
	dts/.code = \xdef\myNODESenderATTdts{\unexpanded{#1}},
	cidx/.code = \xdef\myNODESenderATTcidx{\unexpanded{#1}},
	didx/.code = \xdef\myNODESenderATTdidx{\unexpanded{#1}},
}
\newcommand*\myNODESenderNAME{}\newcommand*\myNODESenderSFSexpand{}\newcommand*\myNODESenderATTnum{}\newcommand*\myNODESenderATTcts{}\newcommand*\myNODESenderATTdts{}\newcommand*\myNODESenderATTcidx{}\newcommand*\myNODESenderATTdidx{}

\newcommand{\NODESenderX}[1]{\gdef\myNODESenderNAME{}\gdef\myNODESenderSFSexpand{}\gdef\myNODESenderATTnum{}\gdef\myNODESenderATTcts{}\gdef\myNODESenderATTdts{}\gdef\myNODESenderATTcidx{}\gdef\myNODESenderATTdidx{}\pgfqkeys{/NODESender}{#1}\NODEcontentWithAttributes{\strut\NODEnameWithType{\myNODESenderNAME}{Sender}\myNODESenderSFSexpand}{\def\NewLineNeeded{0}\ifthenelse{\equal{\myNODESenderATTnum}{}}{}{\ifthenelse{\equal{\NewLineNeeded}{1}}{\\}{}\def\NewLineNeeded{1}\strut\ATTRIBUTEnameWithValue{num}{\myNODESenderATTnum}}\ifthenelse{\equal{\myNODESenderATTcts}{}}{}{\ifthenelse{\equal{\NewLineNeeded}{1}}{\\}{}\def\NewLineNeeded{1}\strut\ATTRIBUTEnameWithValue{cts}{\myNODESenderATTcts}}\ifthenelse{\equal{\myNODESenderATTdts}{}}{}{\ifthenelse{\equal{\NewLineNeeded}{1}}{\\}{}\def\NewLineNeeded{1}\strut\ATTRIBUTEnameWithValue{dts}{\myNODESenderATTdts}}\ifthenelse{\equal{\myNODESenderATTcidx}{}}{}{\ifthenelse{\equal{\NewLineNeeded}{1}}{\\}{}\def\NewLineNeeded{1}\strut\ATTRIBUTEnameWithValue{cidx}{\myNODESenderATTcidx}}\ifthenelse{\equal{\myNODESenderATTdidx}{}}{}{\ifthenelse{\equal{\NewLineNeeded}{1}}{\\}{}\def\NewLineNeeded{1}\strut\ATTRIBUTEnameWithValue{didx}{\myNODESenderATTdidx}}}}

\tikzset{
	styleNODEReceiver/.style={
		line width=1pt,
		minimum width=0cm,
		outer ysep=1pt,
		inner ysep=1pt,
		inner xsep=1pt,
		outer xsep=1pt,
		draw,
		shape=rectangle split,
		rectangle split parts=2,
		align=left,
		rectangle split part fill={green!20!white,white},
		rectangle split every empty part={},
		rectangle split empty part height=0pt,
	}
}

\newcommand{\NODETGReceiverShort}{\NODEcontentWithAttributes{\strut\NODEtype{Receiver\MULT{0}{*}}}{}}

\newcommand{\NODEReceiver}[5]{\NODEcontentWithAttributes{\strut\NODEnameWithType{#1}{Receiver}}{\def\NewLineNeeded{0}\ifthenelse{\equal{#2}{}}{}{\ifthenelse{\equal{\NewLineNeeded}{1}}{\\}{}\def\NewLineNeeded{1}\strut\ATTRIBUTEnameWithValue{cts}{#2}}\ifthenelse{\equal{#3}{}}{}{\ifthenelse{\equal{\NewLineNeeded}{1}}{\\}{}\def\NewLineNeeded{1}\strut\ATTRIBUTEnameWithValue{dts}{#3}}\ifthenelse{\equal{#4}{}}{}{\ifthenelse{\equal{\NewLineNeeded}{1}}{\\}{}\def\NewLineNeeded{1}\strut\ATTRIBUTEnameWithValue{cidx}{#4}}\ifthenelse{\equal{#5}{}}{}{\ifthenelse{\equal{\NewLineNeeded}{1}}{\\}{}\def\NewLineNeeded{1}\strut\ATTRIBUTEnameWithValue{didx}{#5}}}}

\pgfkeys{
	/NODEReceiver/.cd,
	name/.code = \xdef\myNODEReceiverNAME{\unexpanded{#1}},
	appendTitle/.code = \xdef\myNODEReceiverSFSexpand{\unexpanded{#1}},
	cts/.code = \xdef\myNODEReceiverATTcts{\unexpanded{#1}},
	dts/.code = \xdef\myNODEReceiverATTdts{\unexpanded{#1}},
	cidx/.code = \xdef\myNODEReceiverATTcidx{\unexpanded{#1}},
	didx/.code = \xdef\myNODEReceiverATTdidx{\unexpanded{#1}},
}
\newcommand*\myNODEReceiverNAME{}\newcommand*\myNODEReceiverSFSexpand{}\newcommand*\myNODEReceiverATTcts{}\newcommand*\myNODEReceiverATTdts{}\newcommand*\myNODEReceiverATTcidx{}\newcommand*\myNODEReceiverATTdidx{}

\newcommand{\NODEReceiverX}[1]{\gdef\myNODEReceiverNAME{}\gdef\myNODEReceiverSFSexpand{}\gdef\myNODEReceiverATTcts{}\gdef\myNODEReceiverATTdts{}\gdef\myNODEReceiverATTcidx{}\gdef\myNODEReceiverATTdidx{}\pgfqkeys{/NODEReceiver}{#1}\NODEcontentWithAttributes{\strut\NODEnameWithType{\myNODEReceiverNAME}{Receiver}\myNODEReceiverSFSexpand}{\def\NewLineNeeded{0}\ifthenelse{\equal{\myNODEReceiverATTcts}{}}{}{\ifthenelse{\equal{\NewLineNeeded}{1}}{\\}{}\def\NewLineNeeded{1}\strut\ATTRIBUTEnameWithValue{cts}{\myNODEReceiverATTcts}}\ifthenelse{\equal{\myNODEReceiverATTdts}{}}{}{\ifthenelse{\equal{\NewLineNeeded}{1}}{\\}{}\def\NewLineNeeded{1}\strut\ATTRIBUTEnameWithValue{dts}{\myNODEReceiverATTdts}}\ifthenelse{\equal{\myNODEReceiverATTcidx}{}}{}{\ifthenelse{\equal{\NewLineNeeded}{1}}{\\}{}\def\NewLineNeeded{1}\strut\ATTRIBUTEnameWithValue{cidx}{\myNODEReceiverATTcidx}}\ifthenelse{\equal{\myNODEReceiverATTdidx}{}}{}{\ifthenelse{\equal{\NewLineNeeded}{1}}{\\}{}\def\NewLineNeeded{1}\strut\ATTRIBUTEnameWithValue{didx}{\myNODEReceiverATTdidx}}}}

\tikzset{
	styleEDGEnext/.style={
		line width=1pt,
		minimum width=0cm,
		outer ysep=1pt,
		inner ysep=1pt,
		inner xsep=1pt,
		outer xsep=1pt,
		shape=rectangle split,
		rectangle split parts=2,
		align=left,
		rectangle split part fill={white!80!gray,white!60!gray,white},
		rectangle split every empty part={},
		rectangle split empty part height=0pt,
	}
}

\newcommand{\EDGETGnextShort}{\EDGEcontentWithAttributes{\strut\EDGEtype{next\MULT{0}{*}}}{}}

\newcommand{\EDGEnext}[5]{\EDGEcontentWithAttributes{\strut\EDGEnameWithType{#1}{next}}{\def\NewLineNeeded{0}\ifthenelse{\equal{#2}{}}{}{\ifthenelse{\equal{\NewLineNeeded}{1}}{\\}{}\def\NewLineNeeded{1}\strut\ATTRIBUTEnameWithValue{cts}{#2}}\ifthenelse{\equal{#3}{}}{}{\ifthenelse{\equal{\NewLineNeeded}{1}}{\\}{}\def\NewLineNeeded{1}\strut\ATTRIBUTEnameWithValue{dts}{#3}}\ifthenelse{\equal{#4}{}}{}{\ifthenelse{\equal{\NewLineNeeded}{1}}{\\}{}\def\NewLineNeeded{1}\strut\ATTRIBUTEnameWithValue{cidx}{#4}}\ifthenelse{\equal{#5}{}}{}{\ifthenelse{\equal{\NewLineNeeded}{1}}{\\}{}\def\NewLineNeeded{1}\strut\ATTRIBUTEnameWithValue{didx}{#5}}}}

\pgfkeys{
	/EDGEnext/.cd,
	name/.code = \xdef\myEDGEnextNAME{\unexpanded{#1}},
	appendTitle/.code = \xdef\myEDGEnextSFSexpand{\unexpanded{#1}},
	cts/.code = \xdef\myEDGEnextATTcts{\unexpanded{#1}},
	dts/.code = \xdef\myEDGEnextATTdts{\unexpanded{#1}},
	cidx/.code = \xdef\myEDGEnextATTcidx{\unexpanded{#1}},
	didx/.code = \xdef\myEDGEnextATTdidx{\unexpanded{#1}},
}
\newcommand*\myEDGEnextNAME{}\newcommand*\myEDGEnextSFSexpand{}\newcommand*\myEDGEnextATTcts{}\newcommand*\myEDGEnextATTdts{}\newcommand*\myEDGEnextATTcidx{}\newcommand*\myEDGEnextATTdidx{}

\newcommand{\EDGEnextX}[1]{\gdef\myEDGEnextNAME{}\gdef\myEDGEnextSFSexpand{}\gdef\myEDGEnextATTcts{}\gdef\myEDGEnextATTdts{}\gdef\myEDGEnextATTcidx{}\gdef\myEDGEnextATTdidx{}\pgfqkeys{/EDGEnext}{#1}\EDGEcontentWithAttributes{\strut\EDGEnameWithType{\myEDGEnextNAME}{next}\myEDGEnextSFSexpand}{\def\NewLineNeeded{0}\ifthenelse{\equal{\myEDGEnextATTcts}{}}{}{\ifthenelse{\equal{\NewLineNeeded}{1}}{\\}{}\def\NewLineNeeded{1}\strut\ATTRIBUTEnameWithValue{cts}{\myEDGEnextATTcts}}\ifthenelse{\equal{\myEDGEnextATTdts}{}}{}{\ifthenelse{\equal{\NewLineNeeded}{1}}{\\}{}\def\NewLineNeeded{1}\strut\ATTRIBUTEnameWithValue{dts}{\myEDGEnextATTdts}}\ifthenelse{\equal{\myEDGEnextATTcidx}{}}{}{\ifthenelse{\equal{\NewLineNeeded}{1}}{\\}{}\def\NewLineNeeded{1}\strut\ATTRIBUTEnameWithValue{cidx}{\myEDGEnextATTcidx}}\ifthenelse{\equal{\myEDGEnextATTdidx}{}}{}{\ifthenelse{\equal{\NewLineNeeded}{1}}{\\}{}\def\NewLineNeeded{1}\strut\ATTRIBUTEnameWithValue{didx}{\myEDGEnextATTdidx}}}}

\tikzset{
	styleEDGEsnd/.style={
		line width=1pt,
		minimum width=0cm,
		outer ysep=1pt,
		inner ysep=1pt,
		inner xsep=1pt,
		outer xsep=1pt,
		shape=rectangle split,
		rectangle split parts=2,
		align=left,
		rectangle split part fill={white!80!gray,white!60!gray,white},
		rectangle split every empty part={},
		rectangle split empty part height=0pt,
	}
}

\newcommand{\EDGETGsndShort}{\EDGEcontentWithAttributes{\strut\EDGEtype{snd\MULT{0}{*}}}{}}

\newcommand{\EDGEsnd}[5]{\EDGEcontentWithAttributes{\strut\EDGEnameWithType{#1}{snd}}{\def\NewLineNeeded{0}\ifthenelse{\equal{#2}{}}{}{\ifthenelse{\equal{\NewLineNeeded}{1}}{\\}{}\def\NewLineNeeded{1}\strut\ATTRIBUTEnameWithValue{cts}{#2}}\ifthenelse{\equal{#3}{}}{}{\ifthenelse{\equal{\NewLineNeeded}{1}}{\\}{}\def\NewLineNeeded{1}\strut\ATTRIBUTEnameWithValue{dts}{#3}}\ifthenelse{\equal{#4}{}}{}{\ifthenelse{\equal{\NewLineNeeded}{1}}{\\}{}\def\NewLineNeeded{1}\strut\ATTRIBUTEnameWithValue{cidx}{#4}}\ifthenelse{\equal{#5}{}}{}{\ifthenelse{\equal{\NewLineNeeded}{1}}{\\}{}\def\NewLineNeeded{1}\strut\ATTRIBUTEnameWithValue{didx}{#5}}}}

\pgfkeys{
	/EDGEsnd/.cd,
	name/.code = \xdef\myEDGEsndNAME{\unexpanded{#1}},
	appendTitle/.code = \xdef\myEDGEsndSFSexpand{\unexpanded{#1}},
	cts/.code = \xdef\myEDGEsndATTcts{\unexpanded{#1}},
	dts/.code = \xdef\myEDGEsndATTdts{\unexpanded{#1}},
	cidx/.code = \xdef\myEDGEsndATTcidx{\unexpanded{#1}},
	didx/.code = \xdef\myEDGEsndATTdidx{\unexpanded{#1}},
}
\newcommand*\myEDGEsndNAME{}\newcommand*\myEDGEsndSFSexpand{}\newcommand*\myEDGEsndATTcts{}\newcommand*\myEDGEsndATTdts{}\newcommand*\myEDGEsndATTcidx{}\newcommand*\myEDGEsndATTdidx{}

\newcommand{\EDGEsndX}[1]{\gdef\myEDGEsndNAME{}\gdef\myEDGEsndSFSexpand{}\gdef\myEDGEsndATTcts{}\gdef\myEDGEsndATTdts{}\gdef\myEDGEsndATTcidx{}\gdef\myEDGEsndATTdidx{}\pgfqkeys{/EDGEsnd}{#1}\EDGEcontentWithAttributes{\strut\EDGEnameWithType{\myEDGEsndNAME}{snd}\myEDGEsndSFSexpand}{\def\NewLineNeeded{0}\ifthenelse{\equal{\myEDGEsndATTcts}{}}{}{\ifthenelse{\equal{\NewLineNeeded}{1}}{\\}{}\def\NewLineNeeded{1}\strut\ATTRIBUTEnameWithValue{cts}{\myEDGEsndATTcts}}\ifthenelse{\equal{\myEDGEsndATTdts}{}}{}{\ifthenelse{\equal{\NewLineNeeded}{1}}{\\}{}\def\NewLineNeeded{1}\strut\ATTRIBUTEnameWithValue{dts}{\myEDGEsndATTdts}}\ifthenelse{\equal{\myEDGEsndATTcidx}{}}{}{\ifthenelse{\equal{\NewLineNeeded}{1}}{\\}{}\def\NewLineNeeded{1}\strut\ATTRIBUTEnameWithValue{cidx}{\myEDGEsndATTcidx}}\ifthenelse{\equal{\myEDGEsndATTdidx}{}}{}{\ifthenelse{\equal{\NewLineNeeded}{1}}{\\}{}\def\NewLineNeeded{1}\strut\ATTRIBUTEnameWithValue{didx}{\myEDGEsndATTdidx}}}}

\tikzset{
	styleEDGErcv/.style={
		line width=1pt,
		minimum width=0cm,
		outer ysep=1pt,
		inner ysep=1pt,
		inner xsep=1pt,
		outer xsep=1pt,
		shape=rectangle split,
		rectangle split parts=2,
		align=left,
		rectangle split part fill={white!80!gray,white!60!gray,white},
		rectangle split every empty part={},
		rectangle split empty part height=0pt,
	}
}

\newcommand{\EDGETGrcvShort}{\EDGEcontentWithAttributes{\strut\EDGEtype{rcv\MULT{0}{*}}}{}}

\newcommand{\EDGErcv}[5]{\EDGEcontentWithAttributes{\strut\EDGEnameWithType{#1}{rcv}}{\def\NewLineNeeded{0}\ifthenelse{\equal{#2}{}}{}{\ifthenelse{\equal{\NewLineNeeded}{1}}{\\}{}\def\NewLineNeeded{1}\strut\ATTRIBUTEnameWithValue{cts}{#2}}\ifthenelse{\equal{#3}{}}{}{\ifthenelse{\equal{\NewLineNeeded}{1}}{\\}{}\def\NewLineNeeded{1}\strut\ATTRIBUTEnameWithValue{dts}{#3}}\ifthenelse{\equal{#4}{}}{}{\ifthenelse{\equal{\NewLineNeeded}{1}}{\\}{}\def\NewLineNeeded{1}\strut\ATTRIBUTEnameWithValue{cidx}{#4}}\ifthenelse{\equal{#5}{}}{}{\ifthenelse{\equal{\NewLineNeeded}{1}}{\\}{}\def\NewLineNeeded{1}\strut\ATTRIBUTEnameWithValue{didx}{#5}}}}

\pgfkeys{
	/EDGErcv/.cd,
	name/.code = \xdef\myEDGErcvNAME{\unexpanded{#1}},
	appendTitle/.code = \xdef\myEDGErcvSFSexpand{\unexpanded{#1}},
	cts/.code = \xdef\myEDGErcvATTcts{\unexpanded{#1}},
	dts/.code = \xdef\myEDGErcvATTdts{\unexpanded{#1}},
	cidx/.code = \xdef\myEDGErcvATTcidx{\unexpanded{#1}},
	didx/.code = \xdef\myEDGErcvATTdidx{\unexpanded{#1}},
}
\newcommand*\myEDGErcvNAME{}\newcommand*\myEDGErcvSFSexpand{}\newcommand*\myEDGErcvATTcts{}\newcommand*\myEDGErcvATTdts{}\newcommand*\myEDGErcvATTcidx{}\newcommand*\myEDGErcvATTdidx{}

\newcommand{\EDGErcvX}[1]{\gdef\myEDGErcvNAME{}\gdef\myEDGErcvSFSexpand{}\gdef\myEDGErcvATTcts{}\gdef\myEDGErcvATTdts{}\gdef\myEDGErcvATTcidx{}\gdef\myEDGErcvATTdidx{}\pgfqkeys{/EDGErcv}{#1}\EDGEcontentWithAttributes{\strut\EDGEnameWithType{\myEDGErcvNAME}{rcv}\myEDGErcvSFSexpand}{\def\NewLineNeeded{0}\ifthenelse{\equal{\myEDGErcvATTcts}{}}{}{\ifthenelse{\equal{\NewLineNeeded}{1}}{\\}{}\def\NewLineNeeded{1}\strut\ATTRIBUTEnameWithValue{cts}{\myEDGErcvATTcts}}\ifthenelse{\equal{\myEDGErcvATTdts}{}}{}{\ifthenelse{\equal{\NewLineNeeded}{1}}{\\}{}\def\NewLineNeeded{1}\strut\ATTRIBUTEnameWithValue{dts}{\myEDGErcvATTdts}}\ifthenelse{\equal{\myEDGErcvATTcidx}{}}{}{\ifthenelse{\equal{\NewLineNeeded}{1}}{\\}{}\def\NewLineNeeded{1}\strut\ATTRIBUTEnameWithValue{cidx}{\myEDGErcvATTcidx}}\ifthenelse{\equal{\myEDGErcvATTdidx}{}}{}{\ifthenelse{\equal{\NewLineNeeded}{1}}{\\}{}\def\NewLineNeeded{1}\strut\ATTRIBUTEnameWithValue{didx}{\myEDGErcvATTdidx}}}}

\tikzset{
	styleEDGEat/.style={
		line width=1pt,
		minimum width=0cm,
		outer ysep=1pt,
		inner ysep=1pt,
		inner xsep=1pt,
		outer xsep=1pt,
		shape=rectangle split,
		rectangle split parts=2,
		align=left,
		rectangle split part fill={white!80!gray,white!60!gray,white},
		rectangle split every empty part={},
		rectangle split empty part height=0pt,
	}
}

\newcommand{\EDGETGatShort}{\EDGEcontentWithAttributes{\strut\EDGEtype{at\MULT{0}{*}}}{}}

\newcommand{\EDGEat}[5]{\EDGEcontentWithAttributes{\strut\EDGEnameWithType{#1}{at}}{\def\NewLineNeeded{0}\ifthenelse{\equal{#2}{}}{}{\ifthenelse{\equal{\NewLineNeeded}{1}}{\\}{}\def\NewLineNeeded{1}\strut\ATTRIBUTEnameWithValue{cts}{#2}}\ifthenelse{\equal{#3}{}}{}{\ifthenelse{\equal{\NewLineNeeded}{1}}{\\}{}\def\NewLineNeeded{1}\strut\ATTRIBUTEnameWithValue{dts}{#3}}\ifthenelse{\equal{#4}{}}{}{\ifthenelse{\equal{\NewLineNeeded}{1}}{\\}{}\def\NewLineNeeded{1}\strut\ATTRIBUTEnameWithValue{cidx}{#4}}\ifthenelse{\equal{#5}{}}{}{\ifthenelse{\equal{\NewLineNeeded}{1}}{\\}{}\def\NewLineNeeded{1}\strut\ATTRIBUTEnameWithValue{didx}{#5}}}}

\pgfkeys{
	/EDGEat/.cd,
	name/.code = \xdef\myEDGEatNAME{\unexpanded{#1}},
	appendTitle/.code = \xdef\myEDGEatSFSexpand{\unexpanded{#1}},
	cts/.code = \xdef\myEDGEatATTcts{\unexpanded{#1}},
	dts/.code = \xdef\myEDGEatATTdts{\unexpanded{#1}},
	cidx/.code = \xdef\myEDGEatATTcidx{\unexpanded{#1}},
	didx/.code = \xdef\myEDGEatATTdidx{\unexpanded{#1}},
}
\newcommand*\myEDGEatNAME{}\newcommand*\myEDGEatSFSexpand{}\newcommand*\myEDGEatATTcts{}\newcommand*\myEDGEatATTdts{}\newcommand*\myEDGEatATTcidx{}\newcommand*\myEDGEatATTdidx{}

\newcommand{\EDGEatX}[1]{\gdef\myEDGEatNAME{}\gdef\myEDGEatSFSexpand{}\gdef\myEDGEatATTcts{}\gdef\myEDGEatATTdts{}\gdef\myEDGEatATTcidx{}\gdef\myEDGEatATTdidx{}\pgfqkeys{/EDGEat}{#1}\EDGEcontentWithAttributes{\strut\EDGEnameWithType{\myEDGEatNAME}{at}\myEDGEatSFSexpand}{\def\NewLineNeeded{0}\ifthenelse{\equal{\myEDGEatATTcts}{}}{}{\ifthenelse{\equal{\NewLineNeeded}{1}}{\\}{}\def\NewLineNeeded{1}\strut\ATTRIBUTEnameWithValue{cts}{\myEDGEatATTcts}}\ifthenelse{\equal{\myEDGEatATTdts}{}}{}{\ifthenelse{\equal{\NewLineNeeded}{1}}{\\}{}\def\NewLineNeeded{1}\strut\ATTRIBUTEnameWithValue{dts}{\myEDGEatATTdts}}\ifthenelse{\equal{\myEDGEatATTcidx}{}}{}{\ifthenelse{\equal{\NewLineNeeded}{1}}{\\}{}\def\NewLineNeeded{1}\strut\ATTRIBUTEnameWithValue{cidx}{\myEDGEatATTcidx}}\ifthenelse{\equal{\myEDGEatATTdidx}{}}{}{\ifthenelse{\equal{\NewLineNeeded}{1}}{\\}{}\def\NewLineNeeded{1}\strut\ATTRIBUTEnameWithValue{didx}{\myEDGEatATTdidx}}}}

\tikzset{
	styleEDGEdone/.style={
		line width=1pt,
		minimum width=0cm,
		outer ysep=1pt,
		inner ysep=1pt,
		inner xsep=1pt,
		outer xsep=1pt,
		shape=rectangle split,
		rectangle split parts=2,
		align=left,
		rectangle split part fill={white!80!gray,white!60!gray,white},
		rectangle split every empty part={},
		rectangle split empty part height=0pt,
	}
}

\newcommand{\EDGETGdoneNAME}{\ensuremath{\mathit{done}}\xspace }

\newcommand{\EDGETGdoneShort}{\EDGEcontentWithAttributes{\strut\EDGEtype{done\MULT{0}{*}}}{}}

\newcommand{\EDGEdone}[5]{\EDGEcontentWithAttributes{\strut\EDGEnameWithType{#1}{done}}{\def\NewLineNeeded{0}\ifthenelse{\equal{#2}{}}{}{\ifthenelse{\equal{\NewLineNeeded}{1}}{\\}{}\def\NewLineNeeded{1}\strut\ATTRIBUTEnameWithValue{cts}{#2}}\ifthenelse{\equal{#3}{}}{}{\ifthenelse{\equal{\NewLineNeeded}{1}}{\\}{}\def\NewLineNeeded{1}\strut\ATTRIBUTEnameWithValue{dts}{#3}}\ifthenelse{\equal{#4}{}}{}{\ifthenelse{\equal{\NewLineNeeded}{1}}{\\}{}\def\NewLineNeeded{1}\strut\ATTRIBUTEnameWithValue{cidx}{#4}}\ifthenelse{\equal{#5}{}}{}{\ifthenelse{\equal{\NewLineNeeded}{1}}{\\}{}\def\NewLineNeeded{1}\strut\ATTRIBUTEnameWithValue{didx}{#5}}}}

\pgfkeys{
	/EDGEdone/.cd,
	name/.code = \xdef\myEDGEdoneNAME{\unexpanded{#1}},
	appendTitle/.code = \xdef\myEDGEdoneSFSexpand{\unexpanded{#1}},
	cts/.code = \xdef\myEDGEdoneATTcts{\unexpanded{#1}},
	dts/.code = \xdef\myEDGEdoneATTdts{\unexpanded{#1}},
	cidx/.code = \xdef\myEDGEdoneATTcidx{\unexpanded{#1}},
	didx/.code = \xdef\myEDGEdoneATTdidx{\unexpanded{#1}},
}
\newcommand*\myEDGEdoneNAME{}\newcommand*\myEDGEdoneSFSexpand{}\newcommand*\myEDGEdoneATTcts{}\newcommand*\myEDGEdoneATTdts{}\newcommand*\myEDGEdoneATTcidx{}\newcommand*\myEDGEdoneATTdidx{}

\newcommand{\EDGEdoneX}[1]{\gdef\myEDGEdoneNAME{}\gdef\myEDGEdoneSFSexpand{}\gdef\myEDGEdoneATTcts{}\gdef\myEDGEdoneATTdts{}\gdef\myEDGEdoneATTcidx{}\gdef\myEDGEdoneATTdidx{}\pgfqkeys{/EDGEdone}{#1}\EDGEcontentWithAttributes{\strut\EDGEnameWithType{\myEDGEdoneNAME}{done}\myEDGEdoneSFSexpand}{\def\NewLineNeeded{0}\ifthenelse{\equal{\myEDGEdoneATTcts}{}}{}{\ifthenelse{\equal{\NewLineNeeded}{1}}{\\}{}\def\NewLineNeeded{1}\strut\ATTRIBUTEnameWithValue{cts}{\myEDGEdoneATTcts}}\ifthenelse{\equal{\myEDGEdoneATTdts}{}}{}{\ifthenelse{\equal{\NewLineNeeded}{1}}{\\}{}\def\NewLineNeeded{1}\strut\ATTRIBUTEnameWithValue{dts}{\myEDGEdoneATTdts}}\ifthenelse{\equal{\myEDGEdoneATTcidx}{}}{}{\ifthenelse{\equal{\NewLineNeeded}{1}}{\\}{}\def\NewLineNeeded{1}\strut\ATTRIBUTEnameWithValue{cidx}{\myEDGEdoneATTcidx}}\ifthenelse{\equal{\myEDGEdoneATTdidx}{}}{}{\ifthenelse{\equal{\NewLineNeeded}{1}}{\\}{}\def\NewLineNeeded{1}\strut\ATTRIBUTEnameWithValue{didx}{\myEDGEdoneATTdidx}}}}

 \tikzset{
	styleNODEA/.style={
		line width=1pt,
		minimum width=0cm,
		outer ysep=1pt,
		inner ysep=1pt,
		inner xsep=1pt,
		outer xsep=1pt,
		draw,
		shape=rectangle split,
		rectangle split parts=2,
		align=left,
		rectangle split part fill={-red!75!green,white},
		rectangle split every empty part={},
		rectangle split empty part height=0pt,
	}
}

\newcommand{\NODEA}[4]{\NODEcontentWithAttributes{\strut\NODEnameWithType{#1}{A}}{\def\NewLineNeeded{0}\ifthenelse{\equal{#2}{}}{}{\ifthenelse{\equal{\NewLineNeeded}{1}}{\\}{}\def\NewLineNeeded{1}\strut\ATTRIBUTEnameWithValue{id}{#2}}\ifthenelse{\equal{#3}{}}{}{\ifthenelse{\equal{\NewLineNeeded}{1}}{\\}{}\def\NewLineNeeded{1}\strut\ATTRIBUTEnameWithValue{cnt}{#3}}\ifthenelse{\equal{#4}{}}{}{\ifthenelse{\equal{\NewLineNeeded}{1}}{\\}{}\def\NewLineNeeded{1}\strut\ATTRIBUTEnameWithValue{clock}{#4}}}}

\pgfkeys{
	/NODEA/.cd,
	name/.code = \xdef\myNODEANAME{\unexpanded{#1}},
	appendTitle/.code = \xdef\myNODEASFSexpand{\unexpanded{#1}},
	id/.code = \xdef\myNODEAATTid{\unexpanded{#1}},
	cnt/.code = \xdef\myNODEAATTcnt{\unexpanded{#1}},
	clock/.code = \xdef\myNODEAATTclock{\unexpanded{#1}},
}
\newcommand*\myNODEANAME{}\newcommand*\myNODEASFSexpand{}\newcommand*\myNODEAATTid{}\newcommand*\myNODEAATTcnt{}\newcommand*\myNODEAATTclock{}

\newcommand{\NODEAX}[1]{\gdef\myNODEANAME{}\gdef\myNODEASFSexpand{}\gdef\myNODEAATTid{}\gdef\myNODEAATTcnt{}\gdef\myNODEAATTclock{}\pgfqkeys{/NODEA}{#1}\NODEcontentWithAttributes{\strut\NODEnameWithType{\myNODEANAME}{A}\myNODEASFSexpand}{\def\NewLineNeeded{0}\ifthenelse{\equal{\myNODEAATTid}{}}{}{\ifthenelse{\equal{\NewLineNeeded}{1}}{\\}{}\def\NewLineNeeded{1}\strut\ATTRIBUTEnameWithValue{id}{\myNODEAATTid}}\ifthenelse{\equal{\myNODEAATTcnt}{}}{}{\ifthenelse{\equal{\NewLineNeeded}{1}}{\\}{}\def\NewLineNeeded{1}\strut\ATTRIBUTEnameWithValue{cnt}{\myNODEAATTcnt}}\ifthenelse{\equal{\myNODEAATTclock}{}}{}{\ifthenelse{\equal{\NewLineNeeded}{1}}{\\}{}\def\NewLineNeeded{1}\strut\ATTRIBUTEnameWithValue{clock}{\myNODEAATTclock}}}}

\tikzset{
	styleNODEB/.style={
		line width=1pt,
		minimum width=0cm,
		outer ysep=1pt,
		inner ysep=1pt,
		inner xsep=1pt,
		outer xsep=1pt,
		draw,
		shape=rectangle split,
		rectangle split parts=2,
		align=left,
		rectangle split part fill={-red!75!green!50!blue,white},
		rectangle split every empty part={},
		rectangle split empty part height=0pt,
	}
}

\newcommand{\NODEB}[3]{\NODEcontentWithAttributes{\strut\NODEnameWithType{#1}{B}}{\def\NewLineNeeded{0}\ifthenelse{\equal{#2}{}}{}{\ifthenelse{\equal{\NewLineNeeded}{1}}{\\}{}\def\NewLineNeeded{1}\strut\ATTRIBUTEnameWithValue{id}{#2}}\ifthenelse{\equal{#3}{}}{}{\ifthenelse{\equal{\NewLineNeeded}{1}}{\\}{}\def\NewLineNeeded{1}\strut\ATTRIBUTEnameWithValue{done}{#3}}}}

\pgfkeys{
	/NODEB/.cd,
	name/.code = \xdef\myNODEBNAME{\unexpanded{#1}},
	appendTitle/.code = \xdef\myNODEBSFSexpand{\unexpanded{#1}},
	id/.code = \xdef\myNODEBATTid{\unexpanded{#1}},
	done/.code = \xdef\myNODEBATTdone{\unexpanded{#1}},
}
\newcommand*\myNODEBNAME{}\newcommand*\myNODEBSFSexpand{}\newcommand*\myNODEBATTid{}\newcommand*\myNODEBATTdone{}

\newcommand{\NODEBX}[1]{\gdef\myNODEBNAME{}\gdef\myNODEBSFSexpand{}\gdef\myNODEBATTid{}\gdef\myNODEBATTdone{}\pgfqkeys{/NODEB}{#1}\NODEcontentWithAttributes{\strut\NODEnameWithType{\myNODEBNAME}{B}\myNODEBSFSexpand}{\def\NewLineNeeded{0}\ifthenelse{\equal{\myNODEBATTid}{}}{}{\ifthenelse{\equal{\NewLineNeeded}{1}}{\\}{}\def\NewLineNeeded{1}\strut\ATTRIBUTEnameWithValue{id}{\myNODEBATTid}}\ifthenelse{\equal{\myNODEBATTdone}{}}{}{\ifthenelse{\equal{\NewLineNeeded}{1}}{\\}{}\def\NewLineNeeded{1}\strut\ATTRIBUTEnameWithValue{done}{\myNODEBATTdone}}}}

\tikzset{
	styleEDGEeAB/.style={
		line width=1pt,
		minimum width=0cm,
		outer ysep=1pt,
		inner ysep=1pt,
		inner xsep=1pt,
		outer xsep=1pt,
		shape=rectangle split,
		rectangle split parts=1,
		align=left,
		rectangle split part fill={white!80!gray,white!60!gray,white},
		rectangle split every empty part={},
		rectangle split empty part height=0pt,
	}
}

\pgfkeys{
	/EDGEeAB/.cd,
	name/.code = ,
	appendTitle/.code = ,
}

\tikzset{
	styleEDGEeBA/.style={
		line width=1pt,
		minimum width=0cm,
		outer ysep=1pt,
		inner ysep=1pt,
		inner xsep=1pt,
		outer xsep=1pt,
		shape=rectangle split,
		rectangle split parts=1,
		align=left,
		rectangle split part fill={white!80!gray,white!60!gray,white},
		rectangle split every empty part={},
		rectangle split empty part height=0pt,
	}
}

\pgfkeys{
	/EDGEeBA/.cd,
	name/.code = ,
	appendTitle/.code = ,
}

\usepackage{longtable}

\def\SCALEFACTORfigures{.75}
\title{Probabilistic Metric Temporal Graph Logic\thanks{Funded by the Deutsche Forschungsgemeinschaft (DFG, German Research Foundation) - 241885098,
       148420506.}
}

\usepackage{marvosym} 

\titlerunning{Probabilistic Metric Temporal Graph Logic}
\author{
    Sven Schneider\,\orcidID{0000-0001-9828-618X}(\Letter)
    \and
    Maria Maximova\,\orcidID{0000-0001-9275-806X}
    \and
    Holger Giese\,\orcidID{0000-0002-4723-730X}
}
\authorrunning{
    S. Schneider \and M. Maximova \and H. Giese
}
\institute{
    University of Potsdam, Hasso Plattner Institute, Potsdam, Germany\\
    \texttt{\{sven.schneider,maria.maximova,holger.giese\}@hpi.de}
}
\renewcommand{\myref}[1]{\autoref{#1}}

\def\red#1{\begingroup\color{red}#1\endgroup}
\def\HIDEPROOF#1{}

\makeatletter
\RequirePackage[bookmarks,unicode,colorlinks=true]{hyperref}\def\@citecolor{blue}\def\@urlcolor{blue}\def\@linkcolor{blue}
\def\orcidID#1{\smash{\href{http://orcid.org/#1}{\protect\raisebox{-1.25pt}{\protect\includegraphics[draft=false]{orcid_color.eps}}}}}
\makeatother

\usepackage[nonumberlist,nogroupskip]{glossaries}
\makenoidxglossaries
\setacronymstyle{long-short}
\newacronym
    [longplural={Interval Probabilistic Timed Graph Transformation Systems}]
    {IPTGTS}{IPTGTS}{Interval Probabilistic Timed Graph Transformation System}
\newacronym
    [longplural={Probabilistic Timed Graph Transformation Systems}]
    {PTGTS}{PTGTS}{Probabilistic Timed Graph Transformation System}
\newacronym
    [longplural={Timed Graph Transformation Systems}]
    {TGTS}{TGTS}{Timed Graph Transformation System}
\newacronym
    [longplural={Probabilistic Graph Transformation Systems}]
    {PGTS}{PGTS}{Probabilistic Graph Transformation System}
\newacronym
    [longplural={Graph Transformation Systems}]
    {GTS}{GTS}{Graph Transformation System}
\newacronym
    [longplural={Timed Automata},
    \glsshortpluralkey={TA}]
    {TA}{TA}{Timed Automaton}
\newacronym
    [longplural={Probabilistic Timed Automata},
    \glsshortpluralkey={PTA}]
    {PTA}{PTA}{Probabilistic Timed Automaton}
\newacronym
    [longplural={Probabilistic Automata},
    \glsshortpluralkey={PA}]
    {PA}{PA}{Probabilistic Automaton}
\newacronym
    [longplural={Interval Probabilistic Timed Automata},
    \glsshortpluralkey={IPTA}]
    {IPTA}{IPTA}{Interval Probabilistic Timed Automaton}
\newacronym
    [longplural={Interval Probabilistic Automata},
    \glsshortpluralkey={IPA}]
    {IPA}{IPA}{Interval Probabilistic Automaton}
\newacronym
    {BMC}{BMC}{Bounded Model Checking}
\newacronym
    [longplural={Markov Decision Processes}]
    {MDP}{MDP}{Markov Decision Process}
\newacronym
    [longplural={Probabilistic Timed Systems}]
    {PTS}{PTS}{Probabilistic Timed System}
\newacronym
    [longplural={Interval Probabilistic Timed Systems}]
    {IPTS}{IPTS}{Interval Probabilistic Timed System}
\newacronym
    {PTCTL}{PTCTL}{Probabilistic Timed Computation Tree Logic}
\newacronym
    [longplural={Discrete Probability Distributions}]
    {DPD}{DPD}{Discrete Probability Distribution}
\newacronym
    [longplural={Discrete Interval Probability Distributions}]
    {DIPD}{DIPD}{Discrete Interval Probability Distribution}
\newacronym
    [longplural={Probability Mass Functions}]
    {PMF}{PMF}{Probability Mass Function}
\newacronym
    {DPO}{DPO}{Double Pushout}
\newacronym
    {MTGL}{MTGL}{Metric Temporal Graph Logic}
\newacronym
    {PMTGL}{PMTGL}{Probabilistic Metric Temporal Graph Logic}
\newacronym
    [longplural={Probabilistic Metric Temporal Graph Conditions}]
    {PMTGC}{PMTGC}{Probabilistic Metric Temporal Graph Condition}
\newacronym
    [longplural={Metric Temporal Graph Conditions}]
    {MTGC}{MTGC}{Metric Temporal Graph Condition}
\newacronym
    [longplural={Graph Conditions}]
    {GC}{GC}{Graph Condition}
\newacronym
    [longplural={Atomic Propositions}]
    {AP}{AP}{Atomic Proposition}
\newacronym
    [longplural={Graphs with History}]
    {GH}{GH}{Graph with History}
\newacronym
    [longplural={Attribute Conditions}]
    {AC}{AC}{Attribute Condition}
\newacronym
    {FOL}{FOL}{First-Order Logic}
\newacronym
    {MFOTL}{MFOTL}{Metric First-Order Temporal Logic}
\newacronym
    {GL}{GL}{Graph Logic}

\setkeys{glslink}{hyper=false}

\newcommand{\toolPRISM}{\textsc{Prism}\xspace}

\newcommand{\toolMONPOLY}{\textsc{Monpoly}\xspace}

\newcommand{\DPDsupport}[1]{\ensuremath{\mathsf{supp}(#1)}\xspace}

\newcommand{\PTGTRULEsend}{\ensuremath{\PTGTRULEvar[\textsf{send}]}\xspace}
\newcommand{\GTRULEsendSuccess}{\ensuremath{\GTRULEvar[\textsf{send},\textsf{done}]}\xspace}
\newcommand{\PTGTRULEreceive}{\ensuremath{\PTGTRULEvar[\textsf{receive}]}\xspace}
\newcommand{\GTRULEreceiveSuccess}{\ensuremath{\GTRULEvar[\textsf{receive},\textsf{done}]}\xspace}
\newcommand{\PTGTRULEtransmit}{\ensuremath{\PTGTRULEvar[\textsf{transmit}]}\xspace}
\newcommand{\GTRULEtransmitSuccess}{\ensuremath{\GTRULEvar[\textsf{transmit},\textsf{success}]}\xspace}
\newcommand{\GTRULEtransmitFailure}{\ensuremath{\GTRULEvar[\textsf{transmit},\textsf{failure}]}\xspace}
\newcommand{\PTGTinv}{\ensuremath{\GCvar[\textsf{inv}]}\xspace}
\newcommand{\PTGTap}{\ensuremath{\GCvar[\textsf{fin}]}\xspace}
\newcommand{\PMTGCexample}{\ensuremath{\PMTGCvar[\mathsf{max}]}\xspace}

\begin{document}

\maketitle
\begin{abstract}
Cyber\hyph{}physical systems often encompass complex concurrent behavior with timing constraints and probabilistic failures on demand.
The analysis whether such systems with probabilistic timed behavior adhere to a given specification is essential.
When the states of the system can be represented by graphs, the rule-based formalism of \glspl{PTGTS} can be used to suitably capture structure dynamics as well as probabilistic and timed behavior of the system.
The model checking support for \glspl{PTGTS} w.r.t.\ properties specified using \gls{PTCTL} has been already presented.
Moreover, for timed graph-based runtime monitoring, \gls{MTGL} has been developed for stating metric temporal properties on identified subgraphs and their structural changes over time.

In this paper, we
(a)~extend \gls{MTGL} to the \gls{PMTGL} by allowing for the specification of probabilistic properties,
(b)~adapt our \gls{MTGL} satisfaction checking approach to \glspl{PTGTS},
and
(c)~combine the approaches for \gls{PTCTL} model checking and \gls{MTGL} satisfaction checking to obtain a \gls{BMC} approach for \gls{PMTGL}.
In our evaluation, we apply an implementation of our \gls{BMC} approach in \AUTOGRAPH to a running example.

\keywords{cyber-physical systems, probabilistic timed systems, qualitative analysis, quantitative analysis, bounded model checking}
\end{abstract}
\mysection{
    title={Introduction},
    label={introduction}
}
\POINT{context}
Cyber\hyph{}physical systems often encompass complex concurrent behavior with timing constraints and probabilistic failures on demand~\cite{RailCab,2020_formal_testing_of_timed_graph_transformation_systems_using_metric_temporal_graph_logic}.
Such behavior can then be captured in terms of probabilistic timed state sequences (or spaces) where time may elapse between successive states and where each step in such a sequence has a designated probability.
The analysis whether such systems adhere to a given specification describing admissible or desired system behavior is essential in a model-driven development process.

Graph Transformation Systems (GTSs)~\cite{2006_fundamentals_of_algebraic_graph_transformation} can be used for the modeling of systems when each system state can be represented by a graph and when the changes of such states can be captured by rule-based graph transformation.
Moreover, timing constraints based on clocks, guards, invariants, and clock resets as in \glspl{PTA}~\cite{2004_symbolic_model_checking_for_probabilistic_timed_automata} have been combined with graph transformation in \glspl{TGTS}~\cite{2008_on_safe_service_oriented_real_time_coordination_for_autonomous_vehicles}
and probabilistic aspects have been added to graph transformation in \glspl{PGTS}~\cite{2012_probabilistic_graph_transformation_systems}.
Finally, the formalism of \glspl{PTGTS}~\cite{2018_probabilistic_timed_graph_transformation_systems} integrates both extensions and offers model checking support w.r.t.\ \gls{PTCTL}~\cite{2004_symbolic_model_checking_for_probabilistic_timed_automata,2011_prism_4_0__verification_of_probabilistic_real_time_systems} properties employing the \toolPRISM model checker~\cite{2011_prism_4_0__verification_of_probabilistic_real_time_systems}. 
The usage of \gls{PTCTL} allows for stating probabilistic real-time properties on the induced PTGT state space where each graph in the state space is labeled with a set of \glspl{AP} obtained by evaluating that graph w.r.t.\ e.g. some property specified using \gls{GL}~\cite{2009_correctness_of_high_level_transformation_systems_relative_to_nested_conditions,2020_formal_testing_of_timed_graph_transformation_systems_using_metric_temporal_graph_logic}.

However, structural changes over time in the state space cannot always be directly specified using \glspl{AP} that are \emph{locally} evaluated for each graph.
To express such structural changes over time, we introduced \gls{MTGL}~\cite{2019_metric_temporal_graph_logic_over_typed_attributed_graphs,2020_formal_testing_of_timed_graph_transformation_systems_using_metric_temporal_graph_logic} based on GL.
Using MTGL conditions, an unbounded number of subgraphs can be tracked over timed graph transformation steps in a considered state sequence once bindings have been established for them via graph matching.
Moreover, MTGL conditions allow to identify graphs where certain elements have just been added to (removed from) the current graph.
Similarly to \gls{MTGL}, for runtime monitoring, \gls{MFOTL}~\cite{2015_monitoring_metric_first_order_temporal_properties} (with limited support by the tool \toolMONPOLY) and the non-metric timed logic \textsc{Eagle}~\cite{2004_rule_based_runtime_verification,2015_rule_based_runtime_verification_revisited} (with full tool support) have been introduced operating, instead of graphs, on sets of relations and \textsc{Java} objects as state descriptions, respectively.

\POINT{problem}
Obviously, both logics \gls{PTCTL} and \gls{MTGL} have distinguishing key strengths but also lack bindings on the part of \gls{PTCTL} and an operator for expressing probabilistic requirements on the part of \gls{MTGL}.\footnote{\gls{PTCTL} model checkers such as \toolPRISM do not support the branching capabilities of \gls{PTCTL} as of now due to the complexity of the corresponding algorithms.}
Furthermore, specifications using both, PTCTL and MTGL conditions, are insufficient as they cannot capture phenomena based on probabilistic effects and the tracking of subgraphs at once.
Hence, a more complex combination of both logics is required.
Moreover, realistic systems often induce infinite or intractably large state spaces prohibiting the usage of standard model checking techniques.
Bounded Model Checking (BMC) has been proposed in~\cite{2016_bounded_model_checking_for_probabilistic_programs} for such cases implementing an on-the-fly analysis.
Similarly, reachability analysis w.r.t.\ a bounded number of steps or a bounded duration have been discussed in~\cite{2016_the_probabilistic_model_checking_landscape}.

\POINT{idea/contribution}
To combine the strengths of PTCTL and \gls{MTGL}, we introduce \gls{PMTGL} by enriching \gls{MTGL} with an operator for expressing probabilistic requirements as in PTCTL.
Moreover, we present a \gls{BMC} approach for \glspl{PTGTS} w.r.t.\ \gls{PMTGL} properties by combining the \gls{PTCTL} model checking approach for \glspl{PTGTS} from \cite{2018_probabilistic_timed_graph_transformation_systems} (which is based on a translation of \glspl{PTGTS} into \gls{PTA}) with the satisfaction checking approach for \gls{MTGL} from~\cite{2019_metric_temporal_graph_logic_over_typed_attributed_graphs,2020_formal_testing_of_timed_graph_transformation_systems_using_metric_temporal_graph_logic}.
In our approach, we just support \emph{bounded} model checking since the binding capabilities of \gls{PMTGL} conditions require non-local satisfaction checking taking possibly the entire history of a (finite) path into account as for MTGL conditions.
However, we obtain even \emph{full} model checking support for the case of finite loop-free state spaces and for the case where the given \gls{PMTGL} condition does not need to be evaluated beyond a maximal time bound.

\POINT{running example}
As a running example, we consider a system in which a sender decides to send messages at nondeterministically chosen time points, which have then to be transmitted to a receiver via a network of routers within a given time bound.
For this scenario, we employ MTGL allowing to identify messages that have just been sent, to track them over time, and to check whether their individual deadlines are met.

\POINT{tour}
This paper is structured as follows. 
In \autoref{section:probabilistic_timed_automata}, we recall the formalism of \gls{PTA}. 
In \autoref{section:probabilistic_timed_graph_transformation_systems}, we discuss further preliminaries including graph transformation, graph conditions, and the formalism of \glspl{PTGTS}.
In \autoref{section:probabilistic_metric_temporal_graph_logic}, we recall \gls{MTGL} and present the extension of \gls{MTGL} to \gls{PMTGL} in terms of syntax and semantics. 
In \autoref{section:model_checking_procedure}, we present our \gls{BMC} approach for \glspl{PTGTS} w.r.t.\ \gls{PMTGL} properties.
In \autoref{section:evaluation}, we evaluate our \gls{BMC} approach by applying its implementation in the tool \AUTOGRAPH to our running example. 
Finally, in \autoref{section:conclusion_and_future_work}, we close the paper with a conclusion and an outlook on future work.
\mysection{
    title={Probabilistic Timed Automata},
    label={probabilistic_timed_automata}
}
\begin{myfigureWITHBOX}{PTA $A$, one of its paths, and its symbolic state space.}{example_PTA_ZONE_PTS}
\mysubfigure{PTA $A$.}{example_PTA}{
    \def\SCALEFACTORfigures{0.8}
    \def\DEFDIST{1.2cm}
    \scalebox{\SCALEFACTORfigures}{\begin{tikzpicture}
    \node[draw,circle](l0)at(0,0){$\ell_0$};
    \draw($(l0.north)+(0ex,3ex)$)edge[->](l0);

    \node[fill=black,inner sep=1pt,outer sep=0pt,node distance=\DEFDIST,right=of l0](step3){};
    \node[draw,circle,node distance=\DEFDIST,right=of step3](l3){$\ell_3$};
    \draw(l0)edge[-]node[above]{$a;c_0\geq 3$}(step3);
    \draw(step3)edge[->]node[above]{$1;\emptyset$}(l3);

    \node[fill=black,inner sep=1pt,outer sep=0pt,node distance=\DEFDIST,left=of l0](step1){};
    \node[draw,circle,node distance=\DEFDIST,left=of step1,yshift=.5cm](l1){$\ell_1$};
    \node[draw,circle,node distance=\DEFDIST,left=of step1,yshift=-.5cm](l2){$\ell_2$};
    \draw(l0)edge[-]node[above]{$b;c_0\geq 1$}(step1);
    \draw(step1)edge[->,bend right=20]node[above,xshift=.5ex,yshift=.5ex]{$\num{0.5};\{c_0\}$}(l1);
    \draw(step1)edge[->,bend left=20]node[below,xshift=.5ex,yshift=-.5ex]{$\num{0.5};\emptyset$}(l2);

    \node[fill=black!10!white,node distance=.5ex,below=of l0](invAndAP){$c_0\leq 5;\emptyset$};
    \draw(invAndAP)--(l0);
    \node[fill=black!10!white,node distance=.5ex,left=of l1](invAndAP){$\ACtrue;\{\var{done}\}$};
    \draw(invAndAP)--(l1);
    \node[fill=black!10!white,node distance=.5ex,left=of l2](invAndAP){$\ACtrue;\{\var{error1}\}$};
    \draw(invAndAP)--(l2);
    \node[fill=black!10!white,node distance=.5ex,right=of l3](invAndAP){$\ACtrue;\{\var{error2}\}$};
    \draw(invAndAP)--(l3);
    \end{tikzpicture}
    }
}
\mysubfigure{Path of the PTA $A$ for some adversary.}{example_PTA_path}{
    \def\SCALEFACTORfigures{0.8}
    \scalebox{\SCALEFACTORfigures}{\begin{tikzpicture}
    \node[](s0)at(0,0){$(\ell_0,c_0=0)$};
    \node[right=of s0](s1){$(\ell_0,c_0=1.5)$};
    \node[right=of s1](s2){$(\ell_0,c_0=1.8)$};
    \node[right=of s2](s3){$(\ell_2,c_0=1.8)$};
    \node[right=of s3](s4){$(\ell_2,c_0=2.5)$};
    \draw(s0)edge[->]node[above]{$1;1.5$}(s1);
    \draw(s1)edge[->]node[above]{$1;0.3$}(s2);
    \draw(s2)edge[->]node[above]{$0.5;b$}(s3);
    \draw(s3)edge[->]node[above]{$1;0.7$}(s4);
    \end{tikzpicture}
    }
}
\mysubfigure{Symbolic state space induced by the PTA $A$.}{example_symbolic_state_space}{
    \def\SCALEFACTORfigures{0.8}
    \def\DEFDIST{1.2cm}
    \scalebox{\SCALEFACTORfigures}{\begin{tikzpicture}
    \node[draw,rectangle,rounded corners=1ex](l0)at(0,0){$(\ell_0,c_0\leq 5)$};
    \draw($(l0.north)+(0ex,3ex)$)edge[->](l0);

    \node[fill=black,inner sep=1pt,outer sep=0pt,node distance=\DEFDIST,right=of l0](step3){};
    \node[draw,rectangle,rounded corners=1ex,node distance=\DEFDIST,right=of step3](l3){$(\ell_3,c_0\geq 3)$};
    \draw(l0)edge[-]node[above]{$a;c_0\geq 3$}(step3);
    \draw(step3)edge[->]node[above]{$1;\emptyset$}(l3);

    \node[fill=black,inner sep=1pt,outer sep=0pt,node distance=\DEFDIST,left=of l0](step1){};
    \node[draw,rectangle,rounded corners=1ex,node distance=\DEFDIST,left=of step1,yshift=.5cm](l1){$(\ell_1,\ACtrue)$};
    \node[draw,rectangle,rounded corners=1ex,node distance=\DEFDIST,left=of step1,yshift=-.5cm](l2){$(\ell_2,c_0\geq 1)$};
    \draw(l0)edge[-]node[above]{$b;c_0\geq 1$}(step1);
    \draw(step1)edge[->,bend right=20]node[above,xshift=.5ex,yshift=.5ex]{$\num{0.5};\{c_0\}$}(l1);
    \draw(step1)edge[->,bend left=20]node[below,xshift=.5ex,yshift=-.5ex]{$\num{0.5};\emptyset$}(l2);

    \node[fill=black!10!white,node distance=.5ex,below=of l0](invAndAP){$\emptyset$};
    \draw(invAndAP)--(l0);
    \node[fill=black!10!white,node distance=.5ex,left=of l1](invAndAP){$\{\var{done}\}$};
    \draw(invAndAP)--(l1);
    \node[fill=black!10!white,node distance=.5ex,left=of l2](invAndAP){$\{\var{error1}\}$};
    \draw(invAndAP)--(l2);
    \node[fill=black!10!white,node distance=.5ex,right=of l3](invAndAP){$\{\var{error2}\}$};
    \draw(invAndAP)--(l3);
    \end{tikzpicture}
    }
}
\end{myfigureWITHBOX}
 
In this section, we introduce the syntax and semantics of \gls{PTA}~\cite{2004_symbolic_model_checking_for_probabilistic_timed_automata} and probabilistic timed reachability problems to be solved for \gls{PTA} using \toolPRISM~\cite{2011_prism_4_0__verification_of_probabilistic_real_time_systems}.

For a set of clock variables $X$, clock constraints $\CCvar\in\ClockConstraints{X}$ are finite conjunctions of clock comparisons of the form
$c_1\sim n$ and $c_1-c_2\sim n$ where $c_1,c_2\in X$, ${\sim}\in\{<,>,\leq,\geq\}$, and $n\in\NAT\cup\{\infty\}$.
A clock valuation $v\in\ClockValuations{X}$ of type $v:X\FUN\NONNEGATIVEREALS$ satisfies a clock constraint $\CCvar$, written \ClockSatisfaction{v}{\CCvar}, as expected.
The initial clock valuation \InitialClockValuation{X} maps all clocks to $0$.
For a clock valuation $v$ and a set of clocks $X'$, $\ClockReset{v}{X'}$ is the clock valuation mapping the clocks from $X'$ to $0$ and all other clocks according to $v$.
For a clock valuation $v$ and a duration $\delta\in\NONNEGATIVEREALS$, $\ClockIncrement{v}{\delta}$ is the clock valuation mapping each clock $x$ to $v(x)+\delta$.

For a countable set $A$, $\mu:A\FUN\INTERVALcc{0}{1}$ is a \gls{DPD} over $A$, written $\mu\in\DPD{A}$, if the probabilities assigned to elements add up to $1$, i.e., $\MultiSetSum{\mu(a)}{a\in A}=1$ using summation over multisets.
Moreover, the \emph{support} of $\mu$, written $\DPDsupport{\mu}$, contains all $a\in A$ for which the probability $\mu(a)$ is non-zero.

\glspl{PTA} combine the use of clocks to capture real-time phenomena and probabilism to approximate/describe the likelihood of outcomes of certain steps.
A \gls{PTA} (such as $A$ from \myref{figure:example_PTA}) consists of
    (a)~a set of locations with a distinguished initial location (such as $\ell_0$),
    (b)~a set of clocks (such as $c_0$) which are initially set to $0$,
    (c)~an assignment of a set of \glspl{AP} (such as $\{\var{done}\}$) to each location (for subsequent analysis of e.g.\ reachability properties),
    (d)~an assignment of constraints over clocks to each location as invariants such as ($c_0\leq 5$),
    and
    (e)~a set of probabilistic timed edges.
		Each probabilistic timed edge consists thereby of
        (i)~a single source location,
        (ii)~at least one target location,
				(iii)~an action (such as $a$ or $b$),
        (iv)~a clock constraint (such as $c_0\geq 3$) specifying as a guard when the edge is enabled based on the current values of the clocks,
        and
        (v)~a \gls{DPD} assigning a probability to each pair consisting of a set of clocks to be reset (such as $\{c_0\}$) and a target location to be reached.
\begin{mydefinition}{\gls{PTA}}{probabilistic_timed_automata}
A \emph{probabilistic timed automaton (\gls{PTA})} $\PTAvar$ is a tuple with the following components.
\begin{compactitem}
\item
    $\PTAGetlocations$ is a finite set of locations,
\item
    $\PTAGetinitialLocation$ is the unique initial location from $\IPTAGetlocations$,
\item
    $\PTAGetactions$ is a finite set of actions disjoint from $\NONNEGATIVEREALS$,
\item
    $\PTAGetclocks$ is a finite set of clocks,
\item
    $\PTAGetinvariants:\PTAGetlocations\FUN\ClockConstraints{\PTAGetclocks}$ maps each location to an invariant for that location such that the initial clock valuation satisfies the invariant of the initial location (i.e., $\ClockSatisfaction{\InitialClockValuation{\PTAGetclocks}}{\PTAGetinvariants(\PTAGetinitialLocation)}$),
\item
    $\PTAGetedges\subseteq\PTAGetlocations\times\PTAGetactions\times\ClockConstraints{\PTAGetclocks}\times\DPD{\POWERSET{\PTAGetclocks}\times\PTAGetlocations}$ is a finite set of \gls{PTA} edges of the form $(\ell_1,a,\CCvar,\mu)$ where $\ell_1$ is the source location, $a$ is an action, \CCvar is a guard, and $\mu$ is a \gls{DPD} mapping pairs $(\var{Res},\ell_2)$ of clocks to be reset and target locations to probabilities,
\item
    $\PTAGetatomicPropositions$ is a finite set of \glspl{AP},
    and
\item
    $\PTAGetlabelling:\PTAGetlocations\FUN\POWERSET{\PTAGetatomicPropositions}$ maps each location to a set of \glspl{AP}. 
\end{compactitem}
\end{mydefinition}
The semantics of a \gls{PTA} is given in terms of the induced \gls{PTS}.
The states of the induced \gls{PTS} are pairs of locations and clock valuations. 
The sequences of steps between such states define timed probabilistic paths.
Each successive step in a path (such as the one in \myref{figure:example_PTA_path}) is determined by an adversary which resolves the nondeterminism of the \gls{PTA} by selecting either a duration by which all clocks are advanced in a timed step or a \gls{PTA} edge that is used in a discrete step.
\begin{mydefinition}{\gls{PTS} Induced by \gls{PTA}}{PTAtoPTS}
Every \gls{PTA} $\PTAvar$ induces a unique \emph{probabilistic timed system (\gls{PTS})} $\PTAtoPTS{\PTAvar}=\PTSvar$ consisting of the following components.
\begin{compactitem}
\item
    $\PTSGetstates=\{(\ell,v)\in\IPTAGetlocations\times\ClockValuations{\IPTAGetclocks}\mid \ClockSatisfaction{v}{\IPTAGetinvariants(\ell)}\}$ contains as PTS states pairs of locations and clock valuations satisfying the location's invariant,
\item
    $\PTSGetinitialState=(\PTAGetinitialLocation,\InitialClockValuation{\PTAGetclocks})$ is the unique initial state from $\PTSGetstates$,
\item
    $\PTSGetactions=\PTAGetactions$ is the same set of actions,
\item
    $\PTSGetsteps\subseteq\PTSGetstates\times(\PTSGetactions\cup\NONNEGATIVEREALS)\times\DPD{\PTSGetstates}$ is the set of \gls{PTS} steps.\footnote{See \cite{2004_symbolic_model_checking_for_probabilistic_timed_automata} for a full definition of induced timed and discrete steps.}
    A \gls{PTS} step $((\ell,v),a,\mu)\in\PTSGetsteps$
    contains
        a source state $(\ell,v)$,
        an action from $\PTSGetactions$ for a discrete step or a duration from \NONNEGATIVEREALS for a timed step, and
        a DPD $\mu$ assigning a probability to each possible target state.
\item
    $\PTSGetatomicPropositions=\PTAGetatomicPropositions$ is the same set of \glspl{AP},
    and
\item
    $\PTSGetlabelling(\ell,v)=\PTAGetlabelling(\ell)$ labels states in $P$ according to the location labeling of $\IPTAvar$.
\end{compactitem}
\end{mydefinition}
For model checking \gls{PTA} \cite{2004_symbolic_model_checking_for_probabilistic_timed_automata}, \toolPRISM does not compute the induced \gls{PTS} according to \myref{definition:PTAtoPTS} but instead it computes a \emph{symbolic} state space (as in \myref{figure:example_symbolic_state_space}).
In this symbolic state space, states are given by pairs of locations and clock constraints (called \emph{zones}) where one state $(\ell,\CCvar)$ represents all pairs of states $(\ell,v)$ such that $\ClockSatisfaction{v}{\CCvar}$.
To allow for such a symbolic state space representation, the syntax of clock constraints has been carefully chosen.

In \myref{section:model_checking_procedure}, we will use \toolPRISM to solve the following analysis problems defined for induced PTSs.
\begin{mydefinition}{Min/Max Probabilistic Timed Reachability Problems}{analysis_problems}
Evaluate \PTCTLprob{\var{op}=?}{\PTCTLeventuallyUnbounded{\var{ap}}} for a PTS $\PTSvar$
with $\var{op}\in\{\OPminNAME,\OPmaxNAME\}$ and $\var{ap}\in\PTSGetatomicPropositions$
to obtain the infimal/supremal probability (depending on $\var{op}$) over all adversaries to reach some state  in $\PTSvar$ labeled with $\var{ap}$.
\end{mydefinition}
For example, for the PTS $\PTSvar=\PTAtoPTS{\PTAvar}$ induced by the PTA $\PTAvar$ from \myref{figure:example_PTA}, (a) $\PTCTLprob{\OPmaxNAME=?}{\PTCTLeventuallyUnbounded{\var{done}}}$ is evaluated to probability $0.5$ since a probability maximizing adversary would enable the discrete step using action $b$ at time point $1$ to reach $\ell_1$ with probability $0.5$ and
(b) $\PTCTLprob{\OPminNAME=?}{\PTCTLeventuallyUnbounded{\var{done}}}$ is evaluated to probability $0$ since a probability minimizing adversary would enable the discrete step using action $a$ at time point $3$ to reach $\ell_3$ from which then no location labeled with \var{done} can be reached.

\mysection{
    title={Probabilistic Timed Graph Transformation Systems},
    label={probabilistic_timed_graph_transformation_systems}
}
In this section, we briefly recall graphs, graph transformation, graph conditions, and the formalism of \glspl{PTGTS} in our notation.

Using the variation of symbolic graphs~\cite{2011_symbolic_graphs_for_attributed_graph_constraints} from
\cite{2020_formal_testing_of_timed_graph_transformation_systems_using_metric_temporal_graph_logic}, we consider typed attributed graphs (short graphs) (such as $G_0$ in \myref{figure:running_example_start_graph}), which are typed~over a type graph $\TypeGraph$ (such as $TG$ in \myref{figure:running_example_type_graph}).
In such graphs, attributes~are connected to \emph{local variables} and an \gls{AC} over a many~sorted first-order attribute logic is used to specify the values for these variables.
Morphisms $m:G_1\MOR G_2$ between graphs must ensure that the \gls{AC} of $G_2$ is more restrictive compared to the \gls{AC} of $G_1$ (w.r.t.\ the mapping of~variables~by~$m$). 
Hence, the \gls{AC} $\ACfalse$ (false) in $\TypeGraph$ means that $\TypeGraph$ does~not restrict~attribute values.
Lastly, we denote monomorphisms (short monos) by~$m:G_1\MORmono G_2$.

\glspl{GC}~\cite{2020_formal_testing_of_timed_graph_transformation_systems_using_metric_temporal_graph_logic,2009_correctness_of_high_level_transformation_systems_relative_to_nested_conditions} of \gls{GL} are used to state properties on graphs requiring the presence or absence of certain subgraphs in a host graph.\begin{mydefinition}{\glspl{GC}}{graph_conditions}
For a graph $H$, $\GCvar[H]\in\GCformulas{H}$ is a \emph{graph condition (\gls{GC})} over $H$ defined as follows:
\begin{nscenter}$
\GCvar[H]
    {\;::=\;}
    \GCtrue
    {\;\mid\;}
    \GCneg{\GCvar[H]}
    {\;\mid\;}
    \GCandBinaryArgs{\GCvar[H]}{\GCvar[H]}
    {\;\mid\;}
    \GCexists{f}{\GCvar[H']}
    {\;\mid\;}
    \GCrestrict{g}{\GCvar[H'']}
$\end{nscenter}
where $f:H\MORmono H'$ and $g:H''\MORmono H$ are monos and where additional operators such as $\GCfalse$, $\GCorNAME$, and $\GCforallNAME$ are derived as usual.
\end{mydefinition}
The satisfaction relation~\cite{2020_formal_testing_of_timed_graph_transformation_systems_using_metric_temporal_graph_logic,2009_correctness_of_high_level_transformation_systems_relative_to_nested_conditions} for GL defines when a mono satisfies a GC. 
Intuitively, for a graph $H$, the operator \GCexistsNAME (called \GCexistsNAMEintext) is used to extend a current match of $H$ to a supergraph $H'$ and
the operator \GCrestrictNAME (called \GCrestrictNAMEintext) is used to restrict a current match of $H$ to a subgraph $H''$.
\begin{mydefinition}{Satisfaction of \glspl{GC}}{satisfaction_of_graph_conditions}
A mono $m:H\MORmono G$ \emph{satisfies} a \gls{GC} $\GCvar$ over $H$, written $\GCsatINNER{m}{\GCvar}$\IFaia
\begin{compactitem}
\item
        $\GCvar=\GCtrue$.
\end{compactitem}\begin{compactitem}
\item
        $\GCvar=\GCneg{\GCvar'}$
    and
        $\GCNOTsatINNER{m}{\GCvar'}$.
\end{compactitem}\begin{compactitem}
\item
        $\GCvar=\GCandBinaryArgs{\GCvar[1]}{\GCvar[2]}$,
        $\GCsatINNER{m}{\GCvar[1]}$,
    and
        $\GCsatINNER{m}{\GCvar[2]}$.
\end{compactitem}\begin{compactitem}
\item
        $\GCvar=\GCexists{f:H\MORmono H'}{\GCvar'}$
    and 
        $\exists m':H'\MORmono G.\;m'\circ f=m\wedge\GCsatINNER{m'}{\GCvar'}$.
\end{compactitem}\begin{compactitem}
\item
        $\GCvar=\GCrestrict{g:H''\MORmono H}{\GCvar'}$
    and
        $\GCsatINNER{m\circ g}{\GCvar'}$.
\end{compactitem}
Moreover,
if
    $\GCvar\in\GCformulas{\emptygraph}$ is a \gls{GC} over the empty graph,
    $\MORinitialO{G}:\emptygraph\MORmono G$ is an initial morphism,
and
    $\GCsatINNER{\MORinitialO{G}}{\GCvar}$,
then the host graph $G$ satisfies $\GCvar$, written
    $\GCsatOUTER{G}{\GCvar}$.
\end{mydefinition}
\afterpage{\begin{myfigureWITHBOXHere}{Elements of the \gls{PTGTS} and PMTGC \PMTGCexample for the running example.}{running_example}
\mysubfigure[.35\linewidth]{Type graph \TypeGraph.}{running_example_type_graph}{
\scalebox{\SCALEFACTORfigures}{\begin{tikzpicture}[line width=1pt]
    \SetNodeWidthName{TypeGraph}
\node[mtgc_outer] (nodeL) at (0,0) {\begin{tikzpicture}
        \node[conditionGraphNode] (Graph1) at (0,0) {\VerticalExtend[c]{SharedVerticalTypeGraphStartGraph}{\begin{tikzpicture}[mtgc_inner]
			\node[styleNODERouter] (Router) at (0,0) {\NODETGRouterShort};
            \node[styleNODEReceiver,node distance=\TIKZGRAPHSedgelabelheight,below=of Router.south west,anchor=north west] (Receiver) {\NODETGReceiverShort};
            \node[styleNODEMessage,node distance=1ex,right=of Receiver.north east,anchor=north west] (Message) {\NODETGMessageShort};
            \node[styleNODESender,node distance=1ex,left=of Receiver.north west,anchor=north east] (Sender) {\NODETGSenderShort};
            \SetNodeWidthName{TypeGraphX1}
            \gettikzxy{(Sender.north west)}{\nodeAX}{\nodeAY}
            \gettikzxy{(Router.north west)}{\nodeBX}{\nodeBY}
            \draw[styleEdge] ($(Sender.north west)+(1ex,0ex)$)
                    -- (\nodeAX+1ex,\nodeBY-1ex)
                    --
                    node[above,styleEDGEsnd,outer sep=1ex] (LAB) {\EDGETGsndShort\unskip}
                    ($(Router.north west)+(0ex,-1ex)$);
            \SetNodeWidthName{TypeGraphX2}
            \gettikzxy{(Message.north east)}{\nodeAX}{\nodeAY}
            \gettikzxy{(Router.north east)}{\nodeBX}{\nodeBY}
            \draw[styleEdge] ($(Message.north east)+(-1ex,0ex)$)
                    -- (\nodeAX-1ex,\nodeBY-1ex)
                    --
                    node[above,styleEDGEat,outer sep=1ex] (LAB) {\EDGETGatShort\unskip}
                    ($(Router.north east)+(0ex,-1ex)$);
			\SetNodeWidthName{TypeGraph}
			\GraphEdgeDirectStraight{N}{L}{Router}{styleEDGErcv}{LAB}{\EDGETGrcvShort}{Receiver}
			\SetNodeWidthName{TypeGraphX4}
			\GraphEdgeLoop{N}{Router}{styleEDGEnext}{LAB}{\EDGETGnextShort}
			\SetNodeWidthName{TypeGraphX5}
			\def\GraphEdgeLoopSep{2ex}
			\GraphEdgeLoop{N}{Message}{styleEDGEdone}{LAB}{\EDGETGdoneShort}
			\def\GraphEdgeLoopSep{1ex}
			\SetNodeWidthName{TypeGraph}
            \node[constraintNode,node distance=1ex,below=of Receiver] (constraint) {$\ACfalse$};
            \end{tikzpicture}}};
        \end{tikzpicture}};
\begin{pgfonlayer}{background}
    \node (GRAPHL) [fill=black!5!white,fit=(nodeL)] {};
    \end{pgfonlayer}
\end{tikzpicture}
} }\hfill
\mysubfigure[.63\linewidth]{Initial graph $G_0$.}{running_example_start_graph}{
\scalebox{\SCALEFACTORfigures}{\begin{tikzpicture}[line width=1pt]
    \SetNodeWidthName{RunningExampleGraph}

    \node[mtgc_outer] (nodeL) at (0,0) {\begin{tikzpicture}
        \node[conditionGraphNode] (Graph1) at (0,0) {\eqboxwh{SharedVerticalTypeGraphStartGraph}{\begin{tikzpicture}[mtgc_inner]
			\node[styleNODESender] (NodeSender1) at (0,0) {\NODESenderX{name=S,num=1}};
			\node[styleNODERouter,node distance=\TIKZGRAPHSedgelabelheight,below=of NodeSender1.south west,anchor=north west] (NodeRouter1) {\NODERouterX{name=R_1}};
			\node[styleNODERouter,node distance=\TIKZGRAPHSedgelabelwidth+5ex,right=of NodeRouter1.north east,anchor=north west] (NodeRouter2) {\NODERouterX{name=R_2}};
			\node[styleNODERouter,node distance=\TIKZGRAPHSedgelabelwidth+5ex,right=of NodeRouter2.north east,anchor=north west] (NodeRouter3) {\NODERouterX{name=R_3}};
			\node[styleNODERouter,node distance=\TIKZGRAPHSedgelabelheight+0ex,below=of NodeRouter1.south west,anchor=north west] (NodeRouter4) {\NODERouterX{name=R_4}};
			\node[styleNODERouter,node distance=\TIKZGRAPHSedgelabelheight+0ex,below=of NodeRouter3.south west,anchor=north west] (NodeRouter5) {\NODERouterX{name=R_5}};
			\gettikzxy{(NodeSender1.north west)}{\nodeAX}{\nodeAY}
			\gettikzxy{(NodeRouter3.north west)}{\nodeBX}{\nodeBY}
			\node[styleNODEReceiver,anchor=north west] (NodeReceiver1) at (\nodeBX,\nodeAY) {\NODEReceiverX{name=R}};
			\node[styleNODEMessage,node distance=1ex,right=of NodeSender1.north east,anchor=north west] (NodeMessage1) {\NODEMessageX{name=M_1,id=1,clock=c_1}};
			\node[styleNODEMessage,node distance=1ex,left=of NodeReceiver1.north west,anchor=north east] (NodeMessage3) {\NODEMessageX{name=M_3,id=3,clock=c_3}};
			\node[styleNODEMessage,anchor=north] (NodeMessage2) at ($0.5*(NodeMessage1.north east)+0.5*(NodeMessage3.north west)$) {\NODEMessageX{name=M_2,id=2,clock=c_2}};
			\GraphEdgeDirect{N}{R}{NodeSender1}{styleEDGEsnd}{LAB}{\EDGEsndX{name=e_1}}{NodeRouter1}
			\GraphEdgeDirect{S}{R}{NodeRouter3}{styleEDGErcv}{LAB}{\EDGErcvX{name=e_2}}{NodeReceiver1}
			\GraphEdgeDirect{E}{B}{NodeRouter1}{styleEDGEnext}{LAB}{\EDGEnextX{name=e_3}}{NodeRouter2}
			\GraphEdgeDirect{E}{B}{NodeRouter2}{styleEDGEnext}{LAB}{\EDGEnextX{name=e_4}}{NodeRouter3}
			\GraphEdgeDirect{N}{R}{NodeRouter1}{styleEDGEnext}{LAB}{\EDGEnextX{name=e_5}}{NodeRouter4}
			\GraphEdgeDirect{E}{A}{NodeRouter4}{styleEDGEnext}{LAB}{\EDGEnextX{name=e_6}}{NodeRouter5}
			\GraphEdgeDirect{S}{R}{NodeRouter5}{styleEDGEnext}{LAB}{\EDGEnextX{name=e_7}}{NodeRouter3}
            \end{tikzpicture}}};
        \end{tikzpicture}};

    \begin{pgfonlayer}{background}
    \node (GRAPHL) [fill=black!5!white,fit=(nodeL)] {};
    \end{pgfonlayer}

    \end{tikzpicture}
} }
\mysubfigure[.48\linewidth]{PTGT invariant $\PTGTinv$.}{running_example_invariant}{
\scalebox{\SCALEFACTORfigures}{\begin{tikzpicture}[line width=1pt]
    \SetNodeWidthName{RunningExampleInvariant}

    \node[mtgc_outer] (node1) at (0,0) {\begin{tikzpicture}
        \node[conditionGraphNode] (Graph1) at (0,0) {\eqboxwh{RunningExampleInvariantAPheight}{\begin{tikzpicture}[mtgc_inner]
			\node[styleNODERouter] (NodeRouter1) at (0,0) {\NODERouterX{name=R_1}};
			\node[styleNODEMessage,node distance=\TIKZGRAPHSedgelabelwidth,left=of NodeRouter1.north west,anchor=north east] (NodeMessage1) {\NODEMessageX{name=M,clock=c}};
			\GraphEdgeDirect{E}{B}{NodeMessage1}{styleEDGEat}{LAB}{\EDGEatX{name=e_1}}{NodeRouter1}
			\node[constraintNode,node distance=1ex,below=of NodeRouter1.south,anchor=north] (constraint) {$c>5$};
            \end{tikzpicture}}};
		\OPENINGBRACKET{Graph1}
		\node[outer xsep=0ex,inner xsep=0pt,node distance=.8ex,left=of Graph1,anchor=east] (LEFT)
			{\large\strut$\GCnegNAME\GCexistsNAME$};
		\node[outer xsep=0ex,inner xsep=0pt,node distance=0ex,right=of Graph1,anchor=west] (RIGHT)
			{\large\strut$\COMMA\MTGCtrue$};
		\CLOSINGBRACKET{Graph1}{RIGHT}{}		
        \end{tikzpicture}};

    \begin{pgfonlayer}{background}
    \node (GRAPHL) [fill=black!5!white,fit=(node1)] {};
    \end{pgfonlayer}

    \end{tikzpicture}
} }
\mysubfigure[.48\linewidth]{PTGT AP $\PTGTap$.}{running_example_ap_all_done}{
\scalebox{\SCALEFACTORfigures}{\begin{tikzpicture}[line width=1pt]
    \SetNodeWidthName{RunningExampleAPalldone}

    \node[mtgc_outer] (node1) at (0,0) {\begin{tikzpicture}
        \node[conditionGraphNode] (Graph1) at (0,0) {\VerticalExtend[t]{RunningExampleInvariantAPheight}{\begin{tikzpicture}[mtgc_inner]
			\node[styleNODEMessage] (NodeMessage1) at (0,0) {\NODEMessageX{name=M}};
            \end{tikzpicture}}};
		\OPENINGBRACKET{Graph1}
		\node[outer xsep=0ex,inner xsep=0pt,node distance=.8ex,left=of Graph1,anchor=east] (LEFT)
			{\large\strut$\GCnegNAME\GCexistsNAME$};
		\node[outer xsep=0ex,inner xsep=0pt,node distance=0ex,right=of Graph1,anchor=west] (RIGHT)
			{\large\strut$\COMMA$};
        \end{tikzpicture}};

    \node[mtgc_outer,node distance=0ex,right=of node1] (node2) {\begin{tikzpicture}
        \node[conditionGraphNode] (Graph1) at (0,0) {\VerticalExtend[t]{RunningExampleInvariantAPheight}{\begin{tikzpicture}[mtgc_inner]
			\node[styleNODEMessage] (NodeMessage1) at (0,0) {\NODEMessageX{name=M}};
			\GraphEdgeLoop{E}{NodeMessage1}{styleEDGEdone}{LAB}{\EDGEdoneX{name=e_1}}
            \end{tikzpicture}}};
		\OPENINGBRACKET{Graph1}
		\node[outer xsep=0ex,inner xsep=0pt,node distance=.8ex,left=of Graph1,anchor=east] (LEFT)
			{\large\strut$\GCnegNAME\GCexistsNAME$};
		\node[outer xsep=0ex,inner xsep=0pt,node distance=0ex,right=of Graph1,anchor=west] (RIGHT)
			{\large\strut$\COMMA\MTGCtrue$};
		\CLOSINGBRACKET{Graph1}{RIGHT}{}
		\CLOSINGBRACKET[1ex]{Graph1}{RIGHT}{}
        \end{tikzpicture}};

    \begin{pgfonlayer}{background}
    \node (GRAPHL) [fill=black!5!white,fit=(node1) (node2)] {};
    \end{pgfonlayer}

    \end{tikzpicture}
} }
\mysubfigure[\linewidth]{PTGT rules \PTGTRULEsend, \PTGTRULEreceive, and \PTGTRULEtransmit.}{running_example_rules}{
\begin{tabular}{l}
\scalebox{\SCALEFACTORfigures}{\begin{tikzpicture}[line width=1pt]
    \SetNodeWidthName{RunningExampleRulePush}

    \node[mtgc_outer] (nodeRule1) at (0,0) {\begin{tikzpicture}
        \node[conditionGraphNode] (Graph1) at (0,0) {
            \begin{tikzpicture}[mtgc_inner]
			\node[styleNODESender] (NodeSender1) at (0,0) {\NODESenderX{name=S,num=n}};
			\node[styleNODERouter,node distance=\TIKZGRAPHSedgelabelwidth,right=of NodeSender1.north east,anchor=north west] (NodeRouter1) {\NODERouterX{name=R_1}};
			\node[styleNODEMessage,node distance=\TIKZGRAPHSedgelabelwidth,right=of NodeRouter1.north east,anchor=north west] (NodeMessage1) {\NODEMessageX{name=M,id=i,clock=c}};
			\GraphEdgeDirect{E}{B}{NodeSender1}{styleEDGEsnd}{LAB}{\EDGEsndX{name=e_1}}{NodeRouter1}
			\GraphEdgeDirect{W}{B}{NodeMessage1}{styleEDGEat}{LAB}{\EDGEatX{name=e_2,appendTitle=$\;\pmb{\oplus}$}}{NodeRouter1}
            \end{tikzpicture}};
        \end{tikzpicture}};

    \begin{pgfonlayer}{background}
    \node (GRAPHrule1) [fill=black!5!white,fit=(nodeRule1)] {};
    \node[node distance=1ex,left=of GRAPHrule1,rotate=90,anchor=south] (rule1LAB) {[done]};

    \node[constraintNode,fill=black!5!white,node distance=1ex,right=of GRAPHrule1] (reset) {\strut
        reset: $\{c\}$
        };
    \node[constraintNode,node distance=.1ex,above=of reset.north west,anchor=south west] (attributeEffect) {\strut
        attribute effect:
        $n'=n+1\ACandNAME i'=i$};
    \node[constraintNode,fill=black!5!white,node distance=.1ex,below=of reset.south west,anchor=north west] (probability) {\strut
        probability: $1$};

    \node[draw] (GRAPHruleall) [fit=(nodeRule1) (attributeEffect) (rule1LAB)] {};
    \node[draw,node distance=-0.4pt,above=of GRAPHruleall.north west,anchor=south west](IPTGTrulename){\strut\textsl{send}};

    \node[constraintNode,node distance=1ex,right=of IPTGTrulename.south east,anchor=south west] (constraint) {\strut attribute guard: $n=i$};
    \node[constraintNode,fill=black!5!white,node distance=1ex,right=of constraint.south east,anchor=south west] (clockguard) {\strut
        clock guard: $\ACtrue$
        };
    \node[constraintNode,fill=black!5!white,node distance=1ex,right=of clockguard.south east,anchor=south west] (priority) {\strut
        priority: $0$
        };

    \end{pgfonlayer}

    \end{tikzpicture}
} \\[1ex]
\scalebox{\SCALEFACTORfigures}{\begin{tikzpicture}[line width=1pt]
    \SetNodeWidthName{RunningExampleRulePush}

    \node[mtgc_outer] (nodeRule1) at (0,0) {\begin{tikzpicture}
        \node[conditionGraphNode] (Graph1) at (0,0) {
            \begin{tikzpicture}[mtgc_inner]
			\node[styleNODEReceiver] (NodeSender1) at (0,0) {\NODEReceiverX{name=R}};
			\node[styleNODERouter,node distance=\TIKZGRAPHSedgelabelwidth,right=of NodeSender1.north east,anchor=north west] (NodeRouter1) {\NODERouterX{name=R_1}};
			\node[styleNODEMessage,node distance=\TIKZGRAPHSedgelabelwidth,right=of NodeRouter1.north east,anchor=north west] (NodeMessage1) {\NODEMessageX{name=M}};
			\GraphEdgeDirect{W}{B}{NodeRouter1}{styleEDGEsnd}{LAB}{\EDGErcvX{name=e_1}}{NodeSender1}
			\GraphEdgeDirect{W}{B}{NodeMessage1}{styleEDGEat}{LAB}{\EDGEatX{name=e_2,appendTitle=$\;\pmb{\ominus}$}}{NodeRouter1}
			\GraphEdgeLoop{S}{NodeMessage1}{styleEDGEdone}{LAB}{\EDGEdoneX{name=e_3,appendTitle=$\;\pmb{\oplus}$}}
            \end{tikzpicture}};
        \end{tikzpicture}};

    \begin{pgfonlayer}{background}
    \node (GRAPHrule1) [fill=black!5!white,fit=(nodeRule1)] {};
    \node[node distance=1ex,left=of GRAPHrule1,rotate=90,anchor=south] (rule1LAB) {[done]};

    \node[constraintNode,fill=black!5!white,node distance=1ex,right=of GRAPHrule1] (reset) {\strut
        reset: $\emptyset$
        };
    \node[constraintNode,node distance=.1ex,above=of reset.north west,anchor=south west] (attributeEffect) {\strut
        attribute effect:
        $\ACtrue$};
    \node[constraintNode,fill=black!5!white,node distance=.1ex,below=of reset.south west,anchor=north west] (probability) {\strut
        probability: $1$};

    \node[draw] (GRAPHruleall) [fit=(nodeRule1) (attributeEffect) (rule1LAB)] {};
    \node[draw,node distance=-0.4pt,above=of GRAPHruleall.north west,anchor=south west](IPTGTrulename){\strut\textsl{receive}};

    \node[constraintNode,node distance=1ex,right=of IPTGTrulename.south east,anchor=south west] (constraint) {\strut attribute guard: $\ACtrue$};
    \node[constraintNode,fill=black!5!white,node distance=1ex,right=of constraint.south east,anchor=south west] (clockguard) {\strut
        clock guard: $\ACtrue$
        };
    \node[constraintNode,fill=black!5!white,node distance=1ex,right=of clockguard.south east,anchor=south west] (priority) {\strut
        priority: $1$
        };

    \end{pgfonlayer}

    \end{tikzpicture}
} \\[1ex]
\scalebox{\SCALEFACTORfigures}{\begin{tikzpicture}[line width=1pt]
    \SetNodeWidthName{RunningExampleRuleTransmit}

    \node[mtgc_outer] (nodeRule1) at (0,0) {\begin{tikzpicture}
        \node[conditionGraphNode] (Graph1) at (0,0) {
            \begin{tikzpicture}[mtgc_inner]
			\node[styleNODERouter] (NodeRouter1) at (0,0) {\NODERouterX{name=R_1}};
			\node[styleNODERouter,node distance=\TIKZGRAPHSedgelabelwidth,right=of NodeRouter1.north east,anchor=north west] (NodeRouter2) {\NODERouterX{name=R_2}};
			\node[styleNODEMessage,node distance=\TIKZGRAPHSedgelabelwidth,left=of NodeRouter1.north west,anchor=north east] (NodeMessage1) {\NODEMessageX{name=M,clock=c}};
			\GraphEdgeDirect{E}{B}{NodeMessage1}{styleEDGEat}{LAB}{\EDGEatX{name=e_1,appendTitle=$\;\pmb{\ominus}$}}{NodeRouter1}
			\GraphEdgeDirect{E}{B}{NodeRouter1}{styleEDGEnext}{LAB}{\EDGEnextX{name=e_2}}{NodeRouter2}
            \gettikzxy{(NodeMessage1.north)}{\nodeAX}{\nodeAY}
            \gettikzxy{(NodeRouter2.north)}{\nodeBX}{\nodeBY}
            \draw[styleEdge] ($(NodeMessage1.north)$)
                    --
                    ($(NodeMessage1.north)+(0ex,2ex)$)
                    -- node[above,styleEDGEat,outer sep=1ex] (LAB) {\EDGEatX{name=e_3,appendTitle=$\;\pmb{\oplus}$}}
                    ($(NodeRouter2.north)+(0ex,2ex)$)
                    --
                    ($(NodeRouter2.north)$);
            \end{tikzpicture}};
        \end{tikzpicture}};

    \node[mtgc_outer,node distance=2ex,below=of nodeRule1.south west,anchor=north west] (nodeRule2) {\adjustbox{valign=c}{\vphantom{\rule{1pt}{1.6cm}}}\adjustbox{valign=c}{\begin{tikzpicture}
        \node[conditionGraphNode] (Graph1) at (0,0) {
            \begin{tikzpicture}[mtgc_inner]
			\node[styleNODERouter] (NodeRouter1) at (0,0) {\NODERouterX{name=R_1}};
			\node[styleNODERouter,node distance=\TIKZGRAPHSedgelabelwidth,right=of NodeRouter1.north east,anchor=north west] (NodeRouter2) {\NODERouterX{name=R_2}};
			\node[styleNODEMessage,node distance=\TIKZGRAPHSedgelabelwidth,left=of NodeRouter1.north west,anchor=north east] (NodeMessage1) {\NODEMessageX{name=M,clock=c}};
			\GraphEdgeDirect{E}{B}{NodeMessage1}{styleEDGEat}{LAB}{\EDGEatX{name=e_1}}{NodeRouter1}
			\GraphEdgeDirect{E}{B}{NodeRouter1}{styleEDGEnext}{LAB}{\EDGEnextX{name=e_2}}{NodeRouter2}
            \end{tikzpicture}};
        \end{tikzpicture}}};

    \begin{pgfonlayer}{background}
    \node (GRAPHrule1) [fill=black!5!white,fit=(nodeRule1)] {};
    \node (GRAPHrule2) [fill=black!5!white,fit=(nodeRule2)] {};
    \node[node distance=1ex,left=of GRAPHrule1,rotate=90,anchor=south] (rule1LAB) {[success]};
    \node[node distance=1ex,left=of GRAPHrule2,rotate=90,anchor=south] (rule2LAB) {[failure]};

    \node[constraintNode,fill=black!5!white,node distance=1ex,right=of GRAPHrule1] (reset1) {\strut
        reset: $\{c\}$
        };
    \node[constraintNode,node distance=.1ex,above=of reset1.north west,anchor=south west] (attributeEffect1) {\strut
        attribute effect:
        $\ACtrue$};
    \node[constraintNode,fill=black!5!white,node distance=.1ex,below=of reset1.south west,anchor=north west] (probability1) {\strut
        probability: $0.8$};

    \node[constraintNode,fill=black!5!white,node distance=1ex,right=of GRAPHrule2] (reset2) {\strut
        reset: $\{c\}$
        };
    \node[constraintNode,node distance=.1ex,above=of reset2.north west,anchor=south west] (attributeEffect2) {\strut
        attribute effect:
        $\ACtrue$};
    \node[constraintNode,fill=black!5!white,node distance=.1ex,below=of reset2.south west,anchor=north west] (probability2) {\strut
        probability: $0.2$};

    \node[draw] (GRAPHruleall) [fit=(nodeRule1) (nodeRule2) (attributeEffect1) (attributeEffect2) (rule1LAB) (rule2LAB)] {};
    \gettikzxy{(GRAPHruleall.east)}{\NODEAX}{\NODEnoUse}
    \gettikzxy{(GRAPHruleall.west)}{\NODEBX}{\NODEnoUse}
    \gettikzxy{(nodeRule1.south)}{\NODEnoUse}{\NODECY}
    \gettikzxy{(nodeRule2.north)}{\NODEnoUse}{\NODEDY}
    \draw($(\NODEAX,0)+0.5*(0,\NODECY)+0.5*(0,\NODEDY)$)edge[dashed]($(\NODEBX,0)+0.5*(0,\NODECY)+0.5*(0,\NODEDY)$);
    \node[draw,node distance=-0.4pt,above=of GRAPHruleall.north west,anchor=south west](IPTGTrulename){\strut\textsl{transmit}};

    \node[constraintNode,node distance=1ex,right=of IPTGTrulename.south east,anchor=south west] (constraint) {\strut
                attribute guard:
                $\ACtrue$};
    \node[constraintNode,fill=black!5!white,node distance=1ex,right=of constraint.south east,anchor=south west] (clockguard) {\strut
        clock guard: $c\geq 2$
        };
    \node[constraintNode,fill=black!5!white,node distance=1ex,right=of clockguard.south east,anchor=south west] (priority) {\strut
        priority: $0$
        };

    \end{pgfonlayer}

    \end{tikzpicture}
}

 \end{tabular}
}
\mysubfigure{PMTGC \PMTGCexample where the additional MTGL operator \MTGCforallNewSimpleNAMEintext (written \MTGCforallNewSimpleNAME) is derived from
the operator \MTGCexistsNewSimpleNAMEintext by 
$\MTGCforallSimple{f}{\MTGCvar'}=\MTGCneg{\MTGCexistsNewSimple{f}{\MTGCneg{\MTGCvar'}}}$.}{running_example_pmtgc}{
\scalebox{\SCALEFACTORfigures}{\begin{tikzpicture}[line width=1pt]
    \SetNodeWidthName{RunningExamplePMTGC}

    \node[mtgc_outer] (node1) at (0,0) {\begin{tikzpicture}
        \node[conditionGraphNode] (Graph1) at (0,0) {\eqboxwh{RunningExamplePMTGCHeight}{\begin{tikzpicture}[mtgc_inner]
			\node[styleNODESender] (NodeSender1) at (0,0) {\NODESenderX{name=S}};
			\node[styleNODERouter,node distance=\TIKZGRAPHSedgelabelwidth,right=of NodeSender1.north east,anchor=north west] (NodeRouter1) {\NODERouterX{name=R_1}};
			\node[styleNODEMessage,node distance=\TIKZGRAPHSedgelabelwidth,right=of NodeRouter1.north east,anchor=north west] (NodeMessage1) {\NODEMessageX{name=M}};
			\GraphEdgeDirect{E}{B}{NodeSender1}{styleEDGEsnd}{LAB}{\EDGEsndX{name=e_1}}{NodeRouter1}
			\GraphEdgeDirect{W}{B}{NodeMessage1}{styleEDGEat}{LAB}{\EDGEatX{name=e_2}}{NodeRouter1}
            \end{tikzpicture}}};
		\OPENINGBRACKET{Graph1}
		\node[outer xsep=0ex,inner xsep=0pt,node distance=.8ex,left=of Graph1,anchor=east] (LEFT)
			{\large\strut$\MTGCforallNAME^{\DeltaKindNew}$};
		\node[outer xsep=0ex,inner xsep=0pt,node distance=0ex,right=of Graph1,anchor=west] (RIGHT)
			{\large\strut$\COMMA$};
        \end{tikzpicture}};

    \node[mtgc_outer,node distance=0ex,left=of node1] (node0) {\begin{tikzpicture}
        \node[conditionGraphNode] (Graph1) at (0,0) {\VerticalExtend[t]{RunningExamplePMTGCHeight}{}};
		\OPENINGBRACKET{Graph1}
		\node[outer xsep=0ex,inner xsep=0pt,node distance=.8ex,left=of Graph1,anchor=east] (LEFT)
			{\large\strut$\mathcal{P}_{\OPmaxNAME=?}$};
        \end{tikzpicture}};

    \node[mtgc_outer,node distance=0ex,below=of node1.south west,anchor=north west] (node2) {\begin{tikzpicture}
        \node[conditionGraphNode] (Graph1) at (0,0) {\VerticalExtend[t]{RunningExamplePMTGCHeight}{\begin{tikzpicture}[mtgc_inner]
			\node[styleNODEMessage] (NodeMessage1) at (0,0) {\NODEMessageX{name=M}};
            \end{tikzpicture}}};
		\OPENINGBRACKET{Graph1}
		\node[outer xsep=0ex,inner xsep=0pt,node distance=.8ex,left=of Graph1,anchor=east] (LEFT)
			{\large\strut$\MTGCrestrictNAME$};
		\node[outer xsep=0ex,inner xsep=0pt,node distance=0ex,right=of Graph1,anchor=west] (RIGHT)
			{\large\strut$\COMMA\MTGCtrue\MTGCuntilNAME_{\INTERVALcc{0}{5}}$};
        \end{tikzpicture}};

    \node[mtgc_outer,node distance=0ex,right=of node2] (node3) {\begin{tikzpicture}
        \node[conditionGraphNode] (Graph1) at (0,0) {\VerticalExtend[t]{RunningExamplePMTGCHeight}{\begin{tikzpicture}[mtgc_inner]
			\node[styleNODEMessage] (NodeMessage1) at (0,0) {\NODEMessageX{name=M}};
			\GraphEdgeLoop{E}{NodeMessage1}{styleEDGEdone}{LAB}{\EDGEdoneX{name=e_3}}
            \end{tikzpicture}}};
		\OPENINGBRACKET{Graph1}
		\node[outer xsep=0ex,inner xsep=0pt,node distance=.8ex,left=of Graph1,anchor=east] (LEFT)
			{\large\strut$\GCexistsNAME$};
		\node[outer xsep=0ex,inner xsep=0pt,node distance=0ex,right=of Graph1,anchor=west] (RIGHT)
			{\large\strut$\COMMA\MTGCtrue$};
		\CLOSINGBRACKET{Graph1}{RIGHT}{}
		\CLOSINGBRACKET[1ex]{Graph1}{RIGHT}{}
		\CLOSINGBRACKET[2ex]{Graph1}{RIGHT}{}
		\CLOSINGBRACKET[3ex]{Graph1}{RIGHT}{}
        \end{tikzpicture}};

    \begin{pgfonlayer}{background}
    \node (GRAPHL) [fill=black!5!white,fit=(node0) (node1) (node2) (node3)] {};
    \end{pgfonlayer}

    \end{tikzpicture}
} }
\end{myfigureWITHBOXHere}
\clearpage}
A Graph Transformation (GT) step is performed by applying a GT rule $\GTRULEvar=(\ell:K\MORmono L,r:K\MORmono R,\ACvar)$ for a match $m:L\MORmono G$ on the graph to be transformed (see~\cite{2020_formal_testing_of_timed_graph_transformation_systems_using_metric_temporal_graph_logic} for technical details).
A GT rule specifies that
    (a)~the graph elements in $L-\ell(K)$ are to be deleted and the graph elements in $R-r(K)$ are to be added using the monos $\ell$ and $r$, respectively, according to a Double Pushout (DPO) diagram
    and
    (b)~the values of variables of $R$ are derived from~those of $L$ using the \gls{AC} $\ACvar$ (e.g.\ $x'=x+2$) in which the variables from $L$ and $R$~are used in unprimed and primed form, respectively.
Nested application conditions given by \glspl{GC} are straightforwardly supported by our approach but, to improve readability, not used in the running example and omitted  subsequently.

\glspl{PTGTS} introduced in \cite{2018_probabilistic_timed_graph_transformation_systems} are a probabilistic real-time extension of \glspl{GTS}~\cite{2006_fundamentals_of_algebraic_graph_transformation}.
We have shown in \cite{2018_probabilistic_timed_graph_transformation_systems} that \glspl{PTGTS} can be translated into equivalent \glspl{PTA} and, hence, \glspl{PTGTS} can be understood as a high-level language for \gls{PTA}.

Similarly to \glspl{PTA}, a PTGT state is given by a pair $(G,v)$ of a graph and a clock valuation.
The initial state is given by a distinguished initial graph and a valuation mapping all clocks to $0$.
For our running example, the initial graph (given in \myref{figure:running_example_start_graph}) captures a sender, which is connected via a network of routers to a receiver, and three messages to be send.
The type graph of a PTGTS also identifies attributes representing clocks.\footnote{For a PTGT state $(G,v)$, the values of clocks of $G$ are stored in $v$ and not in $G$.} 
For our running example, the type graph $\TypeGraph$ is given in \myref{figure:running_example_type_graph} where each $\var{clock}$ attribute of a message represents such a clock. 
PTGT invariants are specified using \glspl{GC}.
Their evaluation for reachable graphs then results in clock constraints representing invariants as for PTA.
For our running example, the PTGT invariant $\PTGTinv$ from \myref{figure:running_example_invariant} prevents that time elapses once a message was at one router for $5$ time units.
PTGT \glspl{AP} are also specified using \glspl{GC} but a state $(G,v)$ is labeled by such an PTGT \gls{AP} if the evaluation of the \gls{GC} for $G$ results in a satisfiable clock constraint (i.e., the labeling of $(G,v)$ is independent from $v$).
For our running example, the \gls{AP} $\PTGTap$ from \myref{figure:running_example_ap_all_done} labels states where \emph{each} message has been successfully delivered to the receiver as indicated by the $\EDGETGdoneNAME$ loop.

PTGT rules of a \gls{PTGTS} then correspond to edges of a \gls{PTA} and contain~(a)~a left-hand side graph $L$,
    (b)~an \gls{AC} specifying as an \emph{attribute guard} non-clock attributes of $L$,
    (c)~an \gls{AC} specifying as a \emph{clock guard} clock attributes of $L$,
    (d)~a natural number describing a \emph{priority} where higher numbers denote higher priorities,
    and
    (e)~a nonempty set of tuples of the form $(\ell:K\MORmono L,r:K\MORmono R,\ACvar,C,p)$ where
        $(\ell,r,\ACvar)$ is an underlying GT rule,
        $C$ is a set of clocks contained in $R$ to be reset,
        and
        $p$ is a real-valued probability from $\INTERVALcc{0}{1}$ where the probabilities of all such tuples must add up to $1$.
See \myref{figure:running_example_rules} for the three PTGT rules \PTGTRULEsend, \PTGTRULEreceive, and \PTGTRULEtransmit from our running example where the first two PTGT rules have each a unique underlying GT rule \GTRULEsendSuccess and \GTRULEreceiveSuccess, respectively, and where the last PTGT rule has two underlying GT rules \GTRULEtransmitSuccess and \GTRULEtransmitFailure.
For each of these underlying GT rules, we depict the graphs $L$, $K$, and $R$ in a single graph where graph elements to be removed and to be added are annotated with $\ominus$ and $\oplus$, respectively.
Further information about the PTGT rule (i.e., the attribute guard, clock guard, and priority) and each of its underlying GT rules (i.e., the \emph{attribute effect} $\ACvar$, set of clocks to be reset called \emph{reset}, and \emph{probability}) is given in red (for \glspl{AC}) and gray boxes (for the rest).
The PTGT rule \PTGTRULEsend is used to push the next message into the network by connecting it to the router that is adjacent to the sender. Thereby, the attribute $\var{num}$ of the sender is used to push the messages in the order of their $\var{id}$ attributes.
The PTGT rule \PTGTRULEreceive has the higher priority $1$ and is used to pull a message from the router that is adjacent to the receiver by marking the message with a $\EDGETGdoneNAME$ loop.
Lastly, the PTGT rule \PTGTRULEtransmit is used to transmit a message from one router to the next one.
This transmission is successful with probability \num{0.8} and fails with probability \num{0.2}.
The clock guard of \PTGTRULEtransmit (together with the fact that the clock of the message is reset to $0$ whenever \PTGTRULEtransmit is applied or when the message was pushed into the network using \PTGTRULEsend) ensures that transmission attempts may happen not faster than every $2$ time units.

The semantics of a \gls{PTGTS} is given by its induced \gls{PTS} as in \cite{2018_probabilistic_timed_graph_transformation_systems} using here concrete PTGT states instead of their equivalence classes for brevity.
\begin{mydefinition}{\gls{PTS} Induced by \gls{PTGTS}}{PTGTStoPTS}
Every \gls{PTGTS} $\PTGTSvar$ induces a unique \emph{\gls{PTS}} $\PTGTStoPTS{\PTGTSvar}=\PTSvar$ consisting of the following components.
\begin{compactitem}
\item
    $\PTSGetstates$ contains as \gls{PTS} states pairs $(G,v)$ where $G$ is a graph and $v$ is a valuation of the clocks of $G$ satisfying the PTGT invariants of $\PTGTSvar$,
\item
    $\PTSGetinitialState$
		is the unique initial state from $\PTSGetstates$ consisting of the initial graph of \PTGTSvar and the initial clock valuation of its clocks,
\item
    $\PTSGetactions$ contains tuples of the form $(\PTGTRULEvar,m,\var{sp})$ consisting of the used PTGT rule $\PTGTRULEvar$, the used match $m$, and a mapping $\var{sp}$ of each GT rule $\GTRULEvar$ in $\PTGTRULEGetrules[\PTGTRULEvar]$ to the GT span $(k_1:D\MORmono G,k_2:D\MORmono H)$ constructed for a GT step from $G$ to $H$ using $\GTRULEvar$.
\item
    $\PTSGetsteps\subseteq\PTSGetstates\times(\PTSGetactions\cup\NONNEGATIVEREALS)\times\DPD{\PTSGetstates}$ is the set of \gls{PTS} steps.\footnote{See \cite{2018_probabilistic_timed_graph_transformation_systems} for a full definition of induced timed and discrete steps.}
    A PTS step $((G,v),a,\mu)\in\PTSGetsteps$
    contains
        a source state $(G,v)$,
        an action from $\PTSGetactions$ for a discrete step or a duration from \NONNEGATIVEREALS for a timed step, and
        a DPD $\mu$ assigning a probability to each possible target state.
\item
    $\PTSGetatomicPropositions=\PTGTSGetatomicPropositions$ is the same set of PTGT APs,
    and
\item
    $\PTSGetlabelling(G,v)=\{\GCvar\in\PTGTSGetatomicPropositions \mid \GCsatOUTER{G}{\GCvar}\}$ labels states in $\PTSvar$ with PTGT \glspl{AP} based only on the satisfaction of \glspl{GC} for graphs.
\end{compactitem}
\end{mydefinition}

\mysection{
    title={Probabilistic Metric Temporal Graph Logic},
    label={probabilistic_metric_temporal_graph_logic}
}
Before introducing \gls{PMTGL}, we recall \gls{MTGL}~\cite{2019_metric_temporal_graph_logic_over_typed_attributed_graphs,2020_formal_testing_of_timed_graph_transformation_systems_using_metric_temporal_graph_logic} and adapt it to \glspl{PTGTS}.
To simplify our presentation, we focus on a restricted set of \gls{MTGL} operators and conjecture that the presented adaptations of \gls{MTGL} are compatible with full \gls{MTGL} from \cite{2020_formal_testing_of_timed_graph_transformation_systems_using_metric_temporal_graph_logic} as well as with the orthogonal \gls{MTGL} developments in \cite{2020_optimistic_and_pessimistic_on_the_fly_analysis_for_metric_temporal_graph_logic}.

The \glspl{MTGC} of \gls{MTGL} are specified using (a) the \gls{GC} operators to express properties on a single graph in a path and (b) metric temporal operators to navigate through the path.
For the latter, the operator \MTGCexistsNewSimpleNAME (called \MTGCexistsNewSimpleNAMEintext) is used to extend a current match of a graph $H$ to a supergraph $H'$ in the future such that some additionally matched graph element could not have been matched earlier.
Moreover, the operator \MTGCuntilSimpleNAME (called \MTGCuntilSimpleNAMEintext) is used to check whether an \gls{MTGC} $\MTGCvar[2]$ is eventually satisfied in the future within a given time interval while another \gls{MTGC} $\MTGCvar[1]$ is satisfied until then.\begin{mydefinition}{\glspl{MTGC}}{mtgcs}
For a graph $H$, $\MTGCvar[H]\in\MTGCformulas{H}$ is a \emph{metric temporal graph condition (\gls{MTGC})} over $H$ defined as follows:
\begin{nscenter}
\begin{math}
\MTGCvar[H]
    {\;::=\;}
    \MTGCtrue
    {\;\mid\;}
    \MTGCneg{\MTGCvar[H]}
    {\;\mid\;}
    \MTGCandBinaryArgs{\MTGCvar[H]}{\MTGCvar[H]}
    {\;\mid\;}
    \MTGCexistsSimple{f}{\MTGCvar[H']}
    {\;\mid\;}
    \MTGCrestrictSimple{g}{\MTGCvar[H'']}
    {\;\mid\;}
    \MTGCexistsNewSimple{f}{\MTGCvar[H']}
    {\;\mid\;}
    \MTGCuntilSimple{I}{\MTGCvar[H]}{\MTGCvar[H]}
\end{math}
\end{nscenter}
where $f:H\MORmono H'$ and $g:H''\MORmono H$ are monos and where $I$ is an interval over $\NONNEGATIVEREALS$.
\end{mydefinition}
For our running example, consider the \gls{MTGC} given in \myref{figure:running_example_pmtgc} inside the operator $\PTCTLmax{\cdot}$.
Intuitively, this \gls{MTGC} states that (\MTGCforallNewSimpleNAMEintext) whenever a message has just been sent from the sender to the first router, (\MTGCrestrictSimpleNAMEintext) when only tracking this message (since at least the edge $e_2$ can be assumed to be removed in between), (\MTGCuntilSimpleNAMEintext) eventually within $5$ time units, (\MTGCexistsSimpleNAMEintext) this message is delivered to the receiver as indicated by the $\EDGETGdoneNAME$ loop.

In~\cite{2019_metric_temporal_graph_logic_over_typed_attributed_graphs,2020_formal_testing_of_timed_graph_transformation_systems_using_metric_temporal_graph_logic}, MTGL was defined for timed graph sequences in which only discrete steps are allowed each having a duration $\delta>0$.
We now adapt MTGL to \glspl{PTGTS} in which discrete steps and timed steps are interleaved and where zero time may elapse between two discrete steps.  

To be able to track subgraphs in a \gls{PTS} path $\pi$ over time using matches, we first
identify the graph $\pi(\tau)$ in $\pi$ at a position $\tau=(t,s)\in\NONNEGATIVEREALS\times\NAT$ where $t$ is a total time point  and $s$ is a step index.\footnote{To compare positions, we define $(t,s)<(t',s')$ if either $t<t'$ or $t=t'$ and $s<s'$.}
\begin{mydefinition}{Graph at Position}{state_at_position}
A graph $G$ is at position $\tau=(t,s)$ in a path $\pi$~of PTS \PTSvar , written $\pi(\tau)=G$, if $\mathsf{pos}(\pi,t,s,i)=G$ for some index $i$ is defined~as~follows.
\begin{compactitem}
\item
    If
        $\pi_0=((G,v),a,\mu,(G',v'))$,
    then
        $\mathsf{pos}(\pi,0,0,0)=G$.
\item
    If
        $\pi_i=((G,v),a,\mu,(G',v'))$,
        $\mathsf{pos}(\pi,t,s,i)=G$,
    and
        $a\in\NONNEGATIVEREALSWITHOUTZERO$,
    then\\
        $\mathsf{pos}(\pi,t+\delta,s,i)=G$ for each $\delta\in\INTERVALco{0}{a}$
    and
        $\mathsf{pos}(\pi,t+a,0,i+1)=G'$.
\item
    If
        $\pi_i=((G,v),a,\mu,(G',v'))$,
        $\mathsf{pos}(\pi,t,s,i)=G$,
    and
        $a\not\in\NONNEGATIVEREALSWITHOUTZERO$,
    then\\
        $\mathsf{pos}(\pi,t,s+1,i+1)=G'$.
\end{compactitem}
\end{mydefinition}
A match $m:H\MORmono\pi(\tau)$ into the graph at position $\tau$ can be propagated forwards/backwards over the PTS steps in a path to the graph $\pi(\tau')$.
Such a propagated match $m':H\MORmono\pi(\tau')$, written $m'\in\PIpropagationSimple{\pi}{m}{\tau}{\tau'}$, can be obtained uniquely if \emph{all} matched graph elements $m(H)$ are preserved by the considered PTS steps, which is trivially the case for timed steps.
When some graph element is not preserved, $\PIpropagationSimple{\pi}{m}{\tau}{\tau'}$ is empty.

We now present the semantics of \gls{MTGL} by providing a satisfaction relation, which is defined as for \gls{GL} for the operators inherited from \gls{GL} and as explained above for the operators \MTGCexistsNewSimpleNAMEintext and \MTGCuntilSimpleNAMEintext.
\begin{mydefinition}{Satisfaction of MTGCs}{satisfaction_for_mtgc_for_a_tgs}
An \gls{MTGC} $\MTGCvar\in\MTGCformulas{H}$ over a graph $H$ is \emph{satisfied} by
    a path $\pi$ of the \gls{PTS} \PTSvar,
    a position $\tau\in\NONNEGATIVEREALS\times\NAT$, and
    a mono $m:H\MORmono \pi(\tau)$,
written $\MTGCsatTGSINNER{\pi}{\tau}{m}{\psi}$\IFaia
\begin{compactitem}
\item
        $\MTGCvar=\MTGCtrue$.
\item
        $\MTGCvar=\MTGCneg{\MTGCvar'}$
    and
        $\MTGCNOTsatINNER{\pi}{\tau}{m}{\MTGCvar'}$.
\item
        $\MTGCvar=\MTGCandBinaryArgs{\MTGCvar[1]}{\MTGCvar[2]}$,
        $\MTGCsatTGSINNER{\pi}{\tau}{m}{\MTGCvar[1]}$,
    and
        $\MTGCsatTGSINNER{\pi}{\tau}{m}{\MTGCvar[2]}$.
\item
        $\MTGCvar=\MTGCexistsSimple{f:H\MORmono H'}{\MTGCvar'}$
    and 
        $\exists m':H'\MORmono\pi(\tau).\;m'\circ f=m \wedge\MTGCsatTGSINNER{\pi}{\tau}{m'}{\MTGCvar}$.
\item
        $\MTGCvar=\MTGCrestrictSimple{g:H''\MORmono H}{\MTGCvar'}$
    and
        $\MTGCsatTGSINNER{\pi}{\tau}{m\circ g}{\MTGCvar'}$.
\item
        $\MTGCvar=\MTGCexistsNewSimple{f:H\MORmono H'}{\MTGCvar'}$
    and
        there are
            $\tau'\geq\tau$,
            $m'\in\PIpropagationSimple{\pi}{m}{\tau}{\tau'}$, and
            $m'':H'\MORmono\pi(\tau')$ s.t.
                    $m''\circ f=m'$,
                    $\MTGCsatTGSINNER{\pi}{\tau'}{m''}{\MTGCvar}$,
                    and
                    for each $\tau''<\tau'$ it holds that $\PIpropagationSimple{\pi}{m''}{\tau'}{\tau''}=\emptyset$.
\item
        $\MTGCvar=\MTGCuntilSimple{I}{\MTGCvar[1]}{\MTGCvar[2]}$
    and
        there is $\tau'\in I\times\NAT$ s.t.
            \begin{compactitem}
            \item
                there is $m'\in\PIpropagationSimple{\pi}{m}{\tau}{\tau'}$ s.t. $\MTGCsatTGSINNER{\pi}{\tau'}{m'}{\MTGCvar[2]}$ and
            \item
                for every $\tau\leq \tau''<\tau'$
                there is $m''\in\PIpropagationSimple{\pi}{m}{\tau}{\tau''}$ s.t.
                $\MTGCsatTGSINNER{\pi}{\tau''}{m''}{\MTGCvar[1]}$.
            \end{compactitem}
\end{compactitem}
Moreover, if $\MTGCvar\in\MTGCformulas{\emptygraph}$, $\tau=(0,0)$, and $\MTGCsatTGSINNER{\pi}{\tau}{\MORinitialO{\pi(\tau)}}{\MTGCvar}$, then $\MTGCsatTGSOUTER{\pi}{\MTGCvar}$.
\end{mydefinition}
We now introduce the \glspl{PMTGC} of PMTGL, which are defined based on MTGCs.
\begin{mydefinition}{\glspl{PMTGC}}{pmtgc}
Each \emph{probabilistic metric temporal graph condition} \emph{(PMTGC)}~is of the form
$\PMTGCvar=\PTCTLprob{\sim c}{\MTGCvar}$ where ${\sim}\in\{\leq,<,>,\geq\}$, 
$c\in\INTERVALcc{0}{1}$ is a probability, and $\MTGCvar\in\MTGCformulas{\emptygraph}$ is an \gls{MTGC} over the empty graph.
Moreover, we also call expressions of the form $\PTCTLmin{\MTGCvar}$ and $\PTCTLmax{\MTGCvar}$ \glspl{PMTGC}.
\end{mydefinition}
The satisfaction relation for PMTGL defines when a \gls{PTS} satisfies a \gls{PMTGC}.
\begin{mydefinition}{Satisfaction of \glspl{PMTGC}}{satisfaction_of_pmtgc}
A \gls{PTS} $\PTSvar$ \emph{satisfies} the \gls{PMTGC} $\PMTGCvar=\PTCTLprob{\sim c}{\MTGCvar}$, written $\GCsatINNER{\PTSvar}{\PMTGCvar}$, if, for any adversary $\var{Adv}$, the probability over all paths of $\var{Adv}$ that satisfy $\MTGCvar$ is $\sim c$.
Moreover, $\PTCTLmin{\MTGCvar}$ and $\PTCTLmax{\MTGCvar}$ denote the infimal and supremal expected probabilities over all adversaries to satisfy $\MTGCvar$ (cf. \myref{definition:analysis_problems}).
\end{mydefinition}
For our running example, the evaluation of the \gls{PMTGC} \PMTGCexample from \myref{figure:running_example_pmtgc} for the PTS induced by the \gls{PTGTS} from \myref{figure:running_example} results in the probability of $0.8^6=0.262144$ using a probability maximizing adversary $\var{Adv}$ as follows.
Whenever the first graph of the \gls{PMTGC} can be matched, this is the result of an application of the PTGT rule $\PTGTRULEsend$.
The adversary $\var{Adv}$ ensures then that each message is transmitted as fast as possible to the destination router $R_3$ by
(a)~letting time pass only when this is unavoidable to satisfy some guard and
(b)~never allowing to match the router $R_4$ by the PTGT rule $\PTGTRULEtransmit$ as this leads to a transmission with 3 hops.
For each message, the only transmission requiring at most $5$ time units transmits the message via the router $R_2$ to router $R_3$ using $2$ hops in $2+2$ time units.
The urgently (i.e., without prior delay) applied PTGT rule $\PTGTRULEreceive$ then attaches a $\EDGETGdoneNAME$ loop to the message as required by \PMTGCexample.
Since the transmissions of the messages do not affect each other and messages are successfully transmitted only when both transmission attempts succeeded, the maximal probability to satisfy the inner \gls{MTGC} is~$(0.8\times 0.8)^3$.

\mysection{
    title={Bounded Model Checking Approach},
    label={model_checking_procedure}
}
\def\algoBOUNDinp{\textsl{Time Bound $T$}\xspace}
\def\algoPTGTSinp{\textsl{PTGTS~$\PTGTSvar$}\xspace}
\def\algoPTGTSadj{\textsl{PTGTS~$\PTGTSvar'$}\xspace}
\def\algoPMTGCinp{\textsl{PMTGC~$\PMTGCvar$}\xspace}
\def\algoStateSpacePTA{\textsl{PTA~$\PTAvar$}\xspace}
\def\algoGC{\textsl{GC~$\GCvar$}\xspace}
\def\algoGHmap{\textsl{GH-Map \algoMgh}\xspace}
\def\algoACmap{\textsl{AC-Map \algoMac}\xspace}
\def\algoZonemap{\textsl{Zone-Map \algoMzone}\xspace}
\def\algoAPmap{\textsl{AP-Map \algoMap}\xspace}
\def\algoProbability{\textsl{Probability Interval $I$}\xspace}
\def\algoMgh{\ensuremath{M_{\var{GH}}}\xspace}
\def\algoMac{\ensuremath{M_{\var{AC}}}\xspace}
\def\algoMzone{\ensuremath{M_{\var{Zone}}}\xspace}
\def\algoMap{\ensuremath{M_{\var{AP}}}\xspace}
\newcommand{\OverviewTitle}[2]{\tikz\node[outer sep=0pt,inner sep=-1pt,fill=white]at(0,0){\eqboxwh{Overview#1}{\strut\mycentering{#2}}};}
\newcommand{\EvenLine}[2]{\tikz\node[outer sep=0pt,inner sep=0pt,fill=blue!5!white]at(0,0){\eqboxwh{Overview#1}{\strut\mycentering{#2}}};}
\newcommand{\OddLine}[2]{\tikz\node[outer sep=0pt,inner sep=0pt,fill=orange!5!white]at(0,0){\eqboxwh{Overview#1}{\strut\mycentering{#2}}};}
\begin{mytableWITHBOX}{Overview of the steps of our \gls{BMC} approach.}{overview_of_approach}
\centering
\def\arraystretch{0.2}
\scalebox{.95}{
\begin{tabular}{@{}c|cccc|l@{}}
\toprule
\OverviewTitle{1}{Step}&\multicolumn{4}{c}{Inputs}&\multicolumn{1}{|c}{Outputs}\\
\midrule
\EvenLine{1}{1}
    &\EvenLine{2a}{\algoPTGTSinp}
    &\EvenLine{2b}{\algoBOUNDinp}
    &
    &
    &\EvenLine{3}{\algoPTGTSadj}\\
\OddLine{1}{2}
    &\OddLine{2a}{\algoPTGTSadj}
    &
    &
    &
    &\OddLine{3}{\algoStateSpacePTA}\\
\EvenLine{1}{3}
    &\EvenLine{2a}{\algoStateSpacePTA}
    &
    &
    &
    &\EvenLine{3}{\algoGHmap}\\
\OddLine{1}{4}
    &\OddLine{2a}{\algoPMTGCinp}
    &
    &
    &
    &\OddLine{3}{\algoGC}\\
\EvenLine{1}{5}
    &\EvenLine{2a}{\algoGC}
    &\EvenLine{2b}{\algoGHmap}
    &
    &
    &\EvenLine{3}{\algoACmap}\\
\OddLine{1}{6}
    &\OddLine{2a}{\algoStateSpacePTA}
    &
    &
    &
    &\OddLine{3}{\algoZonemap}\\
\EvenLine{1}{7}
    &\EvenLine{2a}{\algoPMTGCinp}
    &\EvenLine{2b}{\algoGHmap}
    &\EvenLine{2c}{\algoACmap}
    &\EvenLine{2d}{\algoZonemap}
    &\EvenLine{3}{\algoAPmap}\\
\OddLine{1}{8}
    &\OddLine{2a}{\algoStateSpacePTA}
    &\OddLine{2b}{\algoAPmap}
    &
    &
    &\OddLine{3}{\algoProbability}\\
\bottomrule
\end{tabular}}
\end{mytableWITHBOX}
We now present our approach for reducing the
    \gls{BMC} problem for
        a fixed \gls{PTGTS} $\PTGTSvar$,
        a fixed \gls{PMTGC} $\PMTGCvar=\PTCTLprob{\sim c}{\MTGCvar}$,
        and
        an optional time bound~$T\in\NONNEGATIVEREALS\cup\{\infty\}$
    to a
    model checking problem for a PTA and an analysis problem from \myref{definition:analysis_problems}.
Using this approach, we can analyze whether $\PTGTSvar$ satisfies~$\PMTGCvar$ when restricting the discrete behavior of $\PTGTSvar$ to the time interval $\INTERVALco{0}{T}$.
In fact, we only consider PMTGCs of the form \PTCTLmin{\MTGCvar} or \PTCTLmax{\MTGCvar} for computing expected probabilities since they are sufficient to analyze the PMTGC $\PTCTLprob{\sim c}{\MTGCvar}$.\footnote{For example, $\PTCTLmin{\MTGCvar}=c$ implies satisfaction of $\PTCTLprob{\geq c'}{\MTGCvar}$ for any $c'\geq c$.}
See \myref{table:overview_of_approach} for an overview of the subsequently discussed steps of our approach.

\Paragraph{1}{Encoding the Time Bound into the PTGTS}
For the given PTGTS $\PTGTSvar$ and time bound $T$,
we construct an adapted PTGTS $\PTGTSvar'$ into which the time bound $T$ is encoded (for $T=\infty$, we use $\PTGTSvar'=\PTGTSvar$).
In $\PTGTSvar'$, we ensure that all discrete PTGT steps and all PTGT invariants are disabled when time bound $T$ is reached.
For this purpose, we
    (a) add an additional node $b$ of a fresh node type $\var{Bound}$ with a clock $x$ to the initial graph of $S$ and to the graphs $L$, $K$, and $R$ of each underlying GT rule $\GTRULEvar=(\ell:K\MORmono L,r:K\MORmono R,\ACvar)$ of each PTGT rule $\PTGTRULEvar$ of $\PTGTSvar$,
    (b) add a PTGT rule with a priority higher than all other used priorities deleting the node $b$ urgently with a guard $x=T$,
    and
    (c) extend each PTGT invariant $\GCvar$ to $\GCvar\vee\GCneg{\GCexists{b{:}\var{Bound}}{\GCtrue}}$ disabling it for states where the $b$ node has been removed.
For the resulting PTGTS $\PTGTSvar'$, we then solve the model checking problem for the given PMTGC $\PMTGCvar$.

\Paragraph{2}{Construction of an Equivalent PTA}
For the PTGTS $\PTGTSvar'$ from step~1, we now construct an equivalent \gls{PTA} $\PTAvar$ using the operation $\PTGTStoPTANAME$, which is based on a similar operation from~\cite{2018_probabilistic_timed_graph_transformation_systems}.

As a first step, we obtain the underlying \gls{GTS} $(G_0,P)$ of $\PTGTSvar'$ where $G_0$ is the initial graph of $\PTGTSvar'$ and $P$ contains all underlying GT rules $\GTRULEvar$ of all PTGT rules $\PTGTRULEvar$ of $\PTGTSvar'$ as in \cite{2018_probabilistic_timed_graph_transformation_systems}.
As a second step, we construct for this GTS its GT state space $(Q,E)$ consisting of states $Q$ and edges $E$ as in \cite{2018_probabilistic_timed_graph_transformation_systems} but deviate by not identifying isomorphic states, which results in a tree-shaped GT state space with root $G_0$.\footnote{
Our BMC approach cannot be used if the PTGTS $S'$ results in an infinite $(Q,E)$.}
Note that the paths through $(Q,E)$ symbolically describe all timed probabilistic paths through $\PTGTSvar'$.
As a third step, we again deviate from \cite{2018_probabilistic_timed_graph_transformation_systems} and modify $(Q,E)$ into $(Q',E')$ by adding \emph{time point clocks} throughout the paths of $(Q,E)$ as follows.
If $N$ is the maximal number of graphs in any path $\pi$ of $(Q,E)$, we
(a) create additional time point clocks $\var{tpc}_1$ to $\var{tpc}_N$,
(b) add the $i$ time point clocks $\var{tpc}_1$ to $\var{tpc}_i$ to the $i$th graph in any path of the state space, and
(c) add the clock $\var{tpc}_i$ to the reset set of the step leading to the graph $G_i$ in any path of the state space.
Consequently, the \gls{AC} $\var{tpc}_i-\var{tpc}_j$ for $j\geq i$ expresses the time expired between the graphs $G_i$ and $G_j$.
Finally, as in \cite{2018_probabilistic_timed_graph_transformation_systems}, we construct the resulting \gls{PTA} $\PTAvar$ from the given \gls{PTGTS} $\PTGTSvar'$ and the state space $(Q',E')$ by
    (a) aggregating GT steps with a common source state and a match belonging to one PTGT rule,		
    (b) annotating such aggregated GT steps with the clock-based timing constraints given by the guards and resets of the used PTGT rule,
    and
    (c) adding the clock-based timing constraints given by the PTGT invariants to the resulting PTA.
This PTA construction ensures that the resulting PTA $\PTAvar$ is equivalent to the given PTGTS $\PTGTSvar'$.
\begin{mylemmaNOPROOF}{Soundness of PTA Construction}{sound_construction_of_pta}
If the PTGTS $\PTGTSvar'$ has a finite tree-shaped state space $(Q,E)$, then the two PTSs $\PTAtoPTS{\PTGTStoPTA{\PTGTSvar'}}$ and $\PTGTStoPTS{\PTGTSvar'}$ return the same results for the analysis problems from \myref{definition:analysis_problems}.
See appendix for a proof sketch.
\end{mylemmaNOPROOF}
In step~8, we will apply the \toolPRISM model checker~\cite{2011_prism_4_0__verification_of_probabilistic_real_time_systems} to the obtained PTA $\PTAvar$ and an analysis problem from \myref{definition:analysis_problems} corresponding to the given PMTGC~$\PMTGCvar$.
For this purpose, we obtain in steps~3--7 the set of leaf-locations of the PTA, in which the MTGC $\MTGCvar$ used inside the PMTGC $\PMTGCvar$ is not violated, and then label precisely those locations from that set with an additional AP $\var{success}$.
Employing this AP, the analysis problems from \myref{definition:analysis_problems} can be used to express the minimal/maximal probability to reach no violation.
\Paragraph{3}{Folding of Paths into Graphs with History}
For the given PTA $\PTAvar$, we consider its structural paths $\pi$, which are the paths through the GT state space $(Q',E')$ from which $\PTAvar$ was constructed.
Such paths $\pi$ may have timed realizations $\pi'$ in which timed steps and discrete steps using the PTA edges of $\pi$ are interleaved.
Following the satisfaction checking approach for MTGL from \cite{2019_metric_temporal_graph_logic_over_typed_attributed_graphs,2020_formal_testing_of_timed_graph_transformation_systems_using_metric_temporal_graph_logic}, we translate the \gls{MTGC} satisfaction problem into an equivalent \gls{GC} satisfaction problem using an operation \OPfoldNAME (introduced subsequently) and an operation \OPencodeNAME (introduced in step~4).
Both operations together ensure for each timed realization $\pi'$ of a structural path $\pi$ of the given PTA $\PTAvar$ that $\MTGCsatTGSOUTER{\pi'}{\MTGCvar}$ iff $\GCsatOUTER{G_H'}{\GCvar}$ when $\OPfold{\pi}=G_H$ is a \gls{GH}, the graph $G_H'$ is obtained from $G_H$ by adding the durations of steps in $\pi'$ as ACs over the time point variables of $G_H$, and $\OPencode{\MTGCvar}=\GCvar$.

The operation \OPfoldNAME is applied to each structural path $\pi$ of the given PTA $\PTAvar$ aggregating the information about the nature and timing of all GT steps into a single resulting \gls{GH}.
As a first step, we construct the colimit $G_H$ for the diagram of the GT spans of $\pi$ (given by the \var{sp} components of step actions according to \myref{definition:PTGTStoPTS}), which contains all graph elements that existed at some time point in $\pi$.
As a second step, each node and edge in $G_H$ is equipped with additional \emph{creation/deletion time stamp attributes} \var{cts}/\var{dts} and \emph{creation/deletion index attributes} \var{cidx}/\var{didx}.
As a third step,
the ACs
$\var{cts}=\var{tpc}_0-\var{tpc}_j$ and $\var{cidx}=j$ are added for each node/edge that appeared first in the graph $G_j$ in the path~$\pi$.
As a fourth step,
the ACs
$\var{dts}=\var{tpc}_0-\var{tpc}_j$ and $\var{didx}=j$ are added for each node/edge that is removed in the step reaching  $G_j$ in the path $\pi$.
Finally, the ACs $\var{dts}=-1$ and $\var{didx}=-1$ are added for nodes/edges that are never removed in $\pi$.\footnote{
The presented operations \OPfoldNAME and \OPencodeNAME are adaptations of the corresponding operations from \cite{2019_metric_temporal_graph_logic_over_typed_attributed_graphs,2020_formal_testing_of_timed_graph_transformation_systems_using_metric_temporal_graph_logic} to the modified MTGL satisfaction relation defined for PTSs (see \myref{definition:satisfaction_for_mtgc_for_a_tgs}).~The adapted operation \OPfoldNAME uses ACs to express clock differences instead of concrete assignments and employs additional index attributes \var{cidx}/\var{didx}.
The adapted operation \OPencodeNAME uses the additional \emph{step index variable} $x_s$ in the \textsf{alive} and \textsf{earliest} \ACs to take not only the time stamp but also the step index into account.
}

As output, we obtain the so-called \emph{GH-restrictions} \algoGHmap mapping all leaf-locations $\ell$ of the PTA \PTAvar to the \gls{GH} constructed for the path ending~in~$\ell$.
\Paragraph{4}{Encoding of an \gls{MTGC} as a \gls{GC}}
We now discuss the operation \OPencodeNAME for translating the \gls{MTGC} $\MTGCvar$ contained in the given \gls{PMTGC} $\PMTGCvar$ into a corresponding \gls{GC} $\GCvar$.
Intuitively, this operation recursively encodes the requirements (see the items of \myref{definition:satisfaction_for_mtgc_for_a_tgs}) expressed using MTGL operators on a timed realization $\pi'$ (of a structural path $\pi$ of the PTA $\PTAvar$ folded in step~3) using GL operators on the \gls{GH} $G_H$ (obtained by folding $\pi$) with additional ACs.
In particular, quantification over positions $\tau=(t,s)$, as for the operators \MTGCexistsNewSimpleNAMEintext and \MTGCuntilSimpleNAMEintext, is encoded by quantifying over additional variables $x_t$ and $x_s$ representing $t$ and $s$, respectively.
Moreover, matching of graphs, as for the operators \MTGCexistsSimpleNAMEintext and \MTGCexistsNewSimpleNAMEintext, is encoded by an additional \gls{AC} \textsf{alive}.
This \gls{AC} requires that each matched graph element in the \gls{GH} $G_H$ has \var{cts}, \var{dts}, \var{cidx}, and \var{didx} attributes implying that this graph element exists for the position $(x_t,x_s)$ in $\pi'$.
Lastly, matching of new graph elements in the \MTGCexistsNewSimpleNAMEintext operator is encoded by an additional \gls{AC} \textsf{earliest}.
This \gls{AC} requires, in addition to \textsf{alive}, that one of the matched graph elements has \var{cts} and \var{cidx} attributes equal to $x_t$ and $x_s$, respectively.\footnotemark[8]

As output, we obtain the GC $\GCvar$, which expresses the MTGC $\MTGCvar$ based on the graph $G_H'$ obtained from the timed realization $\pi'$ in step~3.

\Paragraph{5}{Construction of \gls{AC}-Restrictions of Violations}
For each \gls{GH} $G_H$ (from the given \algoGHmap) obtained in step~3 for some path $\pi$, we evaluate the \emph{negation} of the given \gls{GC} $\GCvar$ obtained in step~4 for this $G_H$.
The result of this evaluation is an \gls{AC} $\ACvar$, which describes valuations of the variables contained in $G_H$.
Each such valuation describes a timed realization $\pi'$ of $\pi$ not satisfying the MTGC $\MTGCvar$ (i.e., a violation) by providing real-valued time points for the additional time point clocks contained in $G_H$.
In the sense of the equivalence discussed in step~3, such a valuation represents the durations of timed steps in $\pi'$, which can be added in the form of an AC to $G_H$ resulting in the graph $G_H'$ such that $\GCNOTsatOUTER{\pi'}{\MTGCvar}$ and $\GCNOTsatOUTER{G_H'}{\GCvar}$.

For our running example, any path $\pi$ ends with all messages being received.
The obtained \gls{AC} $\ACvar$ describes then that a violation has occurred when, for one of the messages, the sum of the timed steps between sending and receiving exceeds $5$ time units.
Certainly, due to possible interleavings of discrete steps and different routes from $R_1$ to $R_3$, there are various structural paths of $\PTAvar$ ending in different \glspl{GH} each resulting in a different \gls{AC}~$\ACvar$.

As output, we obtain the so-called \emph{AC-restrictions} \algoACmap mapping all leaf-locations $\ell$ of the PTA \PTAvar to the AC $\ACvar$ constructed for the GH $G_H$ (which is obtained for the path $\pi$ ending in $\ell$).

\Paragraph{6}{Construction of Zone-Restrictions of Violations}
We adapt the given \gls{PTA} $\PTAvar$ from step~2 to a resulting \gls{PTA} $\PTAvar'$ by adding an additional \gls{AP} $\var{terminal}$ and by labeling all leaf-locations with this AP.
We then construct the symbolic zone-based state space for the \gls{PTA} $\PTAvar'$ by evaluating $\PTCTLmax{\PTCTLeventuallyUnbounded{\var{terminal}}}$ (see \myref{definition:analysis_problems})
using a minor adaptation of the \toolPRISM model checker that outputs the states $s=(\ell,\CCvar)$ labeled with the AP $\var{terminal}$ containing a location $\ell$ and a clock constraint $\CCvar$ as a zone (which is unique due to the tree-shaped form of the PTA $\PTAvar$).
For each structural path $\pi$ of the PTA $\PTAvar$ ending in the location $\ell$, the zone $\CCvar$ symbolically represents all timed realizations $\pi'$ of $\pi$, which respect the timing constraints of the PTA $\PTAvar$, in terms of differences between the additional time point clocks added in step~2.

For our running example, the zone $\CCvar$ obtained for some leaf-location then contains the clock constraints capturing for each message that (a) $2$ to $5$ time units elapsed before each transmission attempt and (b) no time elapsed between the arrival of that message at router $R_3$ and its reception by the receiver.

As output, we obtain the so-called \emph{zone-restrictions} \algoZonemap mapping all leaf-locations $\ell$ of the PTA \PTAvar to the zone $\CCvar$ obtained for $\ell$.

\Paragraph{7}{Construction of Violations}
We now combine the restrictions captured by the given mappings
	\algoGHmap,
	\algoACmap, and
	\algoZonemap to determine the leaf-locations of the PTA $\PTAvar$ representing violations.
A leaf-location $\ell$ represents a violation when it is reached by a structural path $\pi$ of $\PTAvar$ that is realizable in terms of a timed realization $\pi'$ such that the interleaving of timed and discrete steps in $\pi'$ (which depends on the considered adversary) results in a violation when reaching $\ell$.
For this purpose, we compare the AC-restrictions with the zone-restrictions in a way that depends on whether the given PMTGC $\PMTGCvar$ is of the form \PTCTLmax{\MTGCvar} or \PTCTLmin{\MTGCvar}.
In the following, we consider the case for \OPmaxNAME (and the case for \OPminNAME in brackets).
We define the \gls{AC} $\ACvar_{\var{check}}$ as
    $\algoMzone(\ell)\wedge\neg(\algoMac(\ell)\wedge\GraphAC[\algoMgh(\ell)])$
    (for \OPminNAME: $\algoMzone(\ell)\wedge\algoMac(\ell)\wedge\GraphAC[\algoMgh(\ell)]$)
    where \GraphAC[\algoMgh(\ell)] denotes the \gls{AC} of the \gls{GH} $\algoMgh(\ell)$.
This \gls{AC} is satisfiable (for \OPminNAME: unsatisfiable) iff a violation is avoidable (for \OPminNAME: unreachable) for any probability maximizing (for \OPminNAME: probability minimizing) adversary based on interleavings of timed steps.
We use the SMT solver \Zthree~\cite{REFforZ3} to decide whether the obtained \gls{AC} $\ACvar_{\var{check}}$ is satisfiable (for \OPminNAME: unsatisfiable).

As output, we obtain the so-called \algoAPmap, which maps all leaf-locations $\ell$ of the PTA \PTAvar to a set of APs.
The set of APs $\algoMap(\ell)$ contains
    (a) the APs $\var{success}$ and $\var{maybe}$, if \Zthree returns that the checked \gls{AC} $\ACvar_{\var{check}}$ is satisfiable (for \OPminNAME: unsatisfiable)
    and
    (b) the AP $\var{maybe}$, if \Zthree does not return a result.
Hence, structural paths of the PTA $\PTAvar$ ending in locations labeled with the AP $\var{success}$ represent PTGTS paths definitely (for \OPminNAME: possibly) satisfying the considered MTGC whereas PTGTS paths ending in locations labeled with the AP $\var{maybe}$ may or may not represent such paths.

\newpage
\Paragraph{8}{Computation of Resulting Probabilities}
In steps~1--7, we reduced the considered \gls{BMC} problem to one of the analysis problems from \myref{definition:analysis_problems} for which \toolPRISM can be applied.
For this last step, we adapt the given \gls{PTA} $\PTAvar$ from step~2 to a \gls{PTA} $\PTAvar'$ by adding the labeling captured by the given \algoAPmap from step~7.
We compute and output the probability intervals $I=\INTERVALcc{\PTCTLmin{\var{success}}}{\PTCTLmin{\var{maybe}}}$ and $I=\INTERVALcc{\PTCTLmax{\var{success}}}{\PTCTLmax{\var{maybe}}}$ of possible expected probability values for \PTCTLmin{\MTGCvar} and \PTCTLmax{\MTGCvar}, respectively.
If \Zthree always succeeded in step~7, this probability interval $I$ will be a singleton.
Lastly, we state that the presented \gls{BMC} approach is sound (up to the imprecision possibly induced by~\Zthree).
\begin{mytheoremNOPROOF}{Soundness of \gls{BMC} Approach}{correctness_of_model_checking_procedure}
The presented \gls{BMC} approach correctly analyzes (correctly approximates) satisfaction of \glspl{PMTGC} when the returned probability interval $I$ is (is not) a singleton.
See appendix for a proof sketch.
\end{mytheoremNOPROOF}

\mysection{
    title={Evaluation},
    label={evaluation}
}

To evaluate our \gls{BMC} approach, we applied its implementation in the tool \AUTOGRAPH (where \toolPRISM and \Zthree are used as explained before) to our running example given by the \gls{PMTGC} \PMTGCexample from \myref{figure:running_example_pmtgc} and the \gls{PTGTS} from \myref{figure:running_example}.
In this application, we used the time bound $T=\infty$ for which the PTGTS was not adapted in step~1 because it already resulted in a \emph{finite} tree-shaped GT state space $(Q,E)$ in step~2.\footnote{In $(Q,E)$, each of the three messages has either not yet been sent, is at one of the five routers, or has been received resulting in at most $7^3$ states.}
The constraint solver \Zthree was always able to decide all satisfaction problems in step~7, and the probability interval obtained in step~8 using \toolPRISM was $\INTERVALcc{0.262144}{0.262144}$, which is in accordance with our detailed explanations below \myref{definition:satisfaction_of_pmtgc}.

We also applied our \gls{BMC} approach to the same PTGTS (again using the time bound $T=\infty$) and the PMTGC $\PTCTLmin{\MTGCvar}$ where $\MTGCvar$ is the MTGC used in the PMTGC \PMTGCexample from \myref{figure:running_example_pmtgc}.
In this case, we obtained in step~8 the probability interval $\INTERVALcc{0}{0}$ since there is a probability minimizing adversary that sends the first message at time point $0$ and then delays the first two transmission attempts of that message to time points $5$ and $10$ ensuring that the message is not received within $5$ time units as required in the MTGC $\MTGCvar$.

Both discussed applications of our \gls{BMC} approach (where steps~1--7 can be reused for the second application) required negligible runtime and memory.

\mysection{
    title={Conclusion and Future Work},
    label={conclusion_and_future_work}
}
\enlargethispage{1\baselineskip}\glsreset{PMTGL}\glsreset{BMC}In this paper, we introduced the \gls{PMTGL} for the specification of cyber-physical systems with probabilistic timed behavior modeled as \glspl{PTGTS}.
\gls{PMTGL} combines
    (a)~\gls{MTGL} with its binding capabilities for the specification of timed graph sequences and
    (b)~the probabilistic operator from \gls{PTCTL}
to express best-case/worst-case probabilistic timed reachability properties.
Moreover, we presented a novel \gls{BMC} approach for PTGTSs w.r.t.\ PMTGL properties.

In the future, we will consider the case study \cite{RailCab,2018_probabilistic_timed_graph_transformation_systems} of a cyber-physical system where, in accordance with real-time constraints, autonomous shuttles exhibiting probabilistic failures on demand navigate on a track topology.
For this case study, we will evaluate the expressiveness and usability of PMTGL~as well as the performance of our \gls{BMC} approach.
Also, we will integrate our MTGL\hyph{}based approach from \cite{2020_optimistic_and_pessimistic_on_the_fly_analysis_for_metric_temporal_graph_logic} for deriving so-called optimistic violations.

\renewcommand{\doi}[1]{\textsc{doi}: \href{http://dx.doi.org/#1}{\nolinkurl{#1}}}
\newcommand{\urn}[1]{\textsc{urn}: \href{http://nbn-resolving.de/#1}{\nolinkurl{#1}}}

\appendix

\clearpage
\newpage
\begingroup
\printnoidxglossaries
\endgroup

\section{Proofs}

In this appendix, we provide proof sketches omitted in the main body of this paper.

\begin{myproof}{Soundness of PTA Construction}{lemma:sound_construction_of_pta}
The nonidentification of isomorphic states and the addition of time point clocks does not affect the possible steps of the resulting PTA.
This PTA is therefore, following \cite{2018_probabilistic_timed_graph_transformation_systems}, equivalent to the given PTGTS $\PTGTSvar'$ w.r.t.\ the analysis problems from \myref{definition:analysis_problems}.
\end{myproof}

\begin{myproof}{Soundness of \gls{BMC} Approach}{theorem:correctness_of_model_checking_procedure}
We conclude that the presented \gls{BMC} approach computes the correct results
	(a)~by encoding the time bound $T$ properly in step~1, 
	(b)~from the soundness of the operation \PTGTStoPTANAME according to \myref{lemma:sound_construction_of_pta} (following \cite{2018_probabilistic_timed_graph_transformation_systems}),
	(c)~from the soundness of the adapted translation of \gls{MTGC} satisfaction problem into an equivalent \gls{GC} satisfaction problem along the lines of \cite{2019_metric_temporal_graph_logic_over_typed_attributed_graphs,2020_formal_testing_of_timed_graph_transformation_systems_using_metric_temporal_graph_logic}, 
and
	(d)~from the correct computation of zones in \toolPRISM.
\end{myproof}

\section{Details for Simplified Running Example}
In this appendix, we present figures for the steps of our \gls{BMC} approach for a simplified form of our running example where only a single message is transmitted to the receiver.

\begin{myfigureWITHBOX}{Visualization for step 2 of our \gls{BMC} approach:
    A structural path $\pi$ of the \gls{PTA} $\PTAvar$ (using an adapted initial graph with a single message).
}{running_example:structural_path}\newcommand{\MyLabel}[5]{$\stackrel{\begin{array}{l}
    \text{PTGT rule:}~#1\\
    \text{GT rule:}~#2\\
    \text{probability:}~#3\\
    \text{satisfied guard:}~#4\\
    \text{clock resets:}~#5\\
    \end{array}}{\Longrightarrow}$}
\scalebox{\SCALEFACTORfigures}{\begin{tikzpicture}[line width=1pt]
    \SetNodeWidthName{RunningExampleStructuralPath}

    \node[mtgc_outer] (node0) at (0,0) {\begin{tikzpicture}
        \node[conditionGraphNode] (Graph1) at (0,0) {\eqboxwh{RunningExampleStructuralPathHeight}{\begin{tikzpicture}[mtgc_inner]
            \node[styleNODESender] (NodeSender1) at (0,0) {\NODESenderX{name=S,num=1}};
\node[styleNODERouter,node distance=\TIKZGRAPHSedgelabelheight,below=of NodeSender1.south west,anchor=north west] (NodeRouter1) {\NODERouterX{name=R_1}};
\node[styleNODERouter,node distance=\TIKZGRAPHSedgelabelwidth+5ex,right=of NodeRouter1.north east,anchor=north west] (NodeRouter2) {\NODERouterX{name=R_2}};
\node[styleNODERouter,node distance=\TIKZGRAPHSedgelabelwidth+5ex,right=of NodeRouter2.north east,anchor=north west] (NodeRouter3) {\NODERouterX{name=R_3}};
\node[styleNODERouter,node distance=\TIKZGRAPHSedgelabelheight+0ex,below=of NodeRouter1.south west,anchor=north west] (NodeRouter4) {\NODERouterX{name=R_4}};
\node[styleNODERouter,node distance=\TIKZGRAPHSedgelabelheight+0ex,below=of NodeRouter3.south west,anchor=north west] (NodeRouter5) {\NODERouterX{name=R_5}};
\gettikzxy{(NodeSender1.north west)}{\nodeAX}{\nodeAY}
\gettikzxy{(NodeRouter3.north west)}{\nodeBX}{\nodeBY}
\node[styleNODEReceiver,anchor=north west] (NodeReceiver1) at (\nodeBX,\nodeAY) {\NODEReceiverX{name=R}};
\node[styleNODEMessage,anchor=north] (NodeMessage1) at ($0.5*(NodeSender1.north east)+0.5*(NodeReceiver1.north west)$) {\NODEMessageX{name=M_1,id=1,clock=c_1}};
\GraphEdgeDirect{N}{R}{NodeSender1}{styleEDGEsnd}{LAB}{\EDGEsndX{name=e_1}}{NodeRouter1}
\GraphEdgeDirect{S}{R}{NodeRouter3}{styleEDGErcv}{LAB}{\EDGErcvX{name=e_2}}{NodeReceiver1}
\GraphEdgeDirect{E}{B}{NodeRouter1}{styleEDGEnext}{LAB}{\EDGEnextX{name=e_3}}{NodeRouter2}
\GraphEdgeDirect{E}{B}{NodeRouter2}{styleEDGEnext}{LAB}{\EDGEnextX{name=e_4}}{NodeRouter3}
\GraphEdgeDirect{N}{R}{NodeRouter1}{styleEDGEnext}{LAB}{\EDGEnextX{name=e_5}}{NodeRouter4}
\GraphEdgeDirect{E}{A}{NodeRouter4}{styleEDGEnext}{LAB}{\EDGEnextX{name=e_6}}{NodeRouter5}
\GraphEdgeDirect{S}{R}{NodeRouter5}{styleEDGEnext}{LAB}{\EDGEnextX{name=e_7}}{NodeRouter3}
\node[globalVariablesNode,right=of NodeReceiver1,anchor=west] (globalVariables) {$\{\var{tpc}_0\}$};             \end{tikzpicture}}};
        \end{tikzpicture}};
    \node[mtgc_outer,node distance=2ex,below=of node0.south west,anchor=north west] (node1) {\begin{tikzpicture}
        \node[conditionGraphNode] (Graph1) at (0,0) {\eqboxwh{RunningExampleStructuralPathHeight}{\begin{tikzpicture}[mtgc_inner]
            \node[styleNODESender] (NodeSender1) at (0,0) {\NODESenderX{name=S,num=2}};
\node[styleNODERouter,node distance=\TIKZGRAPHSedgelabelheight,below=of NodeSender1.south west,anchor=north west] (NodeRouter1) {\NODERouterX{name=R_1}};
\node[styleNODERouter,node distance=\TIKZGRAPHSedgelabelwidth+5ex,right=of NodeRouter1.north east,anchor=north west] (NodeRouter2) {\NODERouterX{name=R_2}};
\node[styleNODERouter,node distance=\TIKZGRAPHSedgelabelwidth+5ex,right=of NodeRouter2.north east,anchor=north west] (NodeRouter3) {\NODERouterX{name=R_3}};
\node[styleNODERouter,node distance=\TIKZGRAPHSedgelabelheight+0ex,below=of NodeRouter1.south west,anchor=north west] (NodeRouter4) {\NODERouterX{name=R_4}};
\node[styleNODERouter,node distance=\TIKZGRAPHSedgelabelheight+0ex,below=of NodeRouter3.south west,anchor=north west] (NodeRouter5) {\NODERouterX{name=R_5}};
\gettikzxy{(NodeSender1.north west)}{\nodeAX}{\nodeAY}
\gettikzxy{(NodeRouter3.north west)}{\nodeBX}{\nodeBY}
\node[styleNODEReceiver,anchor=north west] (NodeReceiver1) at (\nodeBX,\nodeAY) {\NODEReceiverX{name=R}};
\node[styleNODEMessage,anchor=center] (NodeMessage1) at ($0.5*(NodeSender1.north east)+0.5*(NodeReceiver1.north west)$) {\NODEMessageX{name=M_1,id=1,clock=c_1}};
\GraphEdgeDirect{N}{R}{NodeSender1}{styleEDGEsnd}{LAB}{\EDGEsndX{name=e_1}}{NodeRouter1}
\GraphEdgeDirect{S}{R}{NodeRouter3}{styleEDGErcv}{LAB}{\EDGErcvX{name=e_2}}{NodeReceiver1}
\GraphEdgeDirect{E}{B}{NodeRouter1}{styleEDGEnext}{LAB}{\EDGEnextX{name=e_3}}{NodeRouter2}
\GraphEdgeDirect{E}{B}{NodeRouter2}{styleEDGEnext}{LAB}{\EDGEnextX{name=e_4}}{NodeRouter3}
\GraphEdgeDirect{N}{R}{NodeRouter1}{styleEDGEnext}{LAB}{\EDGEnextX{name=e_5}}{NodeRouter4}
\GraphEdgeDirect{E}{A}{NodeRouter4}{styleEDGEnext}{LAB}{\EDGEnextX{name=e_6}}{NodeRouter5}
\GraphEdgeDirect{S}{R}{NodeRouter5}{styleEDGEnext}{LAB}{\EDGEnextX{name=e_7}}{NodeRouter3}
\draw(NodeMessage1)edge[styleEdge]node[styleEDGEat,above]{\EDGEatX{name=e_8}}(NodeRouter1.north east);
\node[globalVariablesNode,right=of NodeReceiver1,anchor=west] (globalVariables) {$\{\var{tpc}_0,\var{tpc}_1\}$};             \end{tikzpicture}}};
        \end{tikzpicture}};
    \node[mtgc_outer,node distance=2ex,below=of node1.south west,anchor=north west] (node2) {\begin{tikzpicture}
        \node[conditionGraphNode] (Graph1) at (0,0) {\eqboxwh{RunningExampleStructuralPathHeight}{\begin{tikzpicture}[mtgc_inner]
            \node[styleNODESender] (NodeSender1) at (0,0) {\NODESenderX{name=S,num=2}};
\node[styleNODERouter,node distance=\TIKZGRAPHSedgelabelheight,below=of NodeSender1.south west,anchor=north west] (NodeRouter1) {\NODERouterX{name=R_1}};
\node[styleNODERouter,node distance=\TIKZGRAPHSedgelabelwidth+5ex,right=of NodeRouter1.north east,anchor=north west] (NodeRouter2) {\NODERouterX{name=R_2}};
\node[styleNODERouter,node distance=\TIKZGRAPHSedgelabelwidth+5ex,right=of NodeRouter2.north east,anchor=north west] (NodeRouter3) {\NODERouterX{name=R_3}};
\node[styleNODERouter,node distance=\TIKZGRAPHSedgelabelheight+0ex,below=of NodeRouter1.south west,anchor=north west] (NodeRouter4) {\NODERouterX{name=R_4}};
\node[styleNODERouter,node distance=\TIKZGRAPHSedgelabelheight+0ex,below=of NodeRouter3.south west,anchor=north west] (NodeRouter5) {\NODERouterX{name=R_5}};
\gettikzxy{(NodeSender1.north west)}{\nodeAX}{\nodeAY}
\gettikzxy{(NodeRouter3.north west)}{\nodeBX}{\nodeBY}
\node[styleNODEReceiver,anchor=north west] (NodeReceiver1) at (\nodeBX,\nodeAY) {\NODEReceiverX{name=R}};
\node[styleNODEMessage,anchor=center] (NodeMessage1) at ($0.5*(NodeSender1.north east)+0.5*(NodeReceiver1.north west)$) {\NODEMessageX{name=M_1,id=1,clock=c_1}};
\GraphEdgeDirect{N}{R}{NodeSender1}{styleEDGEsnd}{LAB}{\EDGEsndX{name=e_1}}{NodeRouter1}
\GraphEdgeDirect{S}{R}{NodeRouter3}{styleEDGErcv}{LAB}{\EDGErcvX{name=e_2}}{NodeReceiver1}
\GraphEdgeDirect{E}{B}{NodeRouter1}{styleEDGEnext}{LAB}{\EDGEnextX{name=e_3}}{NodeRouter2}
\GraphEdgeDirect{E}{B}{NodeRouter2}{styleEDGEnext}{LAB}{\EDGEnextX{name=e_4}}{NodeRouter3}
\GraphEdgeDirect{N}{R}{NodeRouter1}{styleEDGEnext}{LAB}{\EDGEnextX{name=e_5}}{NodeRouter4}
\GraphEdgeDirect{E}{A}{NodeRouter4}{styleEDGEnext}{LAB}{\EDGEnextX{name=e_6}}{NodeRouter5}
\GraphEdgeDirect{S}{R}{NodeRouter5}{styleEDGEnext}{LAB}{\EDGEnextX{name=e_7}}{NodeRouter3}
\draw(NodeMessage1)edge[styleEdge]node[styleEDGEat,above]{\EDGEatX{name=e_9}}(NodeRouter2.north);
\node[globalVariablesNode,right=of NodeReceiver1,anchor=west] (globalVariables) {$\{\var{tpc}_0,\dots,\var{tpc}_2\}$};             \end{tikzpicture}}};
        \end{tikzpicture}};
    \node[mtgc_outer,node distance=2ex,below=of node2.south west,anchor=north west] (node3) {\begin{tikzpicture}
        \node[conditionGraphNode] (Graph1) at (0,0) {\eqboxwh{RunningExampleStructuralPathHeight}{\begin{tikzpicture}[mtgc_inner]
            \node[styleNODESender] (NodeSender1) at (0,0) {\NODESenderX{name=S,num=2}};
\node[styleNODERouter,node distance=\TIKZGRAPHSedgelabelheight,below=of NodeSender1.south west,anchor=north west] (NodeRouter1) {\NODERouterX{name=R_1}};
\node[styleNODERouter,node distance=\TIKZGRAPHSedgelabelwidth+5ex,right=of NodeRouter1.north east,anchor=north west] (NodeRouter2) {\NODERouterX{name=R_2}};
\node[styleNODERouter,node distance=\TIKZGRAPHSedgelabelwidth+5ex,right=of NodeRouter2.north east,anchor=north west] (NodeRouter3) {\NODERouterX{name=R_3}};
\node[styleNODERouter,node distance=\TIKZGRAPHSedgelabelheight+0ex,below=of NodeRouter1.south west,anchor=north west] (NodeRouter4) {\NODERouterX{name=R_4}};
\node[styleNODERouter,node distance=\TIKZGRAPHSedgelabelheight+0ex,below=of NodeRouter3.south west,anchor=north west] (NodeRouter5) {\NODERouterX{name=R_5}};
\gettikzxy{(NodeSender1.north west)}{\nodeAX}{\nodeAY}
\gettikzxy{(NodeRouter3.north west)}{\nodeBX}{\nodeBY}
\node[styleNODEReceiver,anchor=north west] (NodeReceiver1) at (\nodeBX,\nodeAY) {\NODEReceiverX{name=R}};
\node[styleNODEMessage,anchor=center] (NodeMessage1) at ($0.5*(NodeSender1.north east)+0.5*(NodeReceiver1.north west)$) {\NODEMessageX{name=M_1,id=1,clock=c_1}};
\GraphEdgeDirect{N}{R}{NodeSender1}{styleEDGEsnd}{LAB}{\EDGEsndX{name=e_1}}{NodeRouter1}
\GraphEdgeDirect{S}{R}{NodeRouter3}{styleEDGErcv}{LAB}{\EDGErcvX{name=e_2}}{NodeReceiver1}
\GraphEdgeDirect{E}{B}{NodeRouter1}{styleEDGEnext}{LAB}{\EDGEnextX{name=e_3}}{NodeRouter2}
\GraphEdgeDirect{E}{B}{NodeRouter2}{styleEDGEnext}{LAB}{\EDGEnextX{name=e_4}}{NodeRouter3}
\GraphEdgeDirect{N}{R}{NodeRouter1}{styleEDGEnext}{LAB}{\EDGEnextX{name=e_5}}{NodeRouter4}
\GraphEdgeDirect{E}{A}{NodeRouter4}{styleEDGEnext}{LAB}{\EDGEnextX{name=e_6}}{NodeRouter5}
\GraphEdgeDirect{S}{R}{NodeRouter5}{styleEDGEnext}{LAB}{\EDGEnextX{name=e_7}}{NodeRouter3}
\draw(NodeMessage1)edge[styleEdge]node[styleEDGEat,above]{\EDGEatX{name=e_{10}}}(NodeRouter3.north west);
\node[globalVariablesNode,right=of NodeReceiver1,anchor=west] (globalVariables) {$\{\var{tpc}_0,\dots,\var{tpc}_3\}$};             \end{tikzpicture}}};
        \end{tikzpicture}};
    \node[mtgc_outer,node distance=2ex,below=of node3.south west,anchor=north west] (node4) {\begin{tikzpicture}
        \node[conditionGraphNode] (Graph1) at (0,0) {\eqboxwh{RunningExampleStructuralPathHeight}{\begin{tikzpicture}[mtgc_inner]
            \node[styleNODESender] (NodeSender1) at (0,0) {\NODESenderX{name=S,num=2}};
\node[styleNODERouter,node distance=\TIKZGRAPHSedgelabelheight,below=of NodeSender1.south west,anchor=north west] (NodeRouter1) {\NODERouterX{name=R_1}};
\node[styleNODERouter,node distance=\TIKZGRAPHSedgelabelwidth+5ex,right=of NodeRouter1.north east,anchor=north west] (NodeRouter2) {\NODERouterX{name=R_2}};
\node[styleNODERouter,node distance=\TIKZGRAPHSedgelabelwidth+5ex,right=of NodeRouter2.north east,anchor=north west] (NodeRouter3) {\NODERouterX{name=R_3}};
\node[styleNODERouter,node distance=\TIKZGRAPHSedgelabelheight+0ex,below=of NodeRouter1.south west,anchor=north west] (NodeRouter4) {\NODERouterX{name=R_4}};
\node[styleNODERouter,node distance=\TIKZGRAPHSedgelabelheight+0ex,below=of NodeRouter3.south west,anchor=north west] (NodeRouter5) {\NODERouterX{name=R_5}};
\gettikzxy{(NodeSender1.north west)}{\nodeAX}{\nodeAY}
\gettikzxy{(NodeRouter3.north west)}{\nodeBX}{\nodeBY}
\node[styleNODEReceiver,anchor=north west] (NodeReceiver1) at (\nodeBX,\nodeAY) {\NODEReceiverX{name=R}};
\node[styleNODEMessage,anchor=north] (NodeMessage1) at ($0.5*(NodeSender1.north east)+0.5*(NodeReceiver1.north west)$) {\NODEMessageX{name=M_1,id=1,clock=c_1}};
\GraphEdgeDirect{N}{R}{NodeSender1}{styleEDGEsnd}{LAB}{\EDGEsndX{name=e_1}}{NodeRouter1}
\GraphEdgeDirect{S}{R}{NodeRouter3}{styleEDGErcv}{LAB}{\EDGErcvX{name=e_2}}{NodeReceiver1}
\GraphEdgeDirect{E}{B}{NodeRouter1}{styleEDGEnext}{LAB}{\EDGEnextX{name=e_3}}{NodeRouter2}
\GraphEdgeDirect{E}{B}{NodeRouter2}{styleEDGEnext}{LAB}{\EDGEnextX{name=e_4}}{NodeRouter3}
\GraphEdgeDirect{N}{R}{NodeRouter1}{styleEDGEnext}{LAB}{\EDGEnextX{name=e_5}}{NodeRouter4}
\GraphEdgeDirect{E}{A}{NodeRouter4}{styleEDGEnext}{LAB}{\EDGEnextX{name=e_6}}{NodeRouter5}
\GraphEdgeDirect{S}{R}{NodeRouter5}{styleEDGEnext}{LAB}{\EDGEnextX{name=e_7}}{NodeRouter3}
\GraphEdgeLoop{N}{NodeMessage1}{styleEDGEdone}{LAB}{\EDGEdoneX{name=e_{11}}}{NodeRouter1}
\node[globalVariablesNode,right=of NodeReceiver1,anchor=west] (globalVariables) {$\{\var{tpc}_0,\dots,\var{tpc}_4\}$};             \end{tikzpicture}}};
        \end{tikzpicture}};

    \node[node distance=1ex,left=of node1] {\MyLabel{\PTGTRULEsend}{\GTRULEsendSuccess}{1.0}{\ACtrue}{\{c_1,\var{tpc}_0\}}};
    \node[node distance=1ex,left=of node2] {\MyLabel{\PTGTRULEtransmit}{\GTRULEtransmitSuccess}{0.8}{c_1\geq 2}{\{c_1,\var{tpc}_1\}}};
    \node[node distance=1ex,left=of node3] {\MyLabel{\PTGTRULEtransmit}{\GTRULEtransmitSuccess}{0.8}{c_1\geq 2}{\{c_1,\var{tpc}_2\}}};
    \node[node distance=1ex,left=of node4] {\MyLabel{\PTGTRULEreceive}{\GTRULEreceiveSuccess}{1.0}{\ACtrue}{\{\var{tpc}_3\}}};

    \begin{pgfonlayer}{background}
    \node (GRAPH0) [fill=black!5!white,fit=(node0)] {};
    \node (GRAPH1) [fill=black!5!white,fit=(node1)] {};
    \node (GRAPH2) [fill=black!5!white,fit=(node2)] {};
    \node (GRAPH3) [fill=black!5!white,fit=(node3)] {};
    \node (GRAPH4) [fill=black!5!white,fit=(node4)] {};
    \end{pgfonlayer}

    \end{tikzpicture}
} \end{myfigureWITHBOX}
\begin{myfigureWITHBOX}{Visualization for step 3 of our \gls{BMC} approach:
    \gls{GH} $G_H$ obtained for the structural path $\pi$ from \myref{figure:running_example:structural_path}.
}{running_example:graph_with_history}\scalebox{\SCALEFACTORfigures}{\begin{tikzpicture}[line width=1pt]
    \SetNodeWidthName{RunningExampleGraphWithHistory}

    \node[mtgc_outer] (nodeL) at (0,0) {\begin{tikzpicture}
        \node[conditionGraphNode] (Graph1) at (0,0) {\eqboxwh{RunningExampleGraphWithHistoryHeight}{\begin{tikzpicture}[mtgc_inner]
			\node[styleNODESender] (NodeSender1) at (0,0) {\NODESenderX{name=S,num=2,cts=0,dts=-1,cidx=0,didx=-1}};
			\node[styleNODERouter,node distance=\TIKZGRAPHSedgelabelheight,below=of NodeSender1.south west,anchor=north west] (NodeRouter1) {\NODERouterX{name=R_1,cts=0,dts=-1,cidx=0,didx=-1}};
			\node[styleNODERouter,node distance=\TIKZGRAPHSedgelabelwidth+5ex,right=of NodeRouter1.north east,anchor=north west] (NodeRouter2) {\NODERouterX{name=R_2,cts=0,dts=-1,cidx=0,didx=-1}};
			\node[styleNODERouter,node distance=\TIKZGRAPHSedgelabelwidth+5ex,right=of NodeRouter2.north east,anchor=north west] (NodeRouter3) {\NODERouterX{name=R_3,cts=0,dts=-1,cidx=0,didx=-1}};
			\node[styleNODERouter,node distance=\TIKZGRAPHSedgelabelheight+0ex,below=of NodeRouter1.south west,anchor=north west] (NodeRouter4) {\NODERouterX{name=R_4,cts=0,dts=-1,cidx=0,didx=-1}};
			\node[styleNODERouter,node distance=\TIKZGRAPHSedgelabelheight+0ex,below=of NodeRouter3.south west,anchor=north west] (NodeRouter5) {\NODERouterX{name=R_5,cts=0,dts=-1,cidx=0,didx=-1}};
			\gettikzxy{(NodeSender1.north west)}{\nodeAX}{\nodeAY}
			\gettikzxy{(NodeRouter3.north west)}{\nodeBX}{\nodeBY}
			\node[styleNODEReceiver,anchor=north west] (NodeReceiver1) at (\nodeBX,\nodeAY) {\NODEReceiverX{name=R,cts=0,dts=-1,cidx=0,didx=-1}};
			\node[styleNODEMessage,anchor=south] (NodeMessage1) at ($0.5*(NodeSender1.north east)+0.5*(NodeReceiver1.north west)$) {\NODEMessageX{name=M_1,id=1,clock=c_1,cts=0,dts=-1,cidx=0,didx=-1}};
			\GraphEdgeDirect{N}{R}{NodeSender1}{styleEDGEsnd}{LAB}{\EDGEsndX{name=e_1,cts=0,dts=-1,cidx=0,didx=-1}}{NodeRouter1}
			\GraphEdgeDirect{S}{L}{NodeRouter3}{styleEDGErcv}{LAB}{\EDGErcvX{name=e_2,cts=0,dts=-1,cidx=0,didx=-1}}{NodeReceiver1}
			\GraphEdgeDirect{E}{B}{NodeRouter1}{styleEDGEnext}{LAB}{\EDGEnextX{name=e_3,cts=0,dts=-1,cidx=0,didx=-1}}{NodeRouter2}
			\GraphEdgeDirect{E}{B}{NodeRouter2}{styleEDGEnext}{LAB}{\EDGEnextX{name=e_4,cts=0,dts=-1,cidx=0,didx=-1}}{NodeRouter3}
			\GraphEdgeDirect{N}{R}{NodeRouter1}{styleEDGEnext}{LAB}{\EDGEnextX{name=e_5,cts=0,dts=-1,cidx=0,didx=-1}}{NodeRouter4}
			\GraphEdgeDirect{E}{A}{NodeRouter4}{styleEDGEnext}{LAB}{\EDGEnextX{name=e_6,cts=0,dts=-1,cidx=0,didx=-1}}{NodeRouter5}
			\GraphEdgeDirect{S}{L}{NodeRouter5}{styleEDGEnext}{LAB}{\EDGEnextX{name=e_7,cts=0,dts=-1,cidx=0,didx=-1}}{NodeRouter3}
            \GraphEdgeLoop{N}{NodeMessage1}{styleEDGEdone}{LAB}{\EDGEdoneX{name=e_{11},cts=\var{tpc}_4-\var{tpc}_0,dts=-1,cidx=4,didx=-1}}{NodeRouter1}
            \draw(NodeMessage1)edge[styleEdge]node[styleEDGEat,above]{\EDGEatX{name=e_8,cts=\var{tpc}_1-\var{tpc}_0,dts=\var{tpc}_2-\var{tpc}_0,cidx=1,didx=2}}(NodeRouter1.north east);
            \draw(NodeMessage1)edge[styleEdge]node[styleEDGEat,above]{\EDGEatX{name=e_9,cts=\var{tpc}_2-\var{tpc}_0,dts=\var{tpc}_3-\var{tpc}_0,cidx=2,didx=3}}(NodeRouter2.north);
            \draw(NodeMessage1)edge[styleEdge]node[styleEDGEat,above]{\EDGEatX{name=e_{10},cts=\var{tpc}_3-\var{tpc}_0,dts=\var{tpc}_4-\var{tpc}_0,cidx=3,didx=4}}(NodeRouter3.north west);
            \end{tikzpicture}}};
        \end{tikzpicture}};

    \begin{pgfonlayer}{background}
    \node (GRAPHL) [fill=black!5!white,fit=(nodeL)] {};
    \end{pgfonlayer}

    \end{tikzpicture}
} \end{myfigureWITHBOX}
\begin{myfigureWITHBOX}{Visualization for step 4 of our \gls{BMC} approach:
    \gls{GC} $\GCvar$ obtained by encoding of the \gls{MTGC} from the \gls{PMTGC} \PMTGCexample.
}{running_example:encoded_MTGC}\newcommand{\CTSvar}[1]{t_{#1,\mathsf{c}}}
\newcommand{\DTSvar}[1]{t_{#1,\mathsf{d}}}
\newcommand{\CIDXvar}[1]{s_{#1,\mathsf{c}}}
\newcommand{\DIDXvar}[1]{s_{#1,\mathsf{d}}}
\def\STDINDENT{1ex}
\scalebox{\SCALEFACTORfigures}{\begin{tikzpicture}[line width=1pt]
    \SetNodeWidthName{RunningExampleReducedCondition}

\node[mtgc_outer] (node0) at (0,0) {\begin{tikzpicture}
        \node[] (Graph) at (0,0) {\eqboxwh{RunningExampleReducedConditionHeightLine1}{\begin{tikzpicture}[mtgc_inner]
            \node[globalVariablesNode] (globalVariables) at (0,0) {$\{x_{t,0}{:}\TypeReal,x_{s,0}{:}\TypeInteger\}$};
            \node[constraintNode,node distance=1ex,right=of globalVariables] (Constraint) {$\Theta_0$};
            \end{tikzpicture}}};
        \OPENINGBRACKET{Graph}
        \node[outer xsep=0ex,inner xsep=0pt,node distance=0ex,right=of Graph,anchor=west]
            {\large\strut$\COMMA$};
        \node[outer xsep=0ex,inner xsep=0pt,node distance=.8ex,left=of Graph,anchor=east]
            {\large\strut$\MTGCexistsNAME$};
        \end{tikzpicture}};

\node[mtgc_outer,xshift=\STDINDENT,node distance=0ex,below right=of node0.south west,anchor=north west] (node1) {\begin{tikzpicture}
        \node[] (Graph) at (0,0) {\VerticalExtend[c]{RunningExampleReducedConditionHeightLine2}{\begin{tikzpicture}[mtgc_inner]
			\node[styleNODESender] (NodeSender1) at (0,0) {\NODESenderX{name=S,cts=\CTSvar{S},dts=\DTSvar{S},cidx=\CIDXvar{S},didx=\DIDXvar{S}}};
			\node[styleNODERouter,node distance=\TIKZGRAPHSedgelabelwidth,right=of NodeSender1.north east,anchor=north west] (NodeRouter1) {\NODERouterX{name=R_1,cts=\CTSvar{R_1},dts=\DTSvar{R_1},cidx=\CIDXvar{R_1},didx=\DIDXvar{R_1}}};
			\node[styleNODEMessage,node distance=\TIKZGRAPHSedgelabelwidth,right=of NodeRouter1.north east,anchor=north west] (NodeMessage1) {\NODEMessageX{name=M,cts=\CTSvar{M},dts=\DTSvar{M},cidx=\CIDXvar{M},didx=\DIDXvar{M}}};
			\GraphEdgeDirect{E}{B}{NodeSender1}{styleEDGEsnd}{LAB}{\EDGEsndX{name=e_1,cts=\CTSvar{e_1},dts=\DTSvar{e_1},cidx=\CIDXvar{e_1},didx=\DIDXvar{e_1}}}{NodeRouter1}
			\GraphEdgeDirect{W}{B}{NodeMessage1}{styleEDGEat}{LAB}{\EDGEatX{name=e_2,cts=\CTSvar{e_2},dts=\DTSvar{e_2},cidx=\CIDXvar{e_2},didx=\DIDXvar{e_2}}}{NodeRouter1}
            \node[globalVariablesNode,node distance=1ex,right=of NodeMessage1] (globalVariables) {$\{x_{t,1}{:}\TypeReal,x_{s,1}{:}\TypeInteger\}$};
            \node[constraintNode,node distance=1ex,right=of globalVariables] (Constraint) {$\Theta_1$};
            \end{tikzpicture}}};
        \OPENINGBRACKET{Graph}
        \node[outer xsep=0ex,inner xsep=0pt,node distance=0ex,right=of Graph,anchor=west]
            {\large\strut$\COMMA$};
        \node[outer xsep=0ex,inner xsep=0pt,node distance=.8ex,left=of Graph,anchor=east]
            {\large\strut$\MTGCforallNAME$};
        \end{tikzpicture}};

\node[mtgc_outer,xshift=\STDINDENT,node distance=0ex,below right=of node1.south west,anchor=north west] (node2) {\begin{tikzpicture}
        \node[] (Graph) at (0,0) {\eqboxwh{RunningExampleReducedConditionHeightLine3}{\begin{tikzpicture}[mtgc_inner]
            \node[styleNODEMessage] (NodeMessage1) at (0,0) {\NODEMessageX{name=M,cts=\CTSvar{M},dts=\DTSvar{M},cidx=\CIDXvar{M},didx=\DIDXvar{M}}};
            \node[globalVariablesNode,right=of NodeMessage1] (globalVariables) {$\{x_{t,0}{:}\TypeReal,x_{s,0}{:}\TypeInteger,x_{t,1}{:}\TypeReal,x_{s,1}{:}\TypeInteger\}$};
            \node[constraintNode,node distance=1ex,right=of globalVariables] (Constraint) {$\Theta_2$};
            \end{tikzpicture}}};
        \OPENINGBRACKET{Graph}
        \node[outer xsep=0ex,inner xsep=0pt,node distance=0ex,right=of Graph,anchor=west]
            {\large\strut$\COMMA\;$};
        \node[outer xsep=0ex,inner xsep=0pt,node distance=.8ex,left=of Graph,anchor=east]
            {\large\strut$\MTGCrestrictNAME$};
        \end{tikzpicture}};

\node[mtgc_outer,xshift=\STDINDENT,node distance=0ex,below right=of node2.south west,anchor=north west] (node3) {\begin{tikzpicture}
        \node[] (Graph) at (0,0) {\VerticalExtend[c]{RunningExampleReducedConditionHeightLine4}{\begin{tikzpicture}[mtgc_inner]
            \node[globalVariablesNode] (globalVariables) at (0,0) {$\{x_{t,2}{:}\TypeReal,x_{s,2}{:}\TypeInteger\}$};
            \node[constraintNode,node distance=1ex,right=of globalVariables] (Constraint) {$\Theta_3$};
            \end{tikzpicture}}};
        \OPENINGBRACKET{Graph}
        \node[outer xsep=0ex,inner xsep=0pt,node distance=0ex,right=of Graph,anchor=west]
            {\large\strut$\COMMA$};
        \node[outer xsep=0ex,inner xsep=0pt,node distance=.8ex,left=of Graph,anchor=east]
            {\large\strut$\MTGCexistsNAME$};
        \end{tikzpicture}};

\node[mtgc_outer,xshift=\STDINDENT,node distance=0ex,below right=of node3.south west,anchor=north west] (node4) {\begin{tikzpicture}
        \node[] (Graph) at (0,0) {\eqboxwh{RunningExampleReducedConditionHeightLine5}{\begin{tikzpicture}[mtgc_inner]
			\node[styleNODEMessage] (NodeMessage1) at (0,0) {\NODEMessageX{name=M,cts=\CTSvar{M},dts=\DTSvar{M},cidx=\CIDXvar{M},didx=\DIDXvar{M}}};
			\GraphEdgeLoop{E}{NodeMessage1}{styleEDGEdone}{LAB}{\EDGEdoneX{name=e_3,cts=\CTSvar{e_3},dts=\DTSvar{e_3},cidx=\CIDXvar{e_3},didx=\DIDXvar{e_3}}}{NodeRouter1}
            \node[constraintNode,node distance=1ex,right=of LAB] (Constraint) {$\Theta_4$};
            \end{tikzpicture}}};
        \OPENINGBRACKET{Graph}
        \node[outer xsep=0ex,inner xsep=0pt,node distance=0ex,right=of Graph,anchor=west] (RIGHT) 
            {\large\strut$\COMMA\MTGCtrue$};
        \node[outer xsep=0ex,inner xsep=0pt,node distance=.8ex,left=of Graph,anchor=east] (LEFT9) 
            {\large\strut$\MTGCexistsNAME$};
        \CLOSINGBRACKET{Graph}{RIGHT}{}
        \CLOSINGBRACKET[1ex]{Graph}{RIGHT}{}
        \CLOSINGBRACKET[2ex]{Graph}{RIGHT}{}
        \CLOSINGBRACKET[3ex]{Graph}{RIGHT}{}
        \CLOSINGBRACKET[4ex]{Graph}{RIGHT}{}
        \end{tikzpicture}};

    \begin{pgfonlayer}{background}
    \node (GRAPHL) [fill=black!5!white,fit=(node0) (node1) (node2) (node3) (node4)] {};
    \end{pgfonlayer}

    \node[node distance=2ex,below=of GRAPHL] {$\begin{array}{ll}
        \Theta_0=\{x_{t,0}=0,x_{s,0}=0\}\\
        \Theta_1=\Theta_0\cup\{
            x_{t,0}<x_{t,1}\vee (x_{t,0}=x_{t,1}\wedge x_{s,0}<x_{s,1}),\\
        \phantom{\Theta_1=\Theta_0\cup\{}
            \OPalive{(x_{t,1},x_{s,1})}{\{S,R_1,M,e_1,e_2\}},
            \OPearliest{(x_{t,1},x_{s,1})}{\{S,R_1,M,e_1,e_2\}}
            \}\\						
        \Theta_2=\Theta_0\cup\{
            \OPalive{(x_{t,1},x_{s,1})}{\{M\}}
            \}\\
        \Theta_3=\Theta_2\cup\{
            x_{t,1}<x_{t,2}\vee (x_{t,1}=x_{t,2}\wedge x_{s,1}<x_{s,2}),
            x_{t,2}\leq x_{t,1}+5
        \}\\
        \Theta_4=\Theta_3\cup\{
            \OPalive{(x_{t,2},x_{s,2})}{\{M,e_3\}}
            \}
    \end{array}$};
    \end{tikzpicture}
}

 \end{myfigureWITHBOX}
\begin{myfigureWITHBOX}{Visualization for step 5 of our \gls{BMC} approach:
    AC-restriction of violations (result of evaluating the
     negation of the \gls{GC} $\GCvar$ from \myref{figure:running_example:encoded_MTGC}
     for the \gls{GH} $G_H$ from \myref{figure:running_example:graph_with_history}).
}{running_example:step5}\raggedright
\begin{align*}
&\hspace*{0ex}        \neg\exists x_{t,0}{:}\TypeReal,x_{s,0}{:}\TypeInteger.\;\\
&\hspace*{2ex}            x_{t,0}=0\wedge x_{s,0}=0\\
&\hspace*{2ex}            \wedge\forall x_{t,1}{:}\TypeReal,x_{s,1}{:}\TypeInteger.\;\\
&\hspace*{4ex}                 x_{t,0}<x_{t,1}\vee (x_{t,0}=x_{t,1}\wedge x_{s,0}<x_{s,1})\\
&\hspace*{4ex}                 \wedge\\
&\hspace*{4ex}                 \OPalive{(x_{t,1},x_{s,1})}{\{S,R_1,M_1,e_1,e_8\}}\\
&\hspace*{4ex}                 \wedge\\
&\hspace*{4ex}                 \OPearliest{(x_{t,1},x_{s,1})}{\{S,R_1,M_1,e_1,e_8\}}\\
&\hspace*{4ex}                 \rightarrow\exists x_{t,2}{:}\TypeReal,x_{s,2}{:}\TypeInteger.\;\\
&\hspace*{6ex}                        x_{t,1}<x_{t,2}\vee (x_{t,1}=x_{t,2}\wedge x_{s,1}<x_{s,2})\\
&\hspace*{6ex}                        \wedge\\
&\hspace*{6ex}                        x_{t,2}\leq x_{t,1}+5\\
&\hspace*{6ex}                        \wedge\\
&\hspace*{6ex}                        \OPalive{(x_{t,2},x_{s,2})}{\{M_1,e_{11}\}}\\
\intertext{Intuitively, this expression captures an untimely reception in the sense of:}
&(\var{tpc}_4-\var{tpc}_0)> (\var{tpc}_1-\var{tpc}_0)+5
\intertext{or even in the simplest form:}
&\var{tpc}_4> \var{tpc}_1+5
\end{align*}
Technically, it refers to all attributes of the \gls{GH} $G_H$ (in the \OPalive{}{} and \OPearliest{}{} ACs), which makes the usage of $G_H$ in step~7 necessary. \end{myfigureWITHBOX}
\begin{myfigureWITHBOX}{Visualization for step 6 of our \gls{BMC} approach:
    Zone-restriction of violations (result for the structural path $\pi$ from \myref{figure:running_example:structural_path}).
}{running_example:step6}\raggedright
\begin{align*}
&\var{tpc_1}-\var{tpc_0}\geq 0\\
&\wedge \var{tpc_2}-\var{tpc_1}\geq 2\\
&\wedge \var{tpc_2}-\var{tpc_1}\leq 5\\
&\wedge \var{tpc_3}-\var{tpc_2}\geq 2\\
&\wedge \var{tpc_3}-\var{tpc_2}\leq 5\\
&\wedge \var{tpc_4}-\var{tpc_3}\leq 0\\
&\wedge c_1\geq 0
\end{align*}
Intuitively, the guards and invariants stated for the clock of the message result in a restriction of the time point clock variables.
 \end{myfigureWITHBOX}
\begin{myfigureWITHBOX}{Visualization for step 7 of our \gls{BMC} approach:
    Derivation of labeling.
}{running_example:step7}\raggedright
For the case of \PTCTLmax{\MTGCvar}, we construct the AC $\ACvar_{\var{check}}$ using the AC from \myref{figure:running_example:step5}, the AC from \myref{figure:running_example:step6}, and the AC of the \gls{GH} from \myref{figure:running_example:graph_with_history} (given by the conjunction of all ACs contained in the graph).
$\ACvar_{\var{check}}$ is equivalent to the following simplified AC.
\begin{align*}
\ACvar_{\var{check}}\quad \equiv\quad 4\leq \var{tpc}_4-\var{tpc}_1\leq 10
\wedge
\neg(\var{tpc}_4>\var{tpc}_1+5)
\end{align*}
This AC $\ACvar_{\var{check}}$ is satisfiable.
In fact, it is satisfied by the clock valuation $
\{
\var{tpc}_1\mapsto 0,
\var{tpc}_4\mapsto 4\}$ describing the fastest transmission of the message $M_1$.
From the satisfiability, we obtain the labeling of $G_H$ from \myref{figure:running_example:graph_with_history} using the APs $\var{success}$ and $\var{maybe}$.
 \end{myfigureWITHBOX}
\begin{myfigureWITHBOX}{Visualization for step 8 of our \gls{BMC} approach:
    Derivation of probabilities.
}{running_example:step8}\raggedright
The probability maximizing adversary, will find at least the path to the location given by the GH $G_H$ from \myref{figure:running_example:graph_with_history}.
This path has a probability of $1\times 0.8\times 0.8\times 1$ and is labeled with the APs $\var{success}$ and $\var{maybe}$.
\toolPRISM returns the probability interval $I=\INTERVALcc{0.64}{0.64}$ since all other paths will not be labeled with one of these APs because the timing constraint of at most $5$ time units from the PMTGC \PMTGCexample is not satisfied by the other paths.
 \end{myfigureWITHBOX}

\clearpage
\newpage
\section{Example for Step 7 of the BMC Approach}

In this appendix, we provide a short example on why step 7 is defined as described.
For this purpose, we consider different combinations of zone-restric\-tions and AC\hyph{}restrictions for the two cases of \PTCTLmax{\MTGCvar} and \PTCTLmin{\MTGCvar}.
\begin{myexample}{Computation of Labeling in Step 7}{computation_of_labeling}
We consider a zone-restriction $4\leq x\leq 10$ as well as AC-restrictions $x\geq 3$, $x\geq 5$, and $x\geq 12$.
For the two cases from above, we then determine whether the corresponding leaf-location should be labeled with $\var{success}$ and $\var{maybe}$.
\begin{center}
\begin{tabular}{lll}
\toprule
\OPmaxNAME\\
$(4\leq x\leq 10)\wedge \neg(x\geq 3)$&is unsatisfiable,&hence no labeling\\
$(4\leq x\leq 10)\wedge \neg(x\geq 5)$&is satisfiable,&hence labeling with $\{\var{success},\var{maybe}\}$\\
$(4\leq x\leq 10)\wedge \neg(x\geq 12)$&is satisfiable,&hence labeling with $\{\var{success},\var{maybe}\}$\\
\midrule
\OPminNAME\\
$(4\leq x\leq 10)\wedge (x\geq 3)$&is satisfiable,&hence no labeling\\
$(4\leq x\leq 10)\wedge (x\geq 5)$&is satisfiable,&hence no labeling\\
$(4\leq x\leq 10)\wedge (x\geq 12)$&is unsatisfiable,&hence labeling with $\{\var{success},\var{maybe}\}$\\
\bottomrule
\end{tabular}
\end{center}
For the case of \PTCTLmax{\MTGCvar}, satisfiability means that some interleaving with timed steps does not result in a violation.\\
For the case of \PTCTLmin{\MTGCvar}, unsatisfiability means that each interleaving with timed steps does not result in a violation.
\end{myexample}

\end{document}